\pdfoutput=1
\RequirePackage{ifpdf}
\ifpdf % We~are running pdfTeX in pdf mode
\documentclass[pdftex]{sigma}%,draft
\else
\documentclass{sigma}
\fi

\usepackage{tikz-cd,tikz}

\usepackage[bbgreekl]{mathbbol}

\DeclareFontFamily{OT1}{pzc}{}
\DeclareFontShape{OT1}{pzc}{m}{it}{<-> s * [1.200] pzcmi7t}{}
\DeclareMathAlphabet{\mathpzc}{OT1}{pzc}{m}{it}

\newcommand{\Id}{\mathbb{d}}
\newcommand{\Iomega}{\bbomega}

\numberwithin{equation}{section}

\begin{document}
%\allowdisplaybreaks

\newcommand{\arXivNumber}{1904.07578}

\renewcommand{\PaperNumber}{126}

\FirstPageHeading

\ShortArticleName{Small Gauge Transformations and Universal Geometry in Heterotic Theories}

\ArticleName{Small Gauge Transformations\\ and Universal Geometry in Heterotic Theories}

\Author{Jock MCORIST~$^\dag$ and Roberto SISCA~$^\ddag$}

\AuthorNameForHeading{J.~McOrist and R.~Sisca}

\Address{$^\dag$~Department of Mathematics, School of Science and Technology,\\
\hphantom{$^\dag$}~University of New England, Armidale, 2351, Australia}
\EmailD{\href{mailto:jmcorist@une.edu.au}{jmcorist@une.edu.au}}
%\URLaddressD{\url{https://www.une.edu.au/staff-profiles/science-and-technology/jock-mcorist}}

\Address{$^\ddag$~Department of Mathematics, University of Surrey, UK}
\EmailD{\href{mailto:roberto.sisca@surrey.ac.uk}{roberto.sisca@surrey.ac.uk}}

\ArticleDates{Received July 30, 2020, in final form November 04, 2020; Published online December 02, 2020}

\Abstract{The first part of this paper describes in detail the action of small gauge transformations in heterotic supergravity. We show a convenient gauge fixing is `holomorphic gauge' together with a condition on the holomorphic top form. This gauge fixing, combined with supersymmetry and the Bianchi identity, allows us to determine a set of non-linear PDEs for the terms in the Hodge decomposition. Although solving these in general is highly non-trivial, we give a~prescription for their solution perturbatively in $\alpha^{\backprime}$ and apply this to the moduli space metric. The second part of this paper relates small gauge transformations to a choice of connection on the moduli space. We show holomorphic gauge is related to a~choice of holomorphic structure and Lee form on a `universal bundle'. Connections on the moduli space have field strengths that appear in the second order deformation theory and we point out it is generically the case that higher order deformations do not commute.}


\Classification{53B50; 14D21; 58D27; 83E30}

\section{Introduction}

We continue a programme of work developed in a recent series of papers \cite{Candelas:2016usb,Candelas:2018lib,McOrist:2016cfl} studying the moduli space $M$ of heterotic vacua realising $N=1$ supersymmetry in a ${\mathbb R}^{1,3}$ spacetime. In \cite{Candelas:2016usb} the natural K\"ahler metric for $M$ was constructed using a set of covariant derivatives; in \cite{Candelas:2018lib} it was realised these derivatives implied a natural geometric construction, known as a universal bundle, in which the geometric data of a heterotic vacuum was fibered over the moduli space. The purpose of this paper is two-fold: first to clarify and resolve certain confusions regarding the role of small gauge transformations in heterotic supergravity; and second to relate those outcomes to the universal bundle constructed in~\cite{Candelas:2018lib}. Our analysis is local in nature in the spirit of the earlier papers such as~\cite{Candelas:1990pi}.

We work at large radius, in which the supergravity approximation is valid and so take the heterotic flux $H$ to be subleading in $\alpha^{\backprime}$. These theories are defined by a complex $3$-fold $X$ with $c_1(X) = 0$ and a Hermitian form $\omega$ as well as a holomorphic vector bundle ${\mathcal E}\to X$ with a~connection $A$ satisfying the Hermitian--Yang--Mills (HYM) equation and a~well-defined three-form~$H$. The anomaly relation yields a modified Bianchi identity for $H$,
\begin{gather}
 {\rm d} H =- \tfrac{\alpha^{\backprime}}{4} \big( \operatorname{Tr} \big(F^2\big) - \operatorname{Tr} \big(R^2\big) \big),\label{eq:Anomaly0}
\end{gather}
where $\operatorname{Tr} F^2$ is evaluated in the adjoint representation of $E_8\times E_8$ using a particular normalisation so that the Bianchi identity is compatible with analogous expressions for the ${\rm SO}(32)$ heterotic string.\footnote{That is $\operatorname{Tr} F^2 = \frac{1}{30}\operatorname{Tr}_a F^2$ where $\operatorname{Tr}_a$ is evaluated in the 496-dimensional adjoint representation of $E_8\times E_8$. } This also has the benefit that anomaly cancellation amounts to characteristic classes being identically equal. Our results here apply equally for both strings but for concreteness focus on the $E_8\times E_8$ string. The term $\operatorname{Tr} R^2$ is evaluated in the vector representation of the Lorentz algebra, which for the complex manifolds considered here reduces to ${\rm SU}(3)$. $R$ is the curvature two-form for a connection twisted in a certain way by the field strength $H$ and so the Bianchi identity is a highly non-linear coupled PDE. Supersymmetry implies the manifold is non-K\"ahler\footnote{In the above equations~$R$ is the curvature two-form evaluated with the Hull connection $\Theta^{\rm H} = \Theta^{\rm LC} + \tfrac{1}{2} H$. We denote by $x^m$ the real coordinates of $X$ and its complex coordinates by $\big(x^\mu, x^{\overline\nu}\big)$. The coordinates along~$M$ are denoted by $y^a$, and complex coordinates by $\big(y^\alpha, y^{\overline\beta}\big)$. In the following we will generally omit the wedge product symbol `$\wedge$' between forms, unless doing so would lead to ambiguity.}
\begin{gather*}
 H = {\rm d}^c \omega,\qquad {\rm d}^c \omega = \tfrac{1}{3!} J^m J^n J^p ({\rm d} \omega)_{mnp},
%\label{eq:SusyRelation0}
\end{gather*}
where $J^m = J_n{}^m {\rm d} x^n$ and $J_n{}^m$ is the complex structure of $X$. As this is a large radius expansion of string theory, we have $H=\mathcal{O}(\alpha^{\backprime})$ and so by \cite{Anguelova:2010ed} the background value of the dilaton can be gauged fixed to be a constant up to and including ${\alpha^{\backprime}}^2$ corrections (see Appendix~\ref{s:dilatongauge} for a~summary of this calculation). The data defining heterotic solutions of this type we call a~heterotic structure~$\mathsf{Het}$ and write it as a~tuple $\mathsf{Het} = ([X,\omega,\Omega], [{\mathcal E},A], [{\mathcal T}_X,\Theta], H)$. The heterotic structure includes the connections $A$ and $\Theta$ on the bundles ${\mathcal E}$ and ${\mathcal T}_X$ respectively as well as the Hermitian form $\omega$ and the complex structure~$J$ of~$X$, or equivalently the holomorphic $(3,0)$-form~$\Omega$.

For $X$ and ${\mathcal E}$ having a fixed topology, solutions to the equations of $\alpha^{\backprime}$-corrected heterotic supergravity come with parameters. These are interpreted as coordinates for the moduli space~$M$ of heterotic theories. Each point in~$M$ ought to correspond to a unique heterotic structure~$\mathsf{Het}$. In order for this to be case the case we need to understand the role of gauge symmetries. This is because, roughly speaking, the moduli space is the quotient of the space of solutions to the equations of motion by the action of gauge symmetries. This quotient, in physics, is realised by gauge fixing. The first part of this paper is concerned with understanding this gauge fixing, of which, while bits and pieces appear in the literature, a systematic description is lacking. This is despite the fact the quotient is needed in order to define~$M$.

The gauge symmetries derive from diffeomorphisms of~$X$, `gerbe' transformations of the B-field and gauge transformations of the gauge field~$A$. As we study these theories at large radius and constant dilaton, we can use the language of Wilsonian effective field theory to study the underlying string theory. The result is an $\alpha^{\backprime}$-corrected supergravity theory which is fixed up to and including ${\alpha^{\backprime}}^2$. As noted in \cite{Candelas:2016usb,Candelas:2018lib,McOrist:2016cfl}, within this effective field theory it is convenient to use the background gauge principle. Gauge symmetries are divided into background gauge transformations and small gauge transformations which we describe further below. The role of background gauge transformations was studied in~\cite{Candelas:2016usb} and are accounted for by the introduction of covariant derivatives
\[
\delta A = \delta y^a \mathfrak{D}_a A, \qquad \mathfrak{D}_a A = \partial_a A - {\rm d}_A \Lambda_a,
\]
where $\partial_a A$ is a partial derivative with respect to a parameter $y^a$, $\Lambda=\Lambda_a {\rm d} y^a$ is a connection on $M$ that transforms in a manner parallel to $A$ and ${\rm d}_A \Lambda_a = {\rm d} \Lambda_a + [A, \Lambda_a]$. When this is the case, $\delta A$~transforms in the correct manner. Small gauge transformations act on the deformations of fields and are to be quotiented by in constructing the moduli space~$M$. In particular, they are needed in order to define the relation between coordinates of~$M$ and deformations of the background fields underlying $(X,{\mathcal E}, H)$. The first part of this paper (Sections~\ref{s:smallTransformations}--\ref{s:Hodge}) is concerned with writing down the action of small gauge transformations and describing a complete gauge fixing. The gauge fixing we describe is termed `holomorphic gauge' and is natural from the point of view of supersymmetry.

The second part of this paper (Sections~\ref{s:firstorderUG} and~\ref{s:Second}) builds on the universal bundle constructed in~\cite{Candelas:2018lib}. A key point of the universal bundle is that it geometrises the gauge symmetries and deformations. One presentation of the universal bundle ${\mathcal U}$ (it is really a fibration) is that for each point $y\in M$ in the moduli space we have the total space of a holomorphic vector bundle ${\mathcal E} \to X$ with additional structures such as $\omega$, $\Omega$, $H$ and a connection~$A$ satisfying the string theory equations of motion and Bianchi identity. That is, ${\mathcal U}$ is a double fibration. A presentation of this is as a fibration of holomorphic vector bundles over the moduli space
\begin{equation*}%\label{eq:HeteroticFamily}
\begin{tikzcd}
{\mathcal E} \arrow[r] & {\mathcal U} \arrow[d]\\
& M.
\end{tikzcd}
\end{equation*}
We could denote the fiber, as in~\cite{Candelas:2018lib}, by $\mathsf{Het}$ to remind ourselves that the fibration includes not just the manifold structure of~$E$ but the structures that come with it such as complex structure, Hermitian structure, $H$~and so on.

The fact this is a double fibration lends itself to a second presentation. Define the fibration~${\mathbb X}$ whose fibers are the manifolds~$X$, together with their metric and complex structure and base manifold the moduli space~$M$. This fibration admits an Ehresmann connection $c$ which accounts for the variation of~$X$ over~$M$. The universal bundle is ${\mathcal U}$ whose fibers are the Lie algebra ${\mathfrak g}$ and base manifold ${\mathbb X}$. In fact, more general things are possible. ${\mathcal U}$ could have a structure algebra ${\mathfrak g}_{\mathcal U}$ which, at the very least, contains the structure algebra ${\mathfrak g}$ of ${\mathcal E}$ as a subalgebra and for technical reasons satisfies $[{\mathfrak g}_{\mathcal U}, {\mathfrak g}] = {\mathfrak g}$ where the bracket is the usual Lie algebra bracket. A simple way to satisfy these constraints is to take ${\mathfrak g}_{\mathcal U} = {\mathfrak g} \oplus {\mathfrak g}_b$ for some real Lie algebra ${\mathfrak g}_b$. More complicated things are possible. An example of~${\mathcal U}$ is the tangent bundle ${\mathcal T}_{\mathbb X}$. Its structure algebra contains~${\mathfrak{su}}(3)$ for the manifold $X$ while ${\mathfrak g}_b = {\mathfrak u}(d)$ where $d$ is the complex dimension of the moduli space. This example, in fact, shows that in general ${\mathfrak g}_{\mathcal U}$ is not a simple direct product but is upper triangular. For the purposes of this paper, however, the choice of ${\mathfrak g}_b$ does not appear to affect any physical results. At this point, we could take it to be trivial. We leave studying these possibilities for future work; for now we refer to the structure group as~${\mathfrak g}_{\mathcal U}$.

Diagrammatically, the fibration ${\mathcal U}$ can now be written so that the base manifold is ${\mathbb X}$ and the fibres are ${\mathfrak g}_{\mathcal U}$:
\begin{equation*}%\label{eq:HeteroticFamily2}
\begin{tikzcd}
{\mathfrak g}_{\mathcal U} \arrow[r] & {\mathcal U} \arrow[d]\\
& {\mathbb X}.
\end{tikzcd}
\end{equation*}

The global properties of ${\mathcal U}$ are unexplored, however its local differential geometric structure is described in~\cite{Candelas:2018lib}. A key point is that tensors and structures on $X$ are extended to be defined on ${\mathbb X}$ and structures on ${\mathcal E}$ are extended to structures on ${\mathcal U}$. An important example is the connection~$A$ on the bundle ${\mathcal E}$. We take the connection $A = A_m {\rm d} x^m$ on $X$ and pair it with the connection $\Lambda = \Lambda_a {\rm d} y^a $ defined on $M$ to form a `bigger' connection ${\mathbb A} = A + \Lambda$. This is a connection for ${\mathcal U}$. Demanding it is compatible with the connection $c$, that is the fibration structure of ${\mathbb X}$ is solved by taking $A$ to be a ${\mathfrak g}$-connection and $\Lambda$ a ${\mathfrak g}_{\mathcal U}$ connection. When things are unified in this way, nice things happen. For example, the field strength ${\mathbb F}$ for ${\mathbb A}$ has a mixed term ${\mathbb F}_{am} {\rm d} x^m = \mathfrak{D}_a A$ which contains exactly the covariant derivative defined above. We refer the reader to~\cite{Candelas:2018lib} for further details. In this paper, what we add is the observation that the choice of $\Lambda$ is related to the choice of small gauge fixing. For example, a small deformation $\Lambda_a \to \Lambda_a - \phi_a$ for some adjoint valued~$\phi_a$ results in $\mathfrak{D}_a A \to \mathfrak{D}_aA + {\rm d}_A \phi_a$ which is exactly a small gauge transformation. Conversely, a small gauge fixing corresponds to picking a connection on the moduli space. In Sections~\ref{s:firstorderUG} and~\ref{s:Second} we describe how this works for all the symmetries discussed in Sections~\ref{s:smallTransformations}--\ref{s:Hodge}. We then show, for example, that holomorphic gauge corresponds to a choice of holomorphic structure and Lee form on ${\mathcal U}$. We also point out that in the context of this formalism second order deformations do not generally commute. In particular, the commutator of deformations $[\mathfrak{D}_a, \mathfrak{D}_b] A$ is related to a field strength on~$M$ and that generically this is non-vanishing. It would be fascinating to determine whether there are cases in which we can find a connection on the moduli space that is flat. This is presumably related to the definition of heterotic special geometry, though we postpone this to future work.

The outline for the paper is the following. In Section~\ref{s:smallTransformations}, we describe the complete action of small diffeomorphisms, small gauge transformations and small gerbe transformations on variations of all the fields underlying a heterotic theory. In Section~\ref{s:Gauge} we describe how to fix the action of this symmetry to holomorphic gauge. There is a residual gauge fixing which is related to variations of the holomorphic $(3,0)$-form $\Omega$. In Section~\ref{s:Hodge}, we substitute these results into the supersymmetry equations, Bianchi identity, Hermitian--Yang--Mills equation and balanced equation. This results in a set of Poisson equations which we are not able to explicitly solve due to their non-linear nature. Nonetheless we are able to make some conclusions about solutions in the large radius limit. In Section~\ref{s:firstorderUG}, we build on earlier results to relate the choice of gauge fixing to the universal bundle construction showing holomorphic gauge corresponds to a choice of holomorphic structure. In Section~\ref{s:Second}, we give a flavour of second order deformation theory in heterotic theories, with a complete analysis postponed to future work.

\section{Small gauge transformations and heterotic moduli}\label{s:smallTransformations}

We derive the action of small diffeomorphisms, gauge transformations and small gerbe transformations.
Suppose on $X$ we have a tensor field $T_{m_1\cdots m_p} $ which can undergo a diffeomorphism
\begin{gather*}%\label{eq:DiffonTensor}
T_{n_1\dots n_k}(\widetilde{x}) \frac{\partial
\widetilde{x}{}^{n_1}}{\partial x^{m_1}}\cdots\frac{\partial\widetilde{x}{}^{n_k}}{\partial x^{m_k}} = T_{m_1\dots m_k}(x).
\end{gather*}
For $\widetilde{x} = x + \varepsilon$ with $\varepsilon^m$ small
\begin{gather}
 T_{m_1\dots m_k}(\widetilde{x}) \simeq T_{m_1\dots m_k}(x) + \varepsilon^n \partial_n T_{m_1\dots m_k}(x) + (\partial_{m_1} \varepsilon^n) T_{n\dots m_k}(x) + \cdots + (\partial_{m_k} \varepsilon^p )T_{m_1\dots p}(x) \nonumber\\
\hphantom{T_{m_1\dots m_k}(\widetilde{x})}{} = T_{m_1\dots m_k}(x) + ({\mathcal L}_\varepsilon T)_{m_1\dots m_k}(x) ,\label{eq:SmallDiffonT}\raisetag{.7cm}
\end{gather}
where ${\mathcal L}_{\varepsilon}$ is the Lie derivative taken with respect to the vector $\varepsilon^m$. We have worked to first order in $\varepsilon$. When studying deformations $\delta T$ of the tensor, the small diffeomorphisms are regarded as unphysical and so in the physical theory we identify
\begin{gather}\label{eq:SmallDiffondT}
\delta T \sim \delta T + {\mathcal L}_{\varepsilon} T.
\end{gather}
Appendix~\ref{app:CYGaugeFix} carefully explains the gauge fixing of diffeomorphisms in the study of the moduli space of Calabi--Yau manifolds in three different ways. We now describe how this works for fields of heterotic to first order in deformations.

\subsection[Complex structure $J$]{Complex structure $\boldsymbol{J}$}

On $X$ there is an integrable complex structure $J$ and this facilitates introducing holomorphic coordinates $x^m = \big(x^\mu, x^{\overline\nu}\big)$. Deformations that modify complex structure can be expressed in terms of the undeformed complex structure as
\[
\delta J = \delta J_{\overline\nu}{}^\mu \,{\rm d} x^{\overline\nu}\otimes \partial_\mu + \delta J_\nu{}^{\overline{\mu}} \,{\rm d} x^\nu \otimes\partial_{\overline{\mu}} .
\]
To first order in deformation theory, we demand the Nijenhuis tensor is preserved. Decomposing into type we get
\[
{\overline{\partial}}( \delta J{}^\mu ) = 0, \qquad \partial \big( \delta J{}^{\overline{\mu}} \big) = 0,
\]
where the notation here is as in the introduction $\delta J^\mu = \delta J_{\overline\nu}{}^\mu {\rm d} x^{\overline\nu}$.
Small diffeomorphisms induce an identification $\delta J \sim \delta J + {\mathcal L}_\varepsilon J$ where because $\delta J$ is a real tensor, the vector $\varepsilon$ must also be real. This becomes
\[
\delta J{}^\mu \sim \delta J{}^\mu + 2{\rm i} \overline{\partial} \varepsilon^\mu \qquad \text{and} \qquad \delta J{}^{\overline{\mu}} \sim \delta J{}^{\overline{\mu}} -2{\rm i} \partial \varepsilon^{\overline{\mu}}.
\]
Hence, $\delta J_{{\overline\nu}}{}^\mu\, {\rm d} x^{\overline\nu} \in H_{\overline{\partial}}^{(0,1)} \big(X, {\mathcal T}_X^{(1,0)}\big)$. We expand in a basis for the cohomology group and this defines the Kodaira--Spencer map\footnote{This can also be referred to as a Kuranishi map in the literature.} \cite{Kodaira:1981cx} between 1-forms ${\mathcal T}^*_M$ and field variations
\[
\delta J_{{\overline\nu}}{}^\mu = \delta y^a \big( 2{\rm i} \Delta_{a {\overline\nu}}{}^\mu\big).
\]
The parameters, collectively denoted $y^a$, are also coordinates for a manifold $M$, the moduli space. Once we have fixed small diffeomorphisms -- an issue we will return later~-- a small deformation of parameters $y^a \to y^a+ \delta y^a$ corresponds to a small deformation of fields and this relation is called the Kodaira--Spencer map. In this case, it is a deformation of complex structure $J \to J + \delta J$. As the $y^a$ are also coordinates for a manifold, the $\delta y^a$ form a coordinate basis for~${\mathcal T}^*_M$ in the infinitesimal limit. The pairing $\delta y^a \Delta_a$ is then a first order deformation of complex structure. This will become more complicated once we introduce gauge symmetries.

\subsection[Holomorphic $(3,0)$-form $\Omega$]{Holomorphic $\boldsymbol{(3,0)}$-form $\boldsymbol{\Omega}$}\label{s:holomorphicform}

Consider the ${\rm d}$-closed holomorphic $(3,0)$-form $\Omega$. A first order deformation $\delta \Omega$ obeys three equations
\[
\partial \delta \Omega^{(3,0)} = 0, \qquad {\overline{\partial}}\delta \Omega^{(3,0)} + \partial \delta \Omega^{(2,1)} = 0, \qquad {\overline{\partial}}\delta \Omega^{(2,1)} = 0.
\]
These equations are solved by
\begin{gather}\label{eq:OmPureVarn}
\delta \Omega^{(3,0)} = \delta y^a\big( k_a \Omega + \partial \xi^{(2,0)}_a\big) ,\qquad \delta \Omega^{(2,1)} = \delta y^a \chi_a, \qquad \partial \chi_a = - {\overline{\partial}} \partial\xi_a^{(2,0)},
\end{gather}
where $ \xi^{(2,0)}_a$ are $(2,0)$-forms, the $k_a$ are constant over $X$ and $\chi_a=\Delta_a{}^\mu \Omega_\mu$ are ${\overline{\partial}}$-closed $(2,1)$-forms. The variation $\delta \Omega^{(2,1)}$ is related to a variation of complex structure up to a parameter dependent rescaling~\cite{Candelas:1990pi}.

At this point it is convenient to describe a two-parameter family of connections on ${\mathcal T}_X$ given as follows
\begin{gather*}%\label{eq:GammaFamily}\notag
 \Theta^{(\epsilon,\rho)}{}_\mu{}^\nu{}_\sigma = \Theta^{\rm LC}{}_\mu{}^\nu{}_\sigma+\frac{(\epsilon-\rho)}{2} H_{\mu}{}^\nu{}_\sigma,\\
 \Theta^{(\epsilon,\rho)}{}_\mu{}^{\overline\nu}{}_{\overline{\sigma}} = \Theta^{\rm LC}{}_\mu{}^{\overline\nu}{}_{\overline{\sigma}}+\frac{(\epsilon-\rho)}{2} H_{\mu}{}^{\overline\nu}{}_{\overline{\sigma}},\\
 \Theta^{(\epsilon,\rho)}{}_\mu{}^{\overline\nu}{}_\sigma = 0,\\
 \Theta^{(\epsilon,\rho)}{}_\mu{}^\nu{}_{\overline{\sigma}} = \Theta^{\rm LC}{}_\mu{}^\nu{}_{\overline{\sigma}}+\frac{(\epsilon+\rho)}{2} H_\mu{}^\nu{}_{\overline{\sigma}},
 \end{gather*}
where $\Theta^{\rm LC}$ is the Levi-Civita connection. The Bismut connection is given by $\Theta^{\rm B} = \Theta^{(-1,0)}$, the Hull connection by $\Theta^{\rm H} = \Theta^{(1,0)}$ and the Chern connection by $\Theta^{\rm Ch} = \Theta^{(0,-1)}$.

Recall that we also assume a $g_s$-perturbative string background at large radius in the $\alpha^{\backprime}$-expansion and so by a suitable choice of gauge fixing for the metric supersymmetry implies the dilaton is to constant to ${\alpha^{\backprime}}^3$. We summarise this calculation in Appendix~\ref{s:dilatongauge}. It then follows, by supersymmetry, that $\nabla^{\rm B}_\mu \Omega = \nabla^{\rm B}_{\overline{\mu}} \Omega = 0$ and
\[
H_{\mu\nu}{}^\nu = 0, \qquad \partial_\mu \|\Omega\|^2 = 0.
\]
Using $H = {\rm d}^c \omega$, the first condition means that the Lee form $W(\omega) = \tfrac{1}{2}\omega^{mn}({\rm d} \omega)_{mn}$ vanishes. The Lee form is a one-form measuring the non-primitive part of ${\rm d} \omega$ and manifolds with $W(\omega) = 0$ are called balanced manifolds.

A small diffeomorphism acts as on the $(3,0)$-component\footnote{Due to $H_{\mu\nu}{}^\nu = 0$ divergences of this vector with respect to Levi-Civita and Bismut coincide $\nabla^{\rm LC}_\nu \varepsilon^\nu =\nabla_\nu^{\rm B} \varepsilon^\nu$. Furthermore, with the choice of $\varepsilon^\nu$ we find
\[
\nabla^{\rm LC}_\nu \varepsilon_a{}^\nu=u -\frac{1}{3! \|\Omega\|^2} {\overline{\Omega}}^{\nu\rho\sigma}\big(\partial\xi_a^{(2,0)}\big)_{\nu\rho\sigma}.
\]}
\[
\delta \Omega^{(3,0)} \to \delta \Omega^{(3,0)} + \partial \big(\varepsilon^\nu \Omega_\nu\big) = \delta \Omega^{(3,0)} + \big(\nabla^{\rm LC}_\nu \varepsilon^\nu\big) \Omega.
\]

\subsection[The Hermitian form $\omega$]{The Hermitian form $\boldsymbol{\omega}$}

There is a compatible Hermitian form $\omega$. A real deformation of $\omega$ can be written as
\begin{gather*}
\delta \omega^{(2,0)} = \delta y^a \Delta_a{}^{\overline{\mu}} \omega_{\overline{\mu}}, \qquad \delta \omega^{(1,1)} = \delta y^a (\partial_a \omega)^{(1,1)} , \qquad \delta \omega^{(0,2)} = \delta y^a \Delta_a{}^\mu \omega_{\mu} , \\ \omega_m = \omega_{mn} {\rm d} x^n.
\end{gather*}
It is subject to small diffeomorphisms \eqref{eq:SmallDiffondT}, taking the form
\[
\delta \omega \sim \delta \omega + {\mathcal L}_{\varepsilon} \omega = \delta \omega + \varepsilon{}^m ({\rm d} \omega)_m + {\rm d} \big(\varepsilon{}^m \omega_m\big),
\]
where $\varepsilon$ is a vector on $X$ generating the small diffeomorphism.
If ${\rm d} \omega = 0$ the manifold is K\"ahler and small diffeomorphisms generate ${\rm d}$-exact shifts of $\delta \omega$. In heterotic theories this is not the case.

\subsection[The gauge field $F$]{The gauge field $\boldsymbol{F}$}

The gauge field $A$ and its field strength $F = {\rm d} A + A^2$ also transform under small diffeomorphisms. However this case is complicated by gauge symmetries in which ${}^\Phi F = \Phi F \Phi^{-1}$ and $A$ transforms according to
\begin{gather}
 A\to {}^\Phi A = \Phi A \Phi^{-1} - ({\rm d}\Phi) \Phi^{-1}.\label{eq:Atransf}\end{gather}
So in relating $F( x+ \varepsilon)$ and $F(x)$ we need a covariant generalisation of a Lie derivative, \eqref{eq:SmallDiffonT}, which is
\begin{gather*}%\label{eq:LieDerivF}
F(\widetilde{x}) \simeq F(x) + \varepsilon^m ({\rm d}_A F)_{m} + {\rm d}_A \big(\varepsilon^m F_m\big).
\end{gather*}
Using the Bianchi identity we find an identification
\begin{gather}
\delta F(x) \sim \delta F(x) + {\rm d}_A\big(\varepsilon^m F_m\big).\label{eq:dFvarn}
\end{gather}
We write ${\mathcal A} = A^{(0,1)}$ so that $ A = {\mathcal A} - {\mathcal A}^\dag$. Holomorphy of ${\mathcal E}$ means $F^{(0,2)} = {\overline{\partial}}_{\mathcal A}^{\,2} = 0$. Taking the~$(0,2)$ part of \eqref{eq:dFvarn} and solving for $\delta {\mathcal A}$ gives
\[
\delta {\mathcal A} \sim \delta {\mathcal A} + \varepsilon^\mu F_\mu + {\overline{\partial}}_{\mathcal A} \phi,
\]
where $\phi$ is a section of $\operatorname{End}{\mathcal E}$. In principle the last term could be closed but not exact. However, this would represent a change in moduli space coordinates (see below). This term is interpreted as an independent gauge symmetry, small gauge transformations, and we see we arrived at it for free by studying small diffeomorphisms.

We can expand the deformation in terms of covariant derivatives with respect to parameters
\[
\delta {\mathcal A} = \delta y^a \mathfrak{D}_a {\mathcal A} , \qquad \text{where} \quad {}^\Phi \mathfrak{D}_a {\mathcal A} = \Phi \mathfrak{D}_a {\mathcal A} \Phi^{-1}.
\]
The covariant derivatives both ensure the transformation law~\eqref{eq:gaugeDeltaA} is satisfied and are representation of the Kodaira--Spencer map~\cite{Kodaira:1981cx} relating tangent vectors on the moduli space to deformations of fields~\cite{Candelas:2016usb}.

The fluctuation $\mathfrak{D}_a {\mathcal A}$ satisfies the Atiyah equation
\begin{gather}
{\overline{\partial}}_{\mathcal A}\big( \mathfrak{D}_a {\mathcal A}\big) = \Delta_a{}^\mu F_\mu.\label{eq:Atiyah}
\end{gather}
In terms of covariant derivatives, the gauge symmetry action becomes\footnote{At this point we explain why we can take the last term to be exact. Write $\mathfrak{D}_a {\mathcal A} \sim \widetilde{\mathfrak{D}}_a {\mathcal A} = \mathfrak{D}_a {\mathcal A} + \varepsilon^\mu F_\mu + \gamma_a$ where~${\overline{\partial}}_{\mathcal A} \gamma_a = 0$ and suppose $\gamma_a$ is non-trivial in cohomology. It can be expanded in a basis $\gamma_a = \gamma_a{}^b \mathfrak{D}_b {\mathcal A}$, where~$\mathfrak{D}_b {\mathcal A}$ are basis elements for the cohomology group and $\gamma_a{}^b$ is a parameter dependent matrix. Then,
\begin{gather*}
\mathfrak{D}_a {\mathcal A} \sim \widetilde{\mathfrak{D}}_a {\mathcal A} = \big(\delta_a{}^b + \gamma_a{}^b \big) \mathfrak{D}_b {\mathcal A} + \varepsilon_a{}^\mu F_\mu + {\overline{\partial}}_{\mathcal A} \phi_a.
\end{gather*}
This is a transformation law for a change of parameters and can be absorbed by a redefinition of the $\delta y^a$.}
\begin{gather}\label{eq:fDAtransf}
\mathfrak{D}_a {\mathcal A} \sim\mathfrak{D}_a {\mathcal A} + \varepsilon_a{}^\mu F_\mu + {\overline{\partial}}_{\mathcal A} \phi_a,
\end{gather}
where $\phi_a$ is some adjoint valued field and this equation is a symmetry of \eqref{eq:Atiyah} provided $\Delta_a{}^\mu \sim \Delta_a{}^\mu + {\overline{\partial}} \varepsilon_a{}^\mu$.
This covariant derivative provides a first order definition of the Kodaira--Spencer map~\cite{Kodaira:1981cx} between 1-forms along the moduli space, spanned by $\delta y^a$ deformations of the gauge field
\[
\delta {\mathcal A} = \delta y^a \mathfrak{D}_a {\mathcal A} .
\]

In the physics literature a bundle modulus is typically associated to a fluctuation $\mathfrak{D}_a{\mathcal A} \in H^1(X, \operatorname{End} {\mathcal E})$. From~\eqref{eq:Atiyah} these correspond to $\Delta_a = 0$. While it may be obvious to some readers, we note this is only true in a particular gauge. A more invariant statement is that a bundle modulus satisfies ${\overline{\partial}}_{\mathcal A} (\mathfrak{D}_a {\mathcal A}) = \big({\overline{\partial}} \kappa_a{}^\mu\big) F_\mu$ for any $\kappa_a{}^\mu$. We will see in the heterotic theory that in fact fluctuations $\mathfrak{D}_a{\mathcal A}$ are coupled to deformations of the complex structure and Hermitian structure of~$X$.

\subsection[The three-form $H$]{The three-form $\boldsymbol{H}$}\label{s:threeform}

Consider the three-form $H$
\begin{gather*}
H = {\rm d} B - \tfrac{\alpha^{\backprime}}{4}\big({\rm CS}[A] - {\rm CS}[\Theta]\big),\qquad {\rm CS}[A] = \operatorname{Tr} \big(A{\rm d} A +\tfrac{2}{3} A^3\big),%\label{Hdef}
\end{gather*}
defined so that it satisfies the Bianchi identity~\eqref{eq:Anomaly0}. Here $\Theta$ is the gauge potential for frame transformations, which we suppress for now. Under background gauge transformations
\begin{gather}\label{eq:BTransf}
 {}^\Phi B = B - \tfrac{\alpha^{\backprime}}{4} \big( \operatorname{Tr}(AY) - U \big) , \qquad \tfrac13 \operatorname{Tr}\big(Y^3\big) = {\rm d} U,
\end{gather}
where $Y = \Phi^{-1} {\rm d} \Phi$.

The three-form $H$ is well-defined on $X$. A variation of it is
\begin{gather}\label{eq:deltaH}
\partial_a H = {\rm d} {\mathcal B}_a - \tfrac{\alpha^{\backprime}}{2} \operatorname{Tr} ( \mathfrak{D}_a A F ) ,
\end{gather}
where ${\mathcal B}_a$ is defined as
\begin{gather}\label{eq:ccBdef}
{\mathcal B}_a = \mathfrak{D}_a B +\tfrac{\alpha^{\backprime}}{4}\operatorname{Tr}{(A \mathfrak{D}_a A)}-{\rm d}{\mathbb B}_a ,
\end{gather}
and the field ${\mathbb B}_a$ is a $1$-form on $X$. The definition of $\mathfrak{D}_a B$ and the symmetry properties of ${\mathbb B}_a$ become clearer in universal geometry. The transformation property of $\mathfrak{D}_a B$ under background transformations mimics that of~$B$ in~\eqref{eq:BTransf} and is discussed in~\cite{Candelas:2016usb}:
\begin{gather}\label{eq:ccBmixedBackground}
 {}^\Phi \mathfrak{D}_a B = \mathfrak{D}_a B - \tfrac{\alpha^{\backprime}}{4} \big( \operatorname{Tr}{(\mathfrak{D}_a A Y)} - \mathfrak{U}_a \big) , \qquad {\rm d} \mathfrak{U}_a = 0 .
\end{gather}
We infer the transformation law for ${\mathcal B}_a$ by using \eqref{eq:deltaH} and the transformation law for the gauge field \eqref{eq:fDAtransf} to find
\[
\partial_a H \sim {\rm d} \big(\widetilde {\mathcal B}_a - \tfrac{\alpha^{\backprime}}{2} \operatorname{Tr} ( \phi_a F )\big) - \tfrac{\alpha^{\backprime}}{2} \operatorname{Tr}\big(( \mathfrak{D}_a A + \varepsilon_a{}^m F_m)F \big),
\]
where $\widetilde {\mathcal B}_a$ is to be determined. On the other hand, \eqref{eq:SmallDiffondT} implies
\begin{gather*}%\label{eq:dHTransf}
\partial_a H \sim \partial_a H - \tfrac{\alpha^{\backprime}}{2} \varepsilon_a{}^m \operatorname{Tr} (F_m F) + {\rm d} \big(\varepsilon_a{}^m H_m\big).
\end{gather*}
Comparing the last two equations gives
\begin{gather}\label{eq:ccBtransf}
{\mathcal B}_a \sim\widetilde {\mathcal B}_a = {\mathcal B}_a + \varepsilon_a{}^m H_m + \tfrac{\alpha^{\backprime}}{2} \operatorname{Tr} (\phi_a F ) + {\rm d} {\mathfrak b}_a,
\end{gather}
where ${\mathfrak b}_a$ is a real 1-form. This equation holds up to a gauge invariant ${\rm d}$-closed term. We have taken this to be exact ${\rm d} {\mathfrak b}_a$. In complete analogy with the previous subsection, any non-trivial element of $H^2(X,{\mathbb R})$ can be absorbed by a redefinition of the parameter space coordinates. In the $\alpha^{\backprime}\to 0$ limit, fluctuations of the $B$-field admit a symmetry $\delta B \sim \delta B + {\rm d} {\mathfrak b}$ for some $1$-form~${\mathfrak b}$. The shift by~${\rm d} {\mathfrak b}_a$ is the generalisation to heterotic theories and we label it a small gerbe transformation.

Consider again the covariant derivative $\mathfrak{D}_a B$ in \eqref{eq:ccBdef} constructed so that $\delta B = \delta y^a \mathfrak{D}_a B$. Just as we could infer the small transformation law $\mathfrak{D}_a A \sim \mathfrak{D}_a A + {\rm d}_A \phi_a$ by considering the infinitesimal limit $\Phi = 1 - \phi$ in~\eqref{eq:Atransf}, we could apply the same limit to~\eqref{eq:BTransf} to read off
\begin{gather}\label{eq:CovDerivB}
\mathfrak{D}_a B \sim \mathfrak{D}_a B + \tfrac{\alpha^{\backprime}}{4} \operatorname{Tr} (A \, {\rm d} \phi_a ).
\end{gather}
For future reference, this means the following combination transforms as
\[
\mathfrak{D}_a B + \tfrac{\alpha^{\backprime}}{4} \operatorname{Tr} ( A\mathfrak{D}_a A) \sim \mathfrak{D}_a B + \tfrac{\alpha^{\backprime}}{4} \operatorname{Tr} ( A\mathfrak{D}_a A) + \tfrac{\alpha^{\backprime}}{2} \operatorname{Tr}( \phi_a F) - {\rm d} \big(\tfrac{\alpha^{\backprime}}{2} \operatorname{Tr} (A \phi_a)\big).
\]
We will return to this in the next section.

\subsection{Summary of small transformations on fields}

We now put all of this together. First, it is convenient to form the complexified combinations
\[
{\mathcal Z}_a = {\mathcal B}_a + {\rm i} \mathfrak{D}_a \omega, \qquad {\overline{\mathcal Z}}_a = {\mathcal B}_a - {\rm i} \mathfrak{D}_a \omega,
\]
where we denote $\mathfrak{D}_a \omega^{(p,q)} = (\partial_a \omega)^{(p,q)}$ while on a real form $\mathfrak{D}_a \omega = \partial_a \omega$.

For posterity, we record the action of combined small diffeomorphism, gerbe and gauge transformation on heterotic moduli fields:
\begin{gather}
 \Delta_a{}^\mu \sim \Delta_a{}^\mu + {\overline{\partial}} \varepsilon_a{}^\mu, \qquad \mathfrak{D}_a {\mathcal A} \sim \mathfrak{D}_a {\mathcal A} + \varepsilon_a{}^\mu F_\mu + {\overline{\partial}}_{\mathcal A} \phi_a, \nonumber\\
 {\mathcal Z}_a \sim {\mathcal Z}_a + \varepsilon_a{}^m (H+{\rm i}\, {\rm d} \omega)_m + \tfrac{\alpha^{\backprime}}{2} \operatorname{Tr} ( F \phi_a ) + {\rm d} \big ({\mathfrak b}_a + {\rm i} \varepsilon_a{}^m \omega_m\big),\nonumber\\
 {\overline{\mathcal Z}}_a \sim {\overline{\mathcal Z}}_a + \varepsilon_a{}^m (H-{\rm i}\, {\rm d} \omega)_m + \tfrac{\alpha^{\backprime}}{2} \operatorname{Tr} ( F \phi_a )+ {\rm d} \big({\mathfrak b}_a - {\rm i} \varepsilon_a{}^m \omega_m\big).\label{eq:smallTransf}
\end{gather}

\subsection{Solutions to the supersymmetry equations and Bianchi identity}

To first order in $\alpha^{\backprime}$, it is known that solving the supersymmetry equations and Bianchi identity is sufficient for solving the equations of motion~\cite{Ivanov:2009rh} and so we work with supersymmetry and the Bianchi identity. These amount to a series of equations. First that the complex structure is integrable. Second is the Atiyah equation \eqref{eq:Atiyah}. We have already checked that the small transformations~\eqref{eq:smallTransf} are a symmetry of these first two equations, see below~\eqref{eq:fDAtransf}. Third is $H={\rm d}^c \omega$. We now have to do some work. Using ${\rm d}^c \omega = J^m \partial_m \omega - ({\rm d} J^m) \omega_m$ we have
\begin{gather*}
 (\partial_a {\rm d}^c \omega)^{(0,3)} = - {\rm i} {\overline{\partial}} (\partial_a \omega)^{(0,2)} , \\
 (\partial_a {\rm d}^c \omega)^{(1,2)} = 2{\rm i} \Delta_a{}^\mu( \partial \omega)_\mu - {\rm i} \partial (\partial_a \omega)^{(0,2)} - {\rm i} {\overline{\partial}} (\partial_a \omega)^{(1,1)}.
\end{gather*}
Projecting $\partial_a H$ in \eqref{eq:deltaH} onto type we find
\begin{gather}
 {\overline{\partial}} {\mathcal Z}_a^{(0,2)} = 0, \nonumber\\
\partial {\mathcal Z}_a^{(0,2)} + {\overline{\partial}} {\mathcal Z}_a^{(1,1)} = 2{\rm i} \Delta_a{}^\mu (\partial\omega)_\mu + \frac{\alpha^{\backprime}}{2} \operatorname{Tr} ( \mathfrak{D}_a {\mathcal A} F ).\label{eq:moduliEqnReal}
\end{gather}
There are two other equations given by complex conjugation. It is now straightforward to see that~\eqref{eq:smallTransf} is a~symmetry of~\eqref{eq:moduliEqnReal} provided the Bianchi identity~\eqref{eq:Anomaly0} holds.

For later convenience, we note that a solution to the first equation of~\eqref{eq:moduliEqnReal} with $h^{(0,2)} = 0$ is given by ${\mathcal Z}_a^{(0,2)} = {\overline{\partial}} \beta_a^{(0,1)}$ for some complex $(0,1)$-form $\beta_a^{(0,1)}$. The second equation then becomes
\begin{gather*}%\label{eq:ccZ11}
 {\overline{\partial}} \big({\mathcal Z}_a^{(1,1)} - \partial \beta_a^{(0,1)} \big)= 2{\rm i} \Delta_a{}^\mu (\partial\omega)_\mu + \frac{\alpha^{\backprime}}{2} \operatorname{Tr} ( \mathfrak{D}_a {\mathcal A} F ) .
\end{gather*}

Fourth, is the conformally balanced condition. To this order in $\alpha^{\backprime}$ the dilaton is a constant and so the metric is actually balanced ${\rm d} \omega^2 = 0$. Geometrically, this means the Lee form of $\omega$ vanishes $W(\omega) = 0$. The balanced condition has a variation
\begin{gather}\label{eq:balancedVariation}
 {\rm d} ( \partial_a\omega \omega) = 0 .
\end{gather}
The action of \eqref{eq:smallTransf} on $\partial_a \omega$ is a Lie derivative $\partial_a \omega \sim \partial_a \omega + {\mathcal L}_{\varepsilon_a} \omega$. This is a symmetry of~\eqref{eq:balancedVariation}. To see this, we use that ${\mathcal L}$ satisfies Leibnitz rule and commutes with the exterior derivative~${\rm d}$
\begin{gather*}
 {\rm d} ( {\mathcal L}_{\varepsilon_a}\omega \omega ) = ({\mathcal L}_{\varepsilon_a} \, {\rm d}\omega) \omega + ({\mathcal L}_{\varepsilon_a}\omega) \, {\rm d}\omega = {\mathcal L}_{\varepsilon_a}(\omega \, {\rm d}\omega) = \frac{1}{2} {\mathcal L}_{\varepsilon_a} \, {\rm d} \big(\omega^2\big) = 0 .
\end{gather*}

Finally, the HYM equation $\omega^2 F = 0$ has a variation
\begin{gather*}
 \omega^2 ({\rm d}_A \mathfrak{D}_a A) + 2 F \omega \partial_a\omega = 0 .
\end{gather*}
Note that
\begin{gather*}
 \big({\rm d}^2_A \phi_a\big) \omega^2 = [F, \phi_a] \omega^2 = \big[F \omega^2, \phi_a\big] = 0 .
\end{gather*}
Hence, under \eqref{eq:smallTransf} this is invariant because
\begin{gather*}
 \big({\rm d}_A\big({\rm d}_A \phi_a + \varepsilon_a{}^m \big) F_m\big) \omega^2 + 2 F \omega {\mathcal L}_{\varepsilon_a}\omega = {\mathcal L}_{\varepsilon}\big(F \omega^2\big) = 0 ,
\end{gather*}
where the Lie derivative acting on some gauge group $G$-charged object $\xi$ is defined
\begin{gather*}
 {\mathcal L}_{\varepsilon} \xi = \varepsilon^m ({\rm d}_A \xi)_m + {\rm d}_A \big(\varepsilon^m \xi_m\big) .
\end{gather*}
This confirms is what, of course, obvious: the equations are invariant under gauge transformations. But it also serves as a useful consistency check that \eqref{eq:smallTransf} is the correct transformation law.

\section{Gauge fixing}\label{s:Gauge}
 We now fix the small gauge transformations. The gauge fixing is closely related to holomorphy on the moduli space together with demanding $\delta \Omega^{(3,0)}$ be a harmonic form. In complex coordinates on the moduli space $M$, this involves conditions such as $\mathfrak{D}_{\overline{\alpha}} {\mathcal A} = 0$. It is obvious but often overlooked that this equation partially fixes small gauge transformations. There is a~residual gauge freedom which we use to fix $\delta \Omega^{(3,0)}$ to be harmonic; this also implies $\partial \chi_\alpha = 0$ or equivalently $\nabla_\mu \Delta_\alpha{}^\mu = 0$ for an appropriate choice of $\nabla_\mu$.

\subsection{A warm-up}
Consider a deformation of the Hermitian Yang--Mills equation
\[
\omega^2 F = 0,
\]
where $F = {\rm d} A + A^2$ and $A = {\mathcal A} - {\mathcal A}^\dag$. A variation of this equation on a fixed manifold $X$ results in
\[
{\overline{\partial}}_{\mathcal A}^\dag (\delta {\mathcal A}) = - \partial_{\mathcal A}^\dag \big(\delta {\mathcal A}^\dag\big).
\]
This provides no constraints on $\delta A$, but notice it does depend on both $\delta {\mathcal A}$ and $\delta {\mathcal A}^\dag$ and so is a~real equation. Suppose we can take a holomorphic variation of this equation in which $\delta {\mathcal A}^\dag = 0$. Then we end up with a constraint
\[
{\overline{\partial}}_{\mathcal A}^\dag (\delta {\mathcal A} ) = 0.
\]
It naively looks as though the HYM equation fixed us to harmonic gauge. However, this conclusion is incorrect. The gauge fixing occurred earlier on, in the assumption that we could take a holomorphic variation. Indeed, under a small gauge transformation $\delta {\mathcal A}^\dag \sim \delta {\mathcal A}^\dag + \partial_{{\mathcal A}^\dag} \phi^\dag$, and so it is only true that $\delta {\mathcal A}^\dag = 0$ in a particular gauge. We call this holomorphic gauge. So the correct statement is that in holomorphic gauge the Hermitian Yang--Mills equation further constrains~$\delta {\mathcal A}$ to be in harmonic gauge. The aim of this section is to systematically study gauge fixing, and then describe how these uniquely fix the physical degrees of freedom.

As a toy example to warm-up consider a $d=4$ supersymmetric $U(1)$ gauge theory with $N+1$ chiral multiplets whose bosons are denoted $\phi^i$ can carry charge $+1$ under the gauge symmetry. The scalar potential is $V = \frac{1}{2} D^2$ where the D-term is $D = \phi^i{\overline{\phi}}_i -r$ with $r$ the Fayet--Iliopoulos~(FI) parameter. We have set the coupling constant to unity. The fields $\phi^i$ can be interpreted as complex coordinates on ${\mathbb C}^{N+1}$ with the flat Hermitian metric used in lowering the index ${\overline{\phi}}_i$. The space of classical vacua, therefore, corresponds to the single D-term vanishing modulo gauge transformations and is
\[
 {\mathbb P}^N = {\mathbb C}^{N+1} /\!\!/ U(1),
\]
where the $/\!\!/$ denotes the symplectic quotient: the gauge quotient $\phi^i \sim \phi^i e^{{\rm i} \lambda}$, $\lambda\in{\mathbb R}$ and the moment map $\phi^i {\overline{\phi}}_i - r = 0$ imposed simultaneously.\footnote{It is well-known we can equally view this moduli space as a holomorphic quotient $\big({\mathbb C}^{N+1}-{0}\big)/ {\mathbb C}^*$. This amounts to complexifying the gauge symmetry, e.g., $U(1)\to {\mathbb C}^*$ and forgetting about the D-terms. In a $d=4$ supersymmetric field theory we can study either the real compact gauge group or the complexified gauge group. Whether this holds in string theory is far from clear. That being so, we work with the real compact gauge group in heterotic theories to guarantee self-consistency.}

Given a point $\phi^i \in {\mathbb P}^N$ we can study deformations $\phi^i \to \phi^i + \delta \phi^i$. Deformations $\delta \phi^i$ inherit a~gauge symmetry $\delta \phi^i \sim \delta \phi^i +{\rm i} \phi^i \delta \lambda$ for some small gauge parameter $\delta \lambda$ which is real. These are a simple example of small gauge transformations which we discuss in this paper. The D-terms impose an equation of motion on the fluctuations
\[
 \delta \phi^i {\overline{\phi}}_i + \phi^i \delta {\overline{\phi}}_i = 2 \operatorname{Re} \big(\delta \phi^i {\overline{\phi}}_i\big) = 0.
\]
The D-terms do not fix the $U(1)$ gauge symmetry. This is expected -- after all the equations of motion are gauge invariant. Instead a gauge fixing is an additional condition such as \mbox{$\operatorname{Im} \big(\delta \phi^i {\overline{\phi}}_i \big) = \xi$} for some $\xi\in {\mathbb R}$. The D-term together with the gauge fixing allow us to determine the physical fluctuations. It is not hard to see they are exactly tangent vectors to ${\mathbb P}^N$ about a point. In the heterotic analysis the equations of motion, including the HYM and balanced equation, together with gauge fixing will allow us to determine physical degrees of freedom in the same sort of way.

\subsection{Holomorphy of the moduli space}\label{s:holomorphicgauge}
In Section~\ref{s:smallTransformations}, expressions were written in real coordinates with real parameters. However, an ${\mathcal N} = 1$ supersymmetric theory always comes with a complex parameter space $M$. Hence, its tangent space is complexified (see Appendix~\ref{app:Complexification}) and we gain computational power by introducing holomorphy. This means we need to decide a map between deformations of fields and holomorphic parameters $\delta y^\alpha$. Typically this is done in a way that partially fixes small gauge transformations. For this reason, we first write down the relations without fixing a gauge. This will amount to certain anti-holomorphic combinations being exact. We will then describe a convenient gauge fixing, recovering conventional expressions in the literature. As far as we aware this has not been done for heterotic theories in the literature, so it worth our while to go into detail.

We start with integrable complex structure deformations which obey \hbox{${\overline{\partial}}( \delta y^a \Delta_a{}^\mu) = 0$}. Expanding in complex coordinates, $\delta y^a \Delta_a{}^\mu = \delta y^\alpha \Delta_\alpha{}^\mu + \delta y^{\overline\beta} \Delta_{\overline\beta}{}^\mu$, we take holomorphy to mean the non-trivial elements of $H^1(X, {\mathcal T}_X)$ are associated to holomorphic deformations, $\Delta_\alpha{}^\mu$, while the anti-holomorphic deformations are exact but not necessarily zero
\[
 \Delta_{\overline{\alpha}}{}^\mu = {\overline{\partial}} \kappa_{\overline{\alpha}}{}^\mu.
\]
Under a small diffeomorphism $\Delta_{\overline{\alpha}}{}^\mu \sim \Delta_{\overline{\alpha}}{}^\mu + {\overline{\partial}} \varepsilon_{\overline{\alpha}}{}^\mu$, where $\varepsilon_{\overline{\alpha}}{}^\mu$ is a vector on $X$ and 1-form on ${\mathcal M}$. As a prelude, we immediately see if $\Delta_{\overline{\alpha}}{}^\mu = 0$ then we have partially fixed small diffeomorphisms.

Consider a deformation of the gauge field $\delta {\mathcal A} = \delta y^\alpha \mathfrak{D}_\alpha {\mathcal A} + \delta y^{\overline{\alpha}} \mathfrak{D}_{\overline{\alpha}} {\mathcal A}$. The Atiyah equation~\eqref{eq:Atiyah} implies the anti-holomorphic deformation satisfies ${\overline{\partial}}_{\mathcal A} \mathfrak{D}_{\overline{\alpha}} {\mathcal A} = {\overline{\partial}}_{\mathcal A} (\kappa_{\overline{\alpha}}{}^\mu F_\mu)$. As for complex structure, we take holomorphy to mean
\begin{gather*}%\label{eq:bundleholomorphy}
\mathfrak{D}_{\overline{\alpha}} {\mathcal A} = \kappa_{\overline{\alpha}}{}^\mu F_\mu + {\overline{\partial}}_{\mathcal A} \Phi_{\overline{\alpha}},
\end{gather*}
for some section $\Phi_{\overline{\alpha}}$ of $\operatorname{End}{\mathcal E}$.

Consider the complexified Hermitian form ${\mathcal Z}_a = {\mathcal B}_a + {\rm i} {\mathfrak D}_a \omega$.
Recall from \eqref{eq:moduliEqnReal} that ${\mathcal Z}_a^{(0,2)} = {\overline{\partial}} \beta_a^{(0,2)}$ which we write as
\[
{\mathcal Z}_{\overline{\alpha}}^{(0,2)} = {\overline{\partial}} \beta_{\overline{\alpha}}^{(0,2)}, \qquad {\overline{\mathcal Z}}_{\overline{\alpha}}^{(2,0)} = \partial {\overline{\beta}}_{\overline{\alpha}}^{(2,0)},
\]
while the $(1,1)$-component satisfies
\[
 {\overline{\partial}}\big( {\mathcal Z}_{\overline{\alpha}}^{(1,1)} - 2{\rm i} \kappa_{\overline{\alpha}}{}^\mu (\partial \omega)_\mu - \tfrac{\alpha^{\backprime}}{2} \operatorname{Tr} ( \Phi_{\overline{\alpha}} F ) - \partial \beta_{\overline{\alpha}}{}^{(0,1)}\big) = 0.
\]
As for complex structure and the gauge field, holomorphy here amounts to the solution of the last equation being exact:
\[
 {\mathcal Z}_{\overline{\alpha}}^{(1,1)} = 2{\rm i} \kappa_{\overline{\alpha}}{}^\mu (\partial \omega)_\mu + \tfrac{\alpha^{\backprime}}{2} \operatorname{Tr} ( \Phi_{\overline{\alpha}} F ) + \partial \beta_{\overline{\alpha}}{}^{(0,1)} + {\overline{\partial}} \xi_{\overline{\alpha}}^{(1,0)} .
\]
One might ask, what if we were to take these solution to be not exact? In the example at hand, a non-trivial element of $H^{(1,1)} (X, {\mathbb C})$ can be absorbed by a change in real coordinates on~${\mathcal M}$, see footnote below~\eqref{eq:fDAtransf}. In fact, as described in~\cite{Candelas:2016usb}, it actually amounts to a change in complex structure. So, as expected, choosing the solutions of these equations to be trivial in cohomology amounts to a choice of complex structure on~${\mathcal M}$.

We summarise holomorphy as
\begin{gather}
 \Delta_{\overline{\alpha}}{}^\mu = {\overline{\partial}} \kappa_{\overline{\alpha}}{}^\mu, \qquad \mathfrak{D}_{\overline{\alpha}} {\mathcal A} = \kappa_{\overline{\alpha}}{}^\mu F_\mu + {\overline{\partial}}_{\mathcal A} \Phi_{\overline{\alpha}},\nonumber\\
 {\mathcal Z}_{\overline{\alpha}}^{(1,1)} = 2{\rm i} \kappa_{\overline{\alpha}}{}^\mu (\partial \omega)_\mu + \tfrac{\alpha^{\backprime}}{2} \operatorname{Tr} ( \Phi_{\overline{\alpha}} F ) + \partial \beta_{\overline{\alpha}}{}^{(0,1)} + {\overline{\partial}} \xi_{\overline{\alpha}}^{(1,0)} ,\label{eq:holomorphy1}
\end{gather}
while there are two additional equations derive from \eqref{eq:moduliEqnReal} which we state here as they will be useful shortly:
\begin{gather}\label{eq:holomorphy2}
{\overline{\mathcal Z}}_{\overline{\alpha}}^{(2,0)} = \partial {\overline{\beta}}_{\overline{\alpha}}^{(1,0)},\qquad {\mathcal Z}_{\overline{\alpha}}^{(0,2)} = {\overline{\partial}} \beta_{\overline{\alpha}}^{(0,1)}.
\end{gather}
Using \eqref{eq:smallTransf} the action of small transformations generated by $\varepsilon_{\overline{\alpha}}{}^\mu$, $\varepsilon_{\overline{\alpha}}{}^{\overline{\mu}}$, $\phi_{\overline{\alpha}}$ and ${\mathfrak b}_{\overline{\alpha}}$ is
\begin{gather}
 \Delta_{\overline{\alpha}}{}^\mu \sim \Delta_{\overline{\alpha}}{}^\mu + {\overline{\partial}} \varepsilon_{\overline{\alpha}}{}^\mu,\nonumber\\
 \mathfrak{D}_{\overline{\alpha}} {\mathcal A} \sim \mathfrak{D}_{\overline{\alpha}} {\mathcal A} + \varepsilon_{\overline{\alpha}}{}^\mu F_\mu + {\overline{\partial}}_{\mathcal A} \phi_{\overline{\alpha}},\nonumber\\
 {\mathcal Z}^{(1,1)}_{\overline{\alpha}} \sim {\mathcal Z}^{(1,1)}_{\overline{\alpha}} + 2{\rm i} \varepsilon_{\overline{\alpha}}{}^\mu (\partial\omega)_\mu + \tfrac{\alpha^{\backprime}}{2} \operatorname{Tr}  ( \phi_{\overline{\alpha}} F  ) + \partial \big({\mathfrak b}^{(0,1)}_{\overline{\alpha}} + {\rm i} \varepsilon_{\overline{\alpha}}{}^\mu \omega_\mu\big)+ {\overline{\partial}} \big({\mathfrak b}^{(1,0)}_{\overline{\alpha}} + {\rm i} \varepsilon_{\overline{\alpha}}{}^{\overline{\mu}} \omega_{\overline{\mu}}\big),\nonumber\\
 {\mathcal Z}^{(0,2)}_{\overline{\alpha}} \sim {\mathcal Z}^{(0,2)}_{\overline{\alpha}} + {\overline{\partial}} \big({\mathfrak b}^{(0,1)}_{\overline{\alpha}} + {\rm i} \varepsilon_{\overline{\alpha}}{}^\nu \omega_\nu\big),\qquad
 {\overline{\mathcal Z}}^{(0,2)}_{\overline{\alpha}} \sim {\overline{\mathcal Z}}^{(0,2)}_{\overline{\alpha}} + {\overline{\partial}} \big({\mathfrak b}^{(0,1)}_{\overline{\alpha}} - {\rm i} \varepsilon_{\overline{\alpha}}{}^\nu \omega_\nu\big), \nonumber\\
 {\mathcal Z}^{(2,0)}_{\overline{\alpha}} \sim {\mathcal Z}^{(2,0)}_{\overline{\alpha}} + \partial \big({\mathfrak b}^{(1,0)}_{\overline{\alpha}} + {\rm i} \varepsilon_{\overline{\alpha}}{}^{\overline\nu} \omega_{\overline\nu}\big),\qquad
 {\overline{\mathcal Z}}^{(2,0)}_{\overline{\alpha}} \sim {\overline{\mathcal Z}}^{(2,0)}_{\overline{\alpha}} + \partial \big ({\mathfrak b}^{(1,0)}_{\overline{\alpha}} - {\rm i} \varepsilon_{\overline{\alpha}}{}^{\overline\nu} \omega_{\overline\nu}\big).\label{eq:smallAction}
\end{gather}
Equivalently,
\begin{gather}
 \kappa_{\overline{\alpha}}{}^\mu \sim \kappa_{\overline{\alpha}}{}^\mu + \varepsilon_{\overline{\alpha}}{}^\mu,\qquad \Phi_{\overline{\alpha}} \sim \Phi_{\overline{\alpha}} + \phi_{\overline{\alpha}},\nonumber\\
\beta^{(0,1)}_{\overline{\alpha}} \sim \beta^{(0,1)}_{\overline{\alpha}} + {\mathfrak b}^{(0,1)}_{\overline{\alpha}} + {\rm i} \varepsilon_{\overline{\alpha}}{}^\nu \omega_\nu, \qquad \xi_{\overline{\alpha}}^{(1,0)} \sim \xi_{\overline{\alpha}}^{(1,0)} + {\mathfrak b}^{(1,0)}_{\overline{\alpha}} + {\rm i} \varepsilon_{\overline{\alpha}}{}^{\overline{\mu}} \omega_{\overline{\mu}} ,\nonumber\\
\beta^{(1,0)}_{\overline{\alpha}} \sim \beta^{(1,0)}_{\overline{\alpha}} + {\mathfrak b}^{(1,0)}_{\overline{\alpha}} - {\rm i} \varepsilon_{\overline{\alpha}}{}^{\overline\nu} \omega_{\overline\nu},\label{eq:gaugeActionComp}
\end{gather}
where the last two equations are correct up to ${\overline{\partial}}$- and $\partial$-closed term respectively.

\subsection{Gauge fixing}\label{s:gaugeFixing}
We can partially fix the gauge freedom \eqref{eq:smallAction}--\eqref{eq:gaugeActionComp} by making a choice for the right hand side of each equation within \eqref{eq:holomorphy1}--\eqref{eq:holomorphy2}. A convenient choice is to set as many terms to zero as possible.

We start by setting $\Delta_{\overline{\alpha}}{}^\mu = 0$ and $\mathfrak{D}_{\overline{\alpha}} {\mathcal A} = 0$ by putting $\varepsilon_{\overline{\alpha}}{}^\mu = - \kappa_{\overline{\alpha}}{}^\mu$ and $\phi_{\overline{\alpha}} = -\Phi_{\overline{\alpha}}$ in~\eqref{eq:gaugeActionComp}. We can fix ${\mathcal Z}_{\overline{\alpha}}^{(1,1)} = 0$ by setting
\begin{gather}\label{eq:hologaugefix2}
 {\mathfrak b}^{(1,0)}_{\overline{\alpha}} + {\rm i} \varepsilon_{\overline{\alpha}}{}^{\overline{\mu}} \omega_{\overline{\mu}} = - \xi_{\overline{\alpha}}^{(1,0)} + \partial \psi_{\overline{\alpha}},
\end{gather}
while we can also fix the fields ${\mathcal Z}_{\overline{\alpha}}^{(0,2)} = 0$, ${\overline{\mathcal Z}}_{\overline{\alpha}}^{(2,0)} = 0$ in~\eqref{eq:holomorphy2} by
\[
 {\mathfrak b}^{(0,1)}_{\overline{\alpha}} - {\rm i} \kappa_{\overline{\alpha}}{}^\nu \omega_\nu = - \beta_{\overline{\alpha}}^{(0,1)} +{\overline{\partial}} \psi_{\overline{\alpha}},
 \qquad
 {\mathfrak b}^{(1,0)}_{\overline{\alpha}} - {\rm i} \varepsilon_{\overline{\alpha}}{}^{\overline{\mu}} \omega_{\overline{\mu}} = - \beta_{\overline{\alpha}}^{(1,0)} + \partial {\overline{\psi}}_{\overline{\alpha}}.
\]
Here $\psi_{\overline{\alpha}}$ is a complex $(0,1)$-form on the moduli space ${\mathcal M}$, where ${\overline{\psi}}_{\overline{\alpha}} = (\psi_\alpha)^*$ and it can also depend on the coordinates of~$X$. At this point it is an unfixed quantity and we discuss it further below. As $\Delta_{\overline{\alpha}}{}^\mu = 0$ it follows that $\mathfrak{D}_{\overline{\alpha}} \omega^{(0,2)} = \Delta_{\overline{\alpha}}{}^\mu \omega_\mu = 0$ and because ${\mathcal Z}_{\overline{\alpha}}^{(0,2)} = 0$ it follows that both ${\mathcal B}^{(0,2)}_{\overline{\alpha}} = 0$ and ${\overline{\mathcal Z}}_{\overline{\alpha}}^{(0,2)} = 0$.
Putting these results together
\begin{gather}\label{eq:holomorphicgauge}
 \Delta_{\overline{\alpha}}{}^\mu = 0, \qquad \mathfrak{D}_{\overline{\alpha}} {\mathcal A} = 0, \qquad {\mathcal Z}_{\overline{\alpha}}^{(1,1)} = 0, \qquad {\overline{\mathcal Z}}_{\overline{\alpha}}^{(2,0)} = {\mathcal Z}_{\overline{\alpha}}^{(0,2)} = {\overline{\mathcal Z}}_{\overline{\alpha}}^{(0,2)} = 0.
\end{gather}
The first three equations are often used in the literature as a starting point of holomorphy. What we have shown here is that they are only correct in a certain gauge. We have not been able to gauge fix ${\mathcal Z}_{\overline{\alpha}}^{(2,0)}$ and so ${\overline{\mathcal Z}}_\alpha^{(0,2)} = 2{\mathcal B}_\alpha^{(0,2)} = -2{\rm i} {\mathfrak D}_\alpha \omega^{(0,2)}$ is also unfixed.

To what extent does \eqref{eq:holomorphicgauge} fix gauge transformations? Firstly, any residual gauge transformations must preserve~\eqref{eq:holomorphicgauge} and so for a start must have ${\overline{\partial}} \varepsilon_{\overline{\alpha}}{}^\mu = 0$. As we take $h^{(0,2)} = 0$ there are no non-trivial solutions, as can be seen by contracting with $\Omega_\mu$. So any residual small diffeomorphisms must have form $\varepsilon_\alpha{}^\mu$.

Secondly, \eqref{eq:holomorphicgauge} has completely fixed $\phi_\alpha$. This is because stability of the bundle implies $H^0(X, \operatorname{End}{\mathcal E})$ is trivial and the only solutions to ${\overline{\partial}}_{\mathcal A} \phi_{\overline{\alpha}} = 0$ are constants. The connection~$\Lambda_a {\rm d} y^a$ is antihermitian meaning $\phi_\alpha = - ( \phi_{\overline{\alpha}})^\dag$ is also a constant. Hence, there are no non-trivial small gauge transformations for the bundle.\footnote{If we were considering a complexified gauge group then~$A$ is no longer antihermitian, and so $\delta {\mathcal A}$ and $\delta {\mathcal A}^\dag$ are independent degrees of freedom. That being so, $\phi_\alpha$ is independent from $\phi_{\overline{\alpha}}$ and so $\mathfrak{D}_{\overline{\alpha}} {\mathcal A} = 0$ gauge fixes $\phi_{\overline{\alpha}}$ but does nothing to $\phi_\alpha$. However, we do not know to what extent the theory described by a~complexified gauge group is related to the theory with the physical gauge group.}

Thirdly, there is the unfixed degree of freedom in \eqref{eq:hologaugefix2} parameterised by $\psi_{\overline{\alpha}}$, ${\overline{\psi}}_{\overline{\alpha}}$. We see that performing a gauge transformation~\eqref{eq:smallAction} with
\begin{gather*}%\label{eq:ResidualGauge}
\varepsilon_{\overline{\alpha}}{}^\mu = 0, \qquad {\mathfrak b}^{(0,1)}_{\overline{\alpha}} = {\overline{\partial}} \psi_{\overline{\alpha}}, \qquad {\mathfrak b}^{(1,0)}_{\overline{\alpha}} - {\rm i} \varepsilon_{\overline{\alpha}}{}^{\overline\nu} \omega_{\overline\nu} = \partial {\overline{\psi}}_{\overline{\alpha}}, \qquad {\mathfrak b}^{(1,0)}_{\overline{\alpha}} + {\rm i} \varepsilon_{\overline{\alpha}}{}^{\overline\nu} \omega_{\overline\nu} = \partial \psi_{\overline{\alpha}},
\end{gather*}
 then \eqref{eq:holomorphicgauge} is invariant. Denote $\varepsilon_{\overline{\alpha}} = {\rm i} \varepsilon_{\overline{\alpha}}{}^m\omega_m$. We can invert these equations to parameterise the residual gauge transformations
\begin{gather}\label{eq:ResidualGauge2}
 \varepsilon_{\overline{\alpha}}^{(0,1)} = 0,\qquad {\mathfrak b}^{(0,1)}_{\overline{\alpha}} = {\overline{\partial}} \psi_{\overline{\alpha}},\qquad {\mathfrak b}^{(1,0)}_{\overline{\alpha}} = \tfrac{1}{2} \partial \big(\psi_{\overline{\alpha}} + {\overline{\psi}}_{\overline{\alpha}}\big), \qquad
\varepsilon_{\overline{\alpha}}^{(1,0)} = \tfrac{1}{2} \partial \big(\psi_{\overline{\alpha}} - {\overline{\psi}}_{\overline{\alpha}} \big).
\end{gather}
Before utilising these transformations, we first refer back to Section~\ref{s:holomorphicform} and the role of $\delta \Omega$. From~\eqref{eq:OmPureVarn} we see that $\delta \Omega^{(2,1)} = \delta y^\alpha \Delta_\alpha{}^\mu \Omega_\mu$ because $\Delta_{\overline{\alpha}}{}^\mu = 0$ in this gauge. The fact ${\rm d} \delta \Omega = 0$ implies that ${\overline{\partial}} \partial \xi_{\overline{\alpha}}^{(2,0)} = 0$ and so $\xi_{\overline{\alpha}}^{(2,0)} = \partial \zeta_{\overline{\alpha}}^{(1,0)}$ and hence vanishes in~\eqref{eq:OmPureVarn}. There is a further gauge symmetry coming from $\Omega$ being a section of a line bundle over ${\mathcal M}$. We can use this to set $k_{\overline{\alpha}} = 0$. So in holomorphic gauge, we see that $\delta \Omega$ depends holomorphically on parameters.

Under a small diffeomorphism $\delta\Omega^{(3,0)} \sim \delta\Omega^{(3,0)} + \big(\delta y^\alpha \nabla_\nu \varepsilon_\alpha^\nu\big) \Omega$.\footnote{As $H_{\mu\nu}{}^\nu = 0$, a consequence of the gauge fixing and supersymmetry see Appendix~\ref{s:dilatongauge}, Levi-Civita, Bismut, Hull or Chern connections are the same. That is why the label is left blank. }
If we choose
\begin{gather}\label{eq:gaugefixing}
\varepsilon_\alpha{}^\nu = - \frac{1}{2\|\Omega\|^2} {\overline{\Omega}}^{\nu\rho\sigma} \big(\xi^{(2,0)}_\alpha + \partial \zeta_\alpha^{(1,0)}\big)_{\rho\sigma},
\end{gather}
where $\partial \zeta_\alpha^{(1,0)}$ is a local representative of any closed 2-form, then
\begin{gather*}%\label{eq:residual2}
\nabla_\nu \varepsilon_\alpha{}^\nu = - \frac{1}{3!\|\Omega\|^2} {\overline{\Omega}}^{\nu\rho\sigma} \big(\partial \xi^{(2,0)}_\alpha\big)_{\nu\rho\sigma}.
\end{gather*}
It is now easy to see, that using \eqref{eq:gaugefixing} we can kill $\partial \xi_\alpha^{(2,0)}$ in \eqref{eq:OmPureVarn}. Furthermore, noting that $\nabla_\nu \varepsilon_\alpha{}^\nu = g^{\mu{\overline\nu}} \partial_\mu \varepsilon_{\alpha{\overline\nu}} = {\overline{\partial}}^\dag \varepsilon_\alpha^{(0,1)}{}$, from~\eqref{eq:ResidualGauge2} such a~small diffeomorphisms preserves \eqref{eq:holomorphicgauge} if there is a solution to a certain Poisson equation
\[
\Box_{\overline{\partial}} \big(\psi_\alpha - {\overline{\psi}}_\alpha\big) = \frac{1}{3\|\Omega\|^2} {\overline{\Omega}}^{\nu\rho\sigma} \big(\partial \xi^{(2,0)}_{\alpha}\big)_{\nu\rho\sigma}.
\]
As discussed in Appendix~\ref{eq:appCYRealDef} the Laplacian is invertible provided the source be orthogonal to zero modes. On $X$ these are constants and so there is never any obstruction
\[
 \frac{1}{3!\|\Omega\|^2} \int_X \operatorname{vol} {\overline{\Omega}}^{\nu\rho\sigma} \big(\partial \xi^{(2,0)}_{\alpha}\big)_{\nu\rho\sigma} \sim \int_X \partial \xi^{(2,0)} {\overline{\Omega}} = 0.
\]
Consequently, we can find a $\psi_\alpha - {\overline{\psi}}_\alpha$ that kills $\xi_\alpha^{(2,0)}$ and so that $\delta \Omega^{(3,0)} = (\delta y^\alpha k_\alpha )\Omega$ is harmonic for some parameter dependent constants $ k_\alpha$. Returning to~\eqref{eq:gaugefixing}, the closed form $\partial\zeta_\alpha^{(1,0)}$ corresponds to $\nabla^{\rm LC}_\nu \varepsilon_\alpha{}^\nu = 0$ which are zero modes on $X$, which are constants, and so $\psi_\alpha = {\overline{\psi}}_\alpha$ and~$\partial \zeta^{(0,1)}$ must vanish. We have fixed small diffeomorphisms.

There is a residual freedom in ${\mathfrak b}_\alpha = {\rm d} \psi_\alpha$. These do not act non-trivially on any of the fields because of the transformation law ${\mathcal B}_\alpha \sim {\mathcal B}_\alpha + {\rm d} {\mathfrak b}_\alpha$. They are similar to ghost-for-ghost transformations.

We have completely fixed the gauge, the aim of this section. We can summarise it as
\begin{gather}
 \Delta_{\overline{\alpha}}{}^\mu = 0, \qquad \delta \Omega^{(3,0)} = \delta y^\alpha k_\alpha \Omega, \qquad \mathfrak{D}_{\overline{\alpha}} {\mathcal A} = 0,\nonumber\\
{\mathcal Z}_{\overline{\alpha}}^{(1,1)} = 0, \qquad {\mathcal Z}_{\overline{\alpha}}^{(0,2)} = {\overline{\mathcal Z}}_{\overline{\alpha}}^{(0,2)} = {\overline{\mathcal Z}}_{\overline{\alpha}}^{(2,0)} = 0, \label{eq:finalGauge}
\end{gather}
For the rest of this paper we refer to this collection of equations as {\it holomorphic gauge}. In this gauge
\[
 \partial \chi_\alpha = 0 , \qquad {\rm where} \quad \chi_\alpha = \Delta_\alpha{}^\mu \Omega_\mu .
\]

The non-vanishing degrees of freedom are
\begin{gather}\label{eq:DOF}
\mathfrak{D}_\alpha {\mathcal A},\qquad \Delta_\alpha{}^\mu,\qquad {\mathcal B}_\alpha^{(1,1)} = {\rm i} \mathfrak{D}_\alpha\omega^{(1,1)},\qquad {\mathcal B}_\alpha^{(0,2)} = -{\rm i} \mathfrak{D}_\alpha\omega^{(0,2)}.
\end{gather}
These are not independent degrees of freedom as we describe below. In the following we will interchangably use ${\mathcal Z}^{(1,1)}_\alpha$ and ${\overline{\mathcal Z}}^{(0,2)}_\alpha$ for the last pair respectively; they are equal up to a numerical factor.

We finish by noting some identities. Using $(\partial \Delta_\alpha{}^\mu) \Omega_\mu = \partial_\mu \Delta_\alpha{}^\mu \Omega $ and that $\Omega = \frac{1}{3!} f \epsilon_{\mu\nu\rho} \, {\rm d} x^{\mu\nu\rho}$ with $f$ a holomorphic function of parameters such that $|f|^2 = \|\Omega\|^2 \sqrt{g}$ we have
\begin{gather*}%\label{eq:delChi}\notag
 \partial(\Delta_\alpha{}^\mu \Omega_\mu) = -\big( \partial_\mu \Delta_\alpha{}^\mu +\Delta_\alpha{}^\mu \partial_\mu \log{\sqrt g} \big) \Omega = -\big( \nabla_\mu^{\rm H/Ch} \Delta_\alpha{}^\mu\big) \Omega = 0,
\end{gather*}
where in the last equality $\nabla_\mu^{\rm H/Ch}$ are a family of connections with $\epsilon-\rho = 1$. This, in particular, includes the Hull and Chern connections. Hence,
\begin{gather}\label{eq:gfDelta}
\nabla_\mu^{\rm H/Ch} \Delta_{\alpha{\overline\nu}}{}^\mu = 0 .
\end{gather}

Another identity, useful for the next section, is as follows. First, the codifferential as given in Appendix~\ref{s:Hodgecomplex}, can be used to write
\[
 {\overline{\partial}}^\dag \Delta_\alpha{}^\mu = - \nabla^{{\rm Ch} \,{\overline\nu}} \Delta_{\alpha{\overline\nu}}{}^\mu = - \nabla^{\rm Ch}_\nu \Delta_\alpha{}^{(\mu\nu)} + \nabla^{\rm Ch}_\nu \Delta_\alpha{}^{[\mu\nu]} .
\]
Second, combine this equation with \eqref{eq:gfDelta} and the relation $\mathfrak{D}_\alpha\omega_{{\overline{\mu}}{\overline\nu}} = 2{\rm i} \Delta_{\alpha[{\overline{\mu}}{\overline\nu}]}$ to give
\begin{gather}\label{eq:GaugeFixDelOmega}
{\overline{\partial}}^\dag \Delta_\alpha{}^\mu = {\rm i} g^{\mu{\overline{\rho}}} \nabla^{{\rm Ch}\,{\overline\nu}} \big(\mathfrak{D}_\alpha \omega_{{\overline\nu}{\overline{\rho}}} \big) .
\end{gather}
This is in addition to the algebraic relation $\mathfrak{D}_\alpha\omega^{(0,2)}=\Delta_\alpha{}^\mu \, \omega_\mu$.

\section{The Hodge decomposition and the moduli space metric}\label{s:Hodge}
We now show how gauge fixing together with supersymmetry, the Bianchi identity, the Hermitian Yang--Mills equation and balanced equation, allow us to solve for the terms in the Hodge decomposition of the heterotic moduli in~\eqref{eq:DOF}.

Mathematically we are determining the Kodaira--Spencer map~\cite{Kodaira:1981cx} which associates parameters with field deformations. We do not a priori assume any structure about the moduli space, for example, we do not label parameters as complex structure or bundle moduli. We find that even at first order in $\alpha^{\backprime}$, the moduli fields are incredibly coupled and there is no invariant distinction between moduli.

When the background is Calabi--Yau at large radius, so that supergravity is guaranteed to be a~good approximation, we show how one may go about solving these coupled differential equations for the field deformations in~\eqref{eq:DOF}. As an application of this work, we give a prescription for how to compute the moduli space metric derived in~\cite{Candelas:2016usb, Candelas:2018lib}.

\subsection{The moduli equations}\label{s:balancedHYM}

The physical degrees of freedom in \eqref{eq:DOF} are $\mathfrak{D}_\alpha \omega^{(0,2)}$, $\mathfrak{D}_\alpha \omega^{(1,1)}$, $\mathfrak{D}_\alpha {\mathcal A}$ and $\Delta_\alpha$. These obey a~collection of equations deriving from varying the Nijenhuis tensor, Atiyah equation and \eqref{eq:moduliEqnReal}. The result is
\begin{gather}
{\overline{\partial}} \Delta_\alpha{}^\mu = 0, \nonumber\\
{\overline{\partial}}_{\mathcal A} \mathfrak{D}_\alpha {\mathcal A} = \Delta_\alpha{}^\mu F_\mu, \nonumber\\
 {\overline{\partial}} {\mathcal Z}_\alpha^{(1,1)} = 2{\rm i} \Delta_\alpha{}^\mu (\partial\omega)_\mu +\tfrac{\alpha^{\backprime}}{2} \operatorname{Tr} \big( \mathfrak{D}_\alpha {\mathcal A} F\big) - \alpha^{\backprime} \big( \nabla_\nu \Delta_\alpha{}^\mu + {\rm i} \nabla^\mu \mathfrak{D}_\alpha\omega_\nu^{(0,1)} \big) R^\nu{}_\mu ,\label{eq:moduliEqnHol}
\end{gather}
where $R^\mu{}_\nu = R_{\rho{\overline{\sigma}}}{}^\mu{}_\nu \, {\rm d} x^{\rho{\overline{\sigma}}}$ is the Riemann tensor. We have now included the contribution of the spin connection $\mathfrak{D}_\alpha \Theta$. This depends, to zeroth order in $\alpha^{\backprime}$, on the metric moduli with the relation derived in~\cite{Candelas:2018lib}:
\begin{gather}\label{eq:spinmoduli}
 \mathfrak{D}_{\alpha}\Theta{}_{\overline{\mu}}{}^\nu{}_{\sigma} = \nabla_\sigma \Delta_{\alpha{\overline{\mu}}}{}^\nu + {\rm i} \nabla^{\nu} \mathfrak{D}_\alpha\omega_{\sigma{\overline{\mu}}} ,
\end{gather}
which together with the symmetry $\mathfrak{D}_{\alpha}\Theta{}_{\overline{\mu}}{}^{\overline\nu}{}_{{\overline{\sigma}}} = - g^{{\overline\nu}\lambda} \mathfrak{D}_\alpha\Theta_{{\overline{\mu}}}{}^\rho{}_{\lambda} g_{\rho{\overline{\sigma}}}$ allows us to write
\[
\operatorname{Tr} \big(\mathfrak{D}_\alpha \Theta^{(0,1)} R \big) = 2 \mathfrak{D}_\alpha\Theta^\mu{}_\nu R^\nu{}_\mu = 2 \big( \nabla_\nu \Delta_\alpha{}^\mu + {\rm i} \nabla^\mu \mathfrak{D}_\alpha\omega_\nu^{(0,1)} \big) R^\nu{}_\mu ,
\]
where we have used $R_{{\overline{\mu}}\nu} = - R_{\nu{\overline{\mu}}}$.
Note that in~\cite{Candelas:2018lib} we checked that the spin connection is holomorphic $\mathfrak{D}_\alpha\Theta_\mu = 0$ and that this holds provided we are in the gauge fixing discussed here, at least to this order in $\alpha^{\backprime}$.

A motivation for this paper is that the equations \eqref{eq:moduliEqnHol}, in particular the last one, are exactly captured by the universal bundle in~\cite{Candelas:2018lib}. We explore this further in Section~\ref{s:firstorderUG}, and understand this result as being the statement that the universal bundle being holomorphic corresponds to field deformations of heterotic string being in holomorphic gauge~\eqref{eq:finalGauge} as constructed in the previous subsection. It is, mathematically speaking, natural for the universal bundle to be holomorphic and justifies a posteriori our choice of holomorphic gauge in studying heterotic supergravity.

The balanced condition means ${\rm d} \big(\omega^2\big) = 0$. Taking a real deformation and decomposing into type there is one non-trivial equation
\[
 \partial\big(\mathfrak{D}_a \omega^{(0,2)} \omega\big) + {\overline{\partial}}\big(\mathfrak{D}_a \omega^{(1,1)} \omega\big) = 0,
\]
with the remaining equation given by complex conjugation. In the gauge in which $\delta \Omega^{(3,0)}$ is harmonic, it follows that $\omega^{\mu{\overline\nu}} \delta \omega_{\mu{\overline\nu}}$ is a function of parameters only. Furthermore, in holomorphic gauge, \eqref{eq:finalGauge}, $\mathfrak{D}_\alpha \omega^{(2,0)} = 0$ and ${\overline{\mathcal Z}}_\alpha^{(1,1)} = 0$ meaning ${\mathcal Z}{}_\alpha^{(1,1)} = 2{\rm i} \mathfrak{D}_\alpha \omega^{(1,1)}$.

By definition of the adjoint operators and using Hodge dual relations in Appendix~\ref{s:appendixHodge}, we find two conditions derive from the balanced equation
\begin{gather}\label{eq:Balanced2}
 {\overline{\partial}}^\dag \mathfrak{D}_\alpha \omega^{(1,1)} = 0, \qquad {\overline{\partial}}^\dag \mathfrak{D}_\alpha \omega^{(0,2)} = \partial^\dag \mathfrak{D}_\alpha \omega^{(1,1)} .
\end{gather}
In virtue of the first equation, the Hodge decomposition for $\mathfrak{D}_\alpha\omega^{(1,1)}$ with respect to the ${\overline{\partial}}$-operator is
\begin{gather}\label{eq:HodgeOmega}
\mathfrak{D}_\alpha \omega^{(1,1)} = \mathfrak{D}_\alpha\omega^{(1,1)}{}^{\rm harm} + {\overline{\partial}}^\dag\xi_\alpha^{(1,2)},
\end{gather}
that is, there is no ${\overline{\partial}}$-exact term. The first term is ${\overline{\partial}}$-harmonic and is completely determined via Hodge theory of~$X$ up to a parameter dependent matrix. This matrix can be accounted for by a holomorphic change of coordinates on the moduli space. The ${\overline{\partial}}$-coexact bit is determined by the moduli equation~\eqref{eq:moduliEqnHol}. To see this note that ${\mathcal Z}{}_\alpha^{(1,1)} = 2{\rm i} \mathfrak{D}_\alpha \omega^{(1,1)}$ and so
\[
{\overline{\partial}} {\overline{\partial}}^\dag \xi_\alpha^{(1,2)} = \Delta_\alpha{}^\mu (\partial\omega)_\mu - \tfrac{{\rm i}\alpha^{\backprime}}{4} \operatorname{Tr}( \mathfrak{D}_\alpha {\mathcal A} F) + \tfrac{{\rm i} \alpha^{\backprime}}{2} \big( \nabla_\nu \Delta_\alpha{}^\mu + {\rm i} \nabla^\mu \mathfrak{D}_\alpha\omega_\nu^{(0,1)} \big) R^\nu{}_\mu .
\]
The Hodge decomposition of the $(1,2)$-form is
\[
 \xi_\alpha^{(1,2)} = \xi_\alpha^{(1,2)}{}^{\rm harm} + {\overline{\partial}} \rho_\alpha^{(1,1)} + {\overline{\partial}}^\dag \lambda_\alpha^{(1,3)} .
\]
The terms $\xi_\alpha^{(1,2)}{}^{\rm harm}$ and ${\overline{\partial}}^\dag \lambda_\alpha^{(1,3)}$ do not contribute to $\mathfrak{D}_\alpha \omega^{(1,1)}$ and so are not physical degree of freedom. The remaining term is determined by substituting into the previous equation, and inverting the Laplacian
\begin{gather}
 \xi_\alpha^{(1,2)} = \Box_{\overline{\partial}}^{-1} \big( \Delta_\alpha{}^\mu (\partial\omega)_\mu - \tfrac{{\rm i}\alpha^{\backprime}}{4} \operatorname{Tr}( \mathfrak{D}_\alpha {\mathcal A} F) \notag\\
\hphantom{\xi_\alpha^{(1,2)} =}{} + \tfrac{{\rm i} \alpha^{\backprime}}{2} \big( \nabla_\nu \Delta_\alpha{}^\mu + {\rm i} \nabla^\mu \mathfrak{D}_\alpha\omega_\nu^{(0,1)} \big) R^\nu{}_\mu \big) + {\overline{\partial}}^\dag\text{-closed}.\label{eq:HodgeOmega2}
\end{gather}
The term in the bracket is not a zero mode of Laplacian by assumption and we have not written a~co-closed term as it does not contribute to \eqref{eq:HodgeOmega}. We will come back to this equation in a~moment.

Returning to \eqref{eq:Balanced2}, the second equation can be combined with \eqref{eq:GaugeFixDelOmega} to gain further information about $\Delta_\alpha$. Start with the left hand side
\[
 {\overline{\partial}}^\dag \mathfrak{D}_\alpha\omega^{(0,2)} = - \star \partial \big(\mathfrak{D}_\alpha\omega^{(0,2)} \omega\big) = - \star \big( \partial\mathfrak{D}_\alpha\omega^{(0,2)} \big) \omega -\star\big( \mathfrak{D}_\alpha\omega^{(0,2)} \partial\omega \big) ,
\]
and make use of \eqref{eq:Hodge23} to express this as
\[
 {\overline{\partial}}^\dag \mathfrak{D}_\alpha\omega^{(0,2)} = - \nabla^{{\rm Ch}\,{\overline\nu}} \big(\mathfrak{D}_\alpha\omega_{{\overline\nu}}^{(0,1)}\big) - \tfrac{{\rm i}}{2} \mathfrak{D}_\alpha\omega_{{\overline{\mu}}{\overline\nu}} (\partial\omega)^{{\overline{\mu}}{\overline\nu}} .
\]
The second equation in \eqref{eq:Balanced2} can now be written as
\begin{gather*}%\label{eq:balanced2}
 \nabla^{{\rm Ch}\,{\overline\nu}}\big(\mathfrak{D}_\alpha\omega_{{\overline\nu}}^{(0,1)}\big) = - \partial^\dag\mathfrak{D}_\alpha\omega^{(1,1)} - \tfrac{{\rm i}}{2} \mathfrak{D}_\alpha\omega_{{\overline{\mu}}{\overline\nu}} (\partial\omega)^{{\overline{\mu}}{\overline\nu}} .
\end{gather*}
The gauge fixing in \eqref{eq:GaugeFixDelOmega} becomes
 \begin{gather}
 {\overline{\partial}}^\dag \Delta_\alpha{}^\mu = - {\rm i} \big( \partial^\dag \mathfrak{D}_\alpha\omega^{(1,1)} \big)_{{\overline\nu}} g^{{\overline\nu}\mu} + \frac{1}{2} \, \mathfrak{D}_\alpha\omega_{{\overline\nu}{\overline{\rho}}} (\partial\omega)^{{\overline\nu}{\overline{\rho}}\mu} .\label{eq:delbdagDelta}
\end{gather}
The Hodge decomposition of $\Delta_\alpha$ with respect to the ${\overline{\partial}}$ operator is
\begin{gather}\label{eq:HodgeDelta1}
 \Delta_\alpha = \Delta_\alpha{}^{\rm harm} + {\overline{\partial}} \kappa_\alpha , \qquad \kappa_\alpha = \kappa_\alpha{}^\mu \partial_\mu .
\end{gather}
The first term is determined by Hodge theory as was the case for the Hermitian form. We have used that $\Delta_\alpha$ is integrable ${\overline{\partial}}\Delta_\alpha = 0$ and so there is no ${\overline{\partial}}$-coexact term. Substituting into~\eqref{eq:delbdagDelta} and using that the tangent bundle is stable $H^0_{{\overline{\partial}}}(X,{\mathcal T}_X) = 0$ to invert the Laplacian, we can solve for the vector
\begin{gather}\label{eq:HodgeOmega2b}
 \kappa_\alpha{}^\mu = \Box_{{\overline{\partial}}}^{-1} \big( {-} {\rm i} \big( \partial^\dag \mathfrak{D}_\alpha\omega^{(1,1)} \big)_{{\overline\nu}} g^{{\overline\nu}\mu} + \tfrac{1}{2} \mathfrak{D}_\alpha\omega_{{\overline\nu}{\overline{\rho}}} (\partial\omega)^{{\overline\nu}{\overline{\rho}}\mu} \big) .
\end{gather}
We will return to this equation shortly.

Finally, consider the Hermitian--Yang--Mills (HYM) equation. The Donaldson--Uhlenbeck--Yau (DUY) theorem asserts that given the bundle is stable, we find a unique connection $A$ such that
\[
\omega^2 F = 0.
\]
Suppose this is the case. Under a deformation of the gauge field, with $X$ fixed, we end up
\[
{\overline{\partial}}_{\mathcal A}^\dag \mathfrak{D}_a {\mathcal A} = -\partial_{{\mathcal A}^\dag}^\dag \mathfrak{D}_a {\mathcal A}^\dag,
\]
which is an equation of motion of the fluctuations.

Suppose parameters also vary the manifold $X$. In that case a holomorphic variation of the HYM is
\begin{gather*}
 {\rm i} (\mathfrak{D}_\alpha \omega_{\mu{\overline\nu}}) F^{\mu{\overline\nu}} + {\overline{\partial}}_{\mathcal A}^\dag \mathfrak{D}_\alpha {\mathcal A} = 0,
\end{gather*}
where we use \eqref{eq:Hodge2form}. Write a Hodge decomposition
\begin{gather}\label{eq:HodgeDecompA1}
 \mathfrak{D}_\alpha {\mathcal A} = \mathfrak{D}_\alpha {\mathcal A}^{\rm harm} + {\overline{\partial}}_{\mathcal A} \Phi_\alpha + {\overline{\partial}}_{\mathcal A}^\dag \Psi_\alpha^{(0,2)},
\end{gather}
for some $\operatorname{End}{\mathcal E}$-valued $(0,2)$-form $\Psi_\alpha$ and scalar $\Phi_\alpha$. The harmonic representative is determined purely by Hodge theory. The remaining pieces are determined by the Atiyah equation and the HYM equation as we now show.

Firstly, the Atiyah equation gives
\begin{gather*}%\label{eq:AtiyahDef}
 {\overline{\partial}}_{\mathcal A} {\overline{\partial}}_{\mathcal A}^\dag \Psi_\alpha^{(0,2)} = \Delta_\alpha{}^\mu F_\mu.
 \end{gather*}
As we have done previously, the Hodge decomposition of $\Psi_\alpha^{(0,2)} = {\overline{\partial}}_{\mathcal A} \Xi_\alpha^{(0,1)} + {\overline{\partial}}^\dag (\cdots)$, where the second term does not contribute to $\mathfrak{D}_\alpha {\mathcal A}$ and so is not written explicitly. The Atiyah equation becomes a Laplacian which we can invert for unobstructed deformations $ {\overline{\partial}}_{{\mathcal A}} \Xi_\alpha^{(0,1)}= \Box^{-1}_{{\overline{\partial}}_{\mathcal A}} ( \Delta_\alpha{}^\mu F_\mu)$. Hence, we determine the last term in \eqref{eq:HodgeDecompA1}
\begin{gather}\label{eq:HodgeOmega2c}
 \Psi_\alpha^{(0,2)} = \Box^{-1}_{{\overline{\partial}}_{\mathcal A}} ( \Delta_\alpha{}^\mu F_\mu) + {\overline{\partial}}_{\mathcal A}^\dag\text{-closed},
 \end{gather}
where we have not written the ${\overline{\partial}}_{\mathcal A}^\dag$-closed term as it does not contribute to $\mathfrak{D}_a {\mathcal A}$.

Secondly, because $\big({\overline{\partial}}_{\mathcal A}^\dag\big)^2 = 0$ we find
\begin{gather}\label{eq:HYMvariation}
 \Box_{{\overline{\partial}}_{\mathcal A}} \Phi_\alpha = -{\rm i} (\mathfrak{D}_\alpha \omega_{\mu{\overline\nu}}) F^{\mu{\overline\nu}}.
\end{gather}
The bundle satisfies the HYM and so is stable, thence $\Box_{{\overline{\partial}}_{\mathcal A}}$ has trivial kernel. This means we can invert the Laplacian for $\Phi_\alpha$.

\subsection{The Hodge decomposition in holomorphic gauge}\label{s:HodgeHeterotic}

We collect the results \eqref{eq:HodgeOmega}, \eqref{eq:HodgeDelta1}, and \eqref{eq:HodgeDecompA1} together in one place
\begin{gather}
 \mathfrak{D}_\alpha \omega^{(1,1)} = \mathfrak{D}_\alpha \omega^{(1,1)\,{\rm harm}} + {\overline{\partial}}^\dag\xi_\alpha^{(1,2)},\nonumber\\
\Delta_\alpha = \Delta_\alpha{}^{\rm harm} + {\overline{\partial}} \kappa_\alpha ,\nonumber\\
 \mathfrak{D}_\alpha {\mathcal A} = \mathfrak{D}_\alpha {\mathcal A}^{\rm harm} + {\overline{\partial}}_{\mathcal A} \Phi_\alpha + {\overline{\partial}}_{\mathcal A}^\dag \Psi_\alpha^{(0,2)}.\label{eq:HodgeHet1}
\end{gather}
The field $\mathfrak{D}_\alpha \omega^{(0,2)} = \Delta_\alpha{}^\mu \omega_\mu$ obeys the following equation ${\overline{\partial}}^\dag \Delta_\alpha{}^\mu = {\rm i} g^{\mu{\overline{\rho}}} \nabla^{{\rm Ch}\,{\overline\nu}} \big(\mathfrak{D}_\alpha \omega_{{\overline\nu}{\overline{\rho}}} \big)$.

The gauge fixing and equations of motion allow us to determine the exact and co-exact terms. Collecting \eqref{eq:HodgeOmega2}, \eqref{eq:HodgeOmega2b}, \eqref{eq:HodgeOmega2c}, \eqref{eq:HYMvariation} together in one place:
\begin{subequations}\label{eq:HodgeHeterotic2}
\begin{gather}
 \xi_\alpha^{(1,2)} = \Box_{\overline{\partial}}^{-1} \big( \Delta_\alpha{}^\mu (\partial\omega)_\mu - \tfrac{{\rm i}\alpha^{\backprime}}{4} \operatorname{Tr} ( \mathfrak{D}_\alpha {\mathcal A} F )\notag\\
\hphantom{\xi_\alpha^{(1,2)} =}{} + \tfrac{{\rm i}\alpha^{\backprime}}{2} \big( \nabla_\nu \Delta_\alpha{}^\mu + {\rm i} \nabla^\mu \mathfrak{D}_\alpha\omega_\nu^{(0,1)} \big) R^\nu{}_\mu \big) + {\overline{\partial}}^\dag\text{-closed} ,\label{eq:HodgeHeterotic2a}\\
 \kappa_\alpha{}^\mu = \Box_{{\overline{\partial}}}^{-1}\big( {-} {\rm i} \big( \partial^\dag \mathfrak{D}_\alpha\omega^{(1,1)} \big)^\mu + \tfrac{1}{2}( \mathfrak{D}_\alpha \omega_{{\overline\nu}{\overline{\rho}}})(\partial\omega)^{{\overline\nu}{\overline{\rho}}\mu}\big),\label{eq:HodgeHeterotic2b}\\
 \Psi_\alpha^{(0,2)}= \Box^{-1}_{{\overline{\partial}}_{\mathcal A}} ( \Delta_\alpha{}^\mu F_\mu) + {\overline{\partial}}_{\mathcal A}^\dag\text{-closed} , \qquad \Phi_\alpha = \Box_{{\overline{\partial}}_{\mathcal A}}^{-1} \big( {-} {\rm i} \big(\mathfrak{D}_\alpha \omega^{\mu{\overline\nu}}\big) F_{\mu{\overline\nu}}\big).\label{eq:HodgeHeterotic2c}
 \end{gather}
\end{subequations}
We have assumed the Laplacian is invertible: that is the sources have no zero modes of the Laplacian. The conditions for this to be the case are written in the previous subsection, we repeat here. Consider the first equation of~\eqref{eq:HodgeHeterotic2c}. We have $\Delta_\alpha{}^\mu F_\mu$ is ${\overline{\partial}}_{\mathcal A}$-closed. The Laplacian is invertible if and only if $\Delta_\alpha{}^\mu F_\mu$ is ${\overline{\partial}}_{\mathcal A}$-exact, which is of course, exactly the Atiyah condition~\eqref{eq:Atiyah}. Similarly, the source in \eqref{eq:HodgeHeterotic2a} needs to be ${\overline{\partial}}$-exact which is exactly the last line of \eqref{eq:moduliEqnHol}. On the other Laplacians for $\kappa_\alpha$ and $\Phi_\alpha$ are always invertible provided the bundles are stable. So we really view \eqref{eq:HodgeHeterotic2} as being supplemented by~\eqref{eq:moduliEqnHol}.

We have gauge fixed. Hence, the ${\overline{\partial}}_{\mathcal A}$ and ${\overline{\partial}}$-exact pieces in~\eqref{eq:HodgeHet1} are physical. We could try to discard them by a gauge transformation, but as we have fixed the gauge completely, this would inevitably lead to a condition such as $\mathfrak{D}_\alpha {\mathcal A}^\dag \ne 0$ or $\partial \chi_\alpha \ne 0$. Such terms were set to zero in deriving the equations of above. This is just as in electromagnetism: in say Coulomb gauge, the potential $A$ is physical as the gauge has been fixed, and we no longer have the freedom to perform gauge transformations.

Turning to \eqref{eq:HodgeHeterotic2}, supplemented by \eqref{eq:moduliEqnHol}, these are highly coupled equations, and finding a~direct solution is very hard.\footnote{We would like to thank Mario Garcia-Fernandez for discussion on this point.} These equations are another, rather explicit, echo of the intuition that there is no clear decoupling of parameters between~$X$ and~${\mathcal E}$. One approach is to proceed perturbatively in~$\alpha^{\backprime}$. Expand fields in~$\alpha^{\backprime}$ with a square bracket denoting the order
\[
\mathfrak{D}_\alpha \omega = [\mathfrak{D}_\alpha \omega]_0 +\alpha^{\backprime} [\mathfrak{D}_\alpha \omega]_1 + \cdots, \qquad \xi_\alpha = [\xi_\alpha]_0 + \alpha^{\backprime}[\xi_\alpha]_1 + \cdots.
\]
At zeroth order in $\alpha^{\backprime}$, as we are assuming a large radius limit so that SUGRA embeds into string theory, we have $[{\rm d} \omega]_0 = 0$. The right hand side of the first equation in \eqref{eq:HodgeHeterotic2a} is therefore~$\mathcal{O}(\alpha^{\backprime})$ and so~$ [{\overline{\partial}}^\dag\xi_\alpha^{(1,2)}]_0 = 0$ to this order. Hence, $[\mathfrak{D}_\alpha \omega^{(1,1)}]_0$ is harmonic with respect to the zeroth order Calabi--Yau metric $[g_{\mu{\overline\nu}}]_0 $. Also, $[\mathfrak{D}_\alpha \omega^{(0,2)}]_0=0$ as follows by the calculation in Appendix~\ref{eq:hologaugefix}. Using equivalence of K\"ahler Laplacians, the right hand side of \eqref{eq:HodgeHeterotic2b} also vanishes and so $[\Delta_\alpha{}]_0$ is harmonic with respect to $[g_{\mu{\overline\nu}}]_0$. Finally, the sources in \eqref{eq:HodgeHeterotic2c} completely determines $[\mathfrak{D}_\alpha {\mathcal A}]_0$. In summary, $[\mathfrak{D}_\alpha \omega]_0 $, $[\Delta_\alpha{}]_0$ are both harmonic with respect to the Calabi--Yau metric while $[\mathfrak{D}_\alpha {\mathcal A}]_0$ is harmonic at zeroth order only when $[\Delta_\alpha ]_0=\big[\mathfrak{D}_\alpha \omega^{(1,1)}\big]_0 = 0$.

We now do the first order in $\alpha^{\backprime}$ analysis. Due to $\partial \omega = \mathcal{O}(\alpha^{\backprime})$ the right hand side of the first equation in \eqref{eq:HodgeHeterotic2a} is determined by $[\Delta_\alpha]_0$ and $[\mathfrak{D}_\alpha {\mathcal A}]_0$. Hence, $[\mathfrak{D}_\alpha \omega]_1$ can be computed by inverting the Laplacian and using the $\alpha^{\backprime}$-corrected harmonic representative $\mathfrak{D}_\alpha \omega^{(1,1)\,{\rm harm}}$. We have determined $\big[\mathfrak{D}_\alpha\omega^{(1,1)}\big]_1$. This can then be used in \eqref{eq:HodgeHeterotic2b}. Neglecting the term $(\mathfrak{D}_\alpha \omega_{{\overline{\mu}}{\overline\nu}}) (\partial\omega)^{{\overline{\mu}}{\overline\nu}\rho} = \mathcal{O}\big({\alpha^{\backprime}}^2\big)$, we can solve for $[\kappa_\alpha{}^\mu]_1$ and determine $[\Delta_\alpha{}^\mu]_1$. Finally, substitute $\big[\mathfrak{D}_\alpha\omega^{(1,1)}\big]_1$ and $[\Delta_\alpha{}^\mu]_1$ into \eqref{eq:HodgeHeterotic2c} and solve for $[\mathfrak{D}_\alpha {\mathcal A}]_1$. We are finished to $\mathcal{O}(\alpha^{\backprime})$.

\subsection{Comments}

We pause to make some comments on results so far. First, we now see how quantum corrections modify the field deformations: they do so by the exact and co-exact pieces. These pieces are determined by contractions of zeroth order harmonic representatives with the background fields. The harmonic representatives are a linear combination of elements of $H^1\big(X, {\mathcal T}_X^{(0,1)}\big) \oplus H^1\big(X, {\mathcal T}_X^{(1,0)}\big) \oplus H^1(X, \operatorname{End} V)$. The role of Atiyah equation and anomaly cancellation is to determine which linear combinations to take. For example, we need conditions like
 \begin{gather}
 [\Delta_\alpha{}^\mu]_0 [F_{\mu}]_0 \qquad \text{and} \qquad [\Delta_\alpha{}^\mu]_1 [F_\mu]_0 + [\Delta_\alpha{}^\mu]_0 [F_\mu]_1 = {\overline{\partial}}_{{\mathcal A}}\text{-exact} ,\nonumber\\
 [\Delta_\alpha{}^\mu]_0 [(\partial\omega)_\mu]_1 - \frac{{\rm i}}{4} \big[\operatorname{Tr}( \mathfrak{D}_\alpha {\mathcal A} F) - 2 \big( \nabla_\nu \Delta_\alpha{}^\mu + {\rm i} \nabla^\mu \mathfrak{D}_\alpha\omega_\nu^{(0,1)} \big) R^\nu{}_\mu \big]_0 = {\overline{\partial}}\text{-exact} .\label{eq:obstructions}
\end{gather}
The elements that satisfy these equations form a vector space which is related to a first cohomology of an appropriate bundle, see \cite{Anderson:2014xha,delaOssa:2014msa,delaOssa:2014cia, Garcia-Fernandez:2015hja, Garcia-Fernandez:2018ypt} building on the work of \cite{Melnikov:2011ez} who describe the zeroth order in $\alpha^{\backprime}$ deformations at the level of a non-linear sigma models, taking into account gauge symmetries. However, in constructing these bundles it is convenient to treat deformations of the spin connection $\mathfrak{D}_\alpha \Theta$ as independent degrees of freedom. Doing so gives rise to additional but unphysical directions in the fibres of the bundle and one needs to presumably perform an additional quotient to get the physical degrees of freedom. In~\eqref{eq:obstructions} we have not treated $\mathfrak{D}_\alpha \Theta$ as an independent degree of freedom. Instead we have used~\eqref{eq:spinmoduli} relating $\mathfrak{D}_\alpha \Theta$ to deformations of the metric. This should amount to a quotient of the cohomology but seeing this remains future work.

Second, the analysis performed here is complementary to finding how the background fields are corrected perturbatively in $\alpha^{\backprime}$, see for example~\cite{Anguelova:2010ed,Gillard:2003jh, Witten:1985bz,Witten:1986kg}. In these works the gauge-fixing for the corrections to the background is discussed. Recall that we assume a perturbative string background in which the 10-dimensional dilaton is a constant, see Appendix~\ref{s:dilatongauge}.

Third, the field $\Delta_\alpha$ describes a holomorphic deformation of complex structure. Note that it is perturbatively corrected in $\alpha^{\backprime}$. This is unlike the non-renormalisation that occurs in some type II theories.

Fourth, it is unlikely that one can keep $X$ fixed and vary the bundle at first order in $\alpha^{\backprime}$. To see this, suppose we try to hold~$X$ fixed. Then $[\Delta_\alpha]_0 = [\mathfrak{D}_\alpha \omega]_0 = 0$ and $[\mathfrak{D}_\alpha {\mathcal A}]_0$ are harmonic. At first order in $\alpha^{\backprime}$, \eqref{eq:HodgeHeterotic2a} implies that $\big[\mathfrak{D}_\alpha \omega^{(1,1)}\big]_1$ is turned on and sources \eqref{eq:HodgeHeterotic2b}--\eqref{eq:HodgeHeterotic2c}, unless the following cancellation happens
\begin{gather}\label{eq:cancellationXfixed}
 \big[ \operatorname{Tr}{(\mathfrak{D}_\alpha{\mathcal A}^{\rm harm} F)} \big]_0 = 0 ,
\end{gather}
which is a much stronger condition than \eqref{eq:obstructions}. This phenomenon arose because the conformally balanced and HYM equation, together with gauge fixing holomorphic gauge, \eqref{eq:finalGauge}, coupled these degrees of freedom together.

Fifth, the fixing of small gauge transformations is a necessary part of the map between parameters and physical field deformations. Our choice of holomorphic gauge is natural for an ${\mathcal N} = 1$ supersymmetric theory in which the moduli space is complex. Whether there exists choice of gauge which brings further simplification is far from clear.

\subsection{An application to the moduli space metric}

We can apply these results to the K\"ahler moduli space metric derived in \cite{Candelas:2016usb, Candelas:2018lib}
 \begin{gather*}
 g^\#_{\alpha{\overline{\beta}}} = \frac{1}{V}\int_X \big( \Delta_\alpha{}^\mu\star\Delta_{{\overline{\beta}}}{}^{{\overline\nu}} g_{\mu{\overline\nu}} + \mathfrak{D}_\alpha\omega^{(1,1)}\star\mathfrak{D}_{{\overline{\beta}}}\omega^{(1,1)} + \tfrac{\alpha^{\backprime}}{4}\operatorname{Tr}{ ( \mathfrak{D}_\alpha A \star \mathfrak{D}_{{\overline{\beta}}}A ) } \big)\\
\hphantom{g^\#_{\alpha{\overline{\beta}}} =}{} + \frac{\alpha^{\backprime}}{2V}\int_X \operatorname{vol} \big( \Delta_{\alpha{\overline{\mu}}{\overline\nu}} \Delta_{{\overline{\beta}}\rho\sigma} + \mathfrak{D}_\alpha\omega_{\rho{\overline{\mu}}} \mathfrak{D}_{{\overline{\beta}}}\omega_{\sigma{\overline\nu}} \big) R^{{\overline{\mu}}\rho{\overline\nu}\sigma} + \mathcal{O}\big({\alpha^{\backprime}}^2\big) .
\end{gather*}
This metric is a series of inner products together with an $\alpha^{\backprime}$-correction involving a contraction with the Riemann tensor. The term involving the Riemann tensor appears because of~\eqref{eq:spinmoduli} as mentioned above.

A nice thing that happens is that the exact and co-exact terms in~\eqref{eq:HodgeHet1} either cancel or contribute in a way that is higher order in $\alpha^{\backprime}$. This is due to harmonic, exact and co-exact forms being mutually orthogonal. The $\alpha^{\backprime}$-corrections to the harmonic forms are also orthogonal. For example, writing
\[
\big[\Delta_\alpha{}^{\rm harm}\big] = \big[\Delta_\alpha{}^{\rm harm}\big]_0 + \alpha^{\backprime} \big[\Delta_\alpha{}^{\rm harm}\big]_1.
\]
We may as well take $\big[\Delta_\alpha{}^{\rm harm}\big]_1$ to be orthogonal to $\big[\Delta_\alpha{}^{\rm harm}\big]_0$ as is done in say \cite{Anguelova:2010ed}. Any contribution which is ``parallel'' is just a change of basis and represents a vacuous change. Hence,
\[
\int_X \Delta_\alpha{}^\mu \star \Delta_{\overline\beta}{}^{\overline\nu} g_{\mu{\overline\nu}} = \int_X \big[\Delta_\alpha{}^\mu\big]_0 \star \big[\Delta_{\overline\beta}{}^{\overline\nu}\big]_0 g_{\mu{\overline\nu}} + \mathcal{O}\big({\alpha^{\backprime}}^2\big).
\]
A similar result applies for the other inner products.
Perhaps what was not obvious before the analysis in this section was whether there could be corrections to $\mathfrak{D}_\alpha \omega$ or $\Delta_\alpha$ which would contribute to the metric at $\mathcal{O}(\alpha^{\backprime})$. We now see the answer is no; the $\alpha^{\backprime}$-corrections come either from $\star$ or are written explicitly as above. It would be interesting to see how this structure behaves at next order in $\alpha^{\backprime}$.

We can give a concrete prescription for computing the metric
\begin{gather*}
 g^\#_{\alpha{\overline{\beta}}} = \frac{1}{V} \!\int_X \!\big( \big[\Delta_\alpha{}^{\mu}\big]_0 \star \big[\Delta_{{\overline{\beta}}}{}^{{\overline\nu}}\big]_0 g_{\mu{\overline\nu}} + \big[\mathfrak{D}_\alpha \omega^{(1,1)}\big]_0 \star \big[\mathfrak{D}_{\overline\beta} \omega^{(1,1)}\big]_0\! + \tfrac{\alpha^{\backprime}}{4} \operatorname{Tr}{\big( [\mathfrak{D}_\alpha A]_0 \star [\mathfrak{D}_{\overline\beta} A]_0 \big)} \big)\nonumber\\
\hphantom{g^\#_{\alpha{\overline{\beta}}} =}{} + \frac{\alpha^{\backprime}}{2V} \int_X \big[\operatorname{vol} \big( \Delta_{\alpha {\overline{\mu}}{\overline\nu}} \Delta_{{\overline{\beta}} \rho\sigma} + \mathfrak{D}_\alpha \omega_{\rho{\overline{\mu}}} \mathfrak{D}_{\overline\beta} \omega_{\sigma{\overline\nu}} \big) R^{{\overline{\mu}}\rho{\overline\nu}\sigma} \big]_0
 + \mathcal{O}\big({\alpha^{\backprime}}^2\big) .% \label{eq:modulimetric}
\end{gather*}
The terms $[\Delta_\alpha{}^\mu]_0$ and $[\mathfrak{D}_\alpha \omega^{(1,1)}]_0$ are the zeroth order harmonic forms. The only $\alpha^{\backprime}$-correction in the first two terms comes from background metric $g_{\mu{\overline\nu}}$, and their computation are discussed in \cite{Anguelova:2010ed,Gillard:2003jh, Witten:1986kg}.

Just as the case for special geometry \cite{Candelas:1990pi}, we believe this metric to coincide with the Zamalodchikov metric on the underlying $(0,2)$ SCFT corresponding to heterotic moduli. This forms a~central part of what we would believe would amount to heterotic special geometry and the generalisation of the $tt^*$ equations to~$(0,2)$ theories. The lowest order contribution was determined in~\cite{Candelas:2018lib} and from a different point of view in \cite{Anguelova:2010ed}. Note that this is the physical string theory metric, in which the spin connection is determined in terms of the remaining moduli. It differs from the metric on the twisted cohomology studied in \cite{Anderson:2014xha,delaOssa:2014msa,delaOssa:2014cia, Garcia-Fernandez:2015hja, Garcia-Fernandez:2018ypt} which contains spurious degrees of freedom by treating the spin connection as an independent degree of freedom. The gauge charged matter sector forms another part of the data for heterotic special geometry. At the level of supergravity the matter sector metric is computed in~\cite{McOrist:2016cfl} and classical Yukawa couplings. For theories connected to the standard embedding, these couplings are computed in quasi-topological linear sigma models in \cite{McOrist:2007kp,McOrist:2008ji, McOrist:2011bn}, while a mathematics discussion is in \cite{Donagi:2011va,Donagi:2011uz}; see \cite{McOrist:2010ae,Melnikov:2012hk,Melnikov:2019tpl} for some reviews. A key open question remains: are there any special relations between the gauge charged and neutral sectors as is the case in special geometry?

\section{First order universal geometry and gauge transformations}\label{s:firstorderUG}

In the previous section we gauge-fixed and used the equations of motion to isolate the physical degrees of freedom. In this section, we put this into the context of the universal geometry constructed in \cite{Candelas:2016usb, Candelas:2018lib}. We will find that a small transformation corresponds to deforming a~connection on the moduli space. This means that the choice of connection on~$M$ corresponds to a~choice of gauge. This makes sense: neither the connection on~$M$ nor the choice of gauge are physical and so should not enter the physical theory.

So far we have not been too detailed about parameter space derivatives. They are needed for general covariance of the theory as symmetries can depend on coordinates of~$X$ but also on parameters.

\subsection{Why universal geometry?}\label{s:whyUG}

In the classic works on complex structure moduli space for CY manifolds, such as \cite{Candelas:1990pi, Strominger:1990pd}, the authors benefited from the existence of special coordinates called periods of $\Omega$. In those coordinates $\Omega$ takes a simple form
\[
 \Omega(x,z) = z^a \alpha_a(x) - {\mathcal G}_a(z) \beta^a(x) ,
\]
where $\alpha_a$, $\beta^a$ are a symplectic basis for $H^3(X)$. Within this choice of special coordinates it is sensible to compute deformations through a partial derivative $\partial_a$. This is because the tensorial part of $\Omega$ does not depend on parameters: the forms $\alpha_a$ and $\beta^a$ are `constant' across $M$. This feature remains true as long as diffeomorphisms and symplectic rotations do not depend on parameters.

The same story goes for the complexified K\"ahler form $B+{\rm i} \omega$. There exist special coordinates on $M$ in which this depends linearly on parameters
\[
(B + {\rm i} \omega)(x,t) = t^a \omega_a(x) ,
\]
with $\omega_a$ a choice of `constant' harmonic $(1,1)$-forms.

The existence of coordinates on $M$ of this type is a remarkable feature. One might wonder, given the intricate analysis of the previous section, if similar coordinates exist for heterotic moduli. If one wants to investigate without assuming that special coordinates like above exist, a more covariant formalism is needed. Universal geometry allows to do that.

The caveat~-- or the gain, according to one's view~-- of universal geometry is that one needs to study connections on the moduli space and their associated curvatures, as well as higher order tensors. In particular, we cannot really assume that deformations commute $[\mathfrak{D}_a, \mathfrak{D}_b]=0$ as a starting point. We will see in the next section how important is the role of gauge fixing: curvatures have their own small gauge transformations and these affect commutativity.

\subsection[Small gauge transformations on a fixed manifold $X$]{Small gauge transformations on a fixed manifold $\boldsymbol{X}$}\label{eq:smallgaugeonX}

We now sketch some of the essential details of the universal bundle using a toy example, however we refer the reader to~\cite{Candelas:2018lib} for details of the universal bundle construction. The toy example is the space of heterotic theories in the limit $\alpha^{\backprime}\to 0$, in which the manifold $X$ is kept fixed and we only vary the connection $A$ on the vector bundle ${\mathcal E}$. To be specific, this vector bundle has a structure group $G$ whose Lie algebra is denoted ${\mathfrak g}$. The moduli space of these theories~-- which we will keep denoting by $M$ for simplicity~-- coincides with the space of holomorphic connections on $X$ that solve the HYM equation. This is closely related to the context of the Kobayashi--Hitchin correspondence~\cite{Kobayashi:1987}. We have taken the $\alpha^{\backprime}\to0$ limit because a result of this paper is that, with non-zero $\alpha^{\backprime}$, in order to vary the connection $A$ while leaving the manifold $X$ fixed one needs the highly non-trivial condition~\eqref{eq:cancellationXfixed}. While this setting is not accounting for all the heterotic features, it is good enough to illustrate the relevant points.

We wish to study deformations $\delta A$. The background gauge principle implies two types of transformation properties for $\delta A$:
\begin{gather}\label{eq:gaugeDeltaA}
 \delta A \to {} ^\Phi \delta A = \Phi \delta A \Phi^{-1} , \qquad \delta A\to {} ^{\phi} \delta A = \delta A + {\rm d}_A \phi, \qquad \phi \ll 1.
\end{gather}
The former is a classical symmetry of the background, while the latter is a small gauge transformation. To make progress in understanding the moduli space~$M$ we need a relation between parameters and deformations of fields, which we refer to as the Kodaira--Spencer map~\cite{Kodaira:1981cx}, that respects the transformation laws of~$\delta A$.

A solution implementing the first symmetry of \eqref{eq:gaugeDeltaA} was described in~\cite{Candelas:2016usb}. One introduces an 1-form $\Lambda = {\Lambda}_a {\rm d} y^a$ with legs along the moduli space so that it transforms under~$G$ in a manner parallel to~$A$
\begin{gather}
 \Lambda_a\to {}^\Phi\Lambda_a = \Phi \Lambda_a \Phi^{-1} - (\partial_a\Phi) \Phi^{-1}.
\label{eq:Ltransf}\end{gather}
As described in detail in~\cite{Candelas:2018lib} we can think of $\Lambda$ as a connection for a vector bundle whose base manifold includes~$X$ and~$M$. We will be more precise about this vector bundle shortly. For now however, all we want to observe that with the formal transformation property~\eqref{eq:Ltransf}, we can express the deformation in terms of a covariant derivative, $\delta A = \delta y^a \mathfrak{D}_a A$ where
\begin{gather}\label{eq:covariantderivative1}
 \mathfrak{D}_a A = \partial_a A - {\rm d}_A \Lambda_a, \qquad \text{where}\quad {\rm d}_A \Lambda_a = {\rm d}\Lambda_a + [A, \Lambda_a],
\end{gather}
and it is easy to check that under \eqref{eq:Atransf} and \eqref{eq:Ltransf} the covariant derivative $\mathfrak{D}_a A$ transforms homogeneously as required.

While this takes care of background gauge transformations, small gauge transformations were not studied in \cite{Candelas:2016usb,Candelas:2018lib, McOrist:2016cfl}. Here we explore them in more detail especially in relation to the gauge-fixing of previous sections. The key observation is that by comparing \eqref{eq:gaugeDeltaA}--\eqref{eq:covariantderivative1} we see that a~small gauge transformation precisely corresponds to a~deformation of the connection
\begin{gather}\label{eq:smallgaugeLambda}
\Lambda_a \to \Lambda_a - \phi_a ,
\end{gather}
where we remind the reader the $\phi_a$ is a deformation of a connection and so is adjoint valued. Hence, in fixing small gauge transformations, we must have fixed $\Lambda$, at least to some degree.

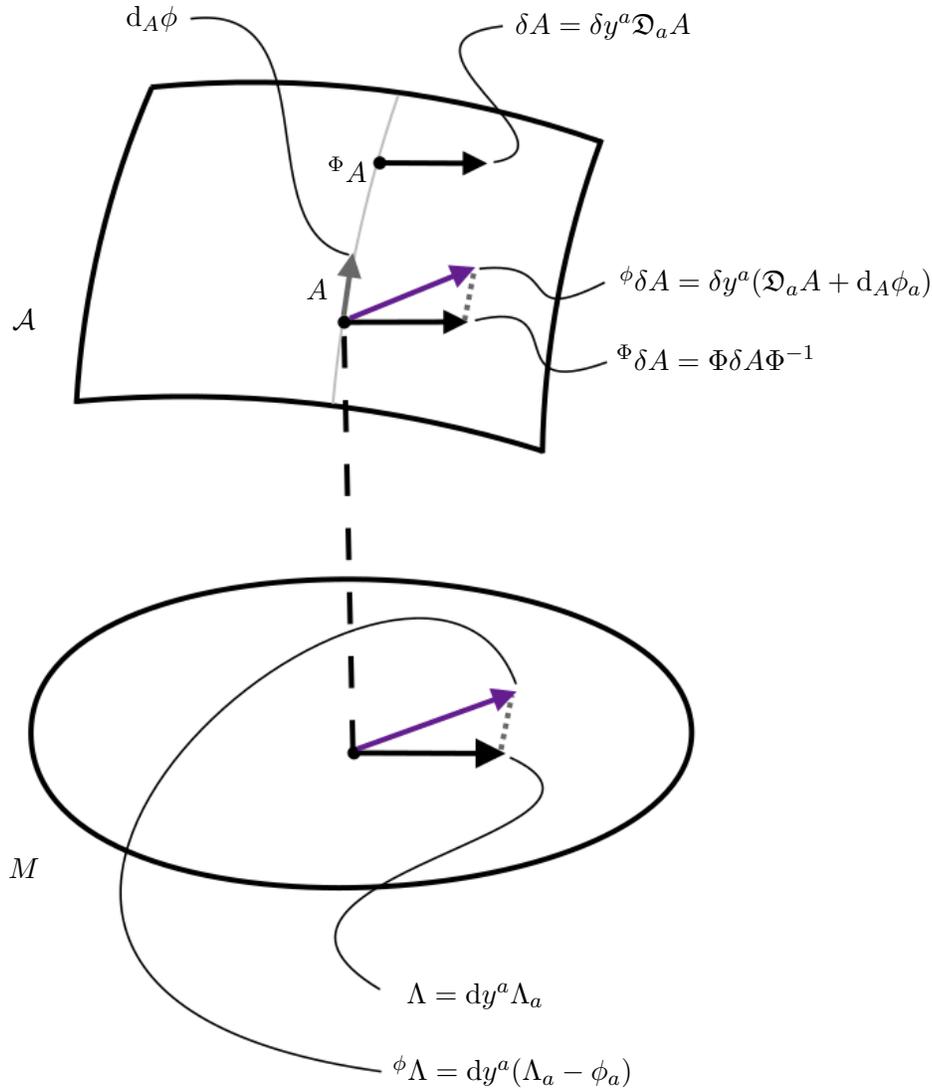
\begin{figure}[th!]\centering

\begin{tikzpicture}
\node at (0,0) {\includegraphics[width=90mm]{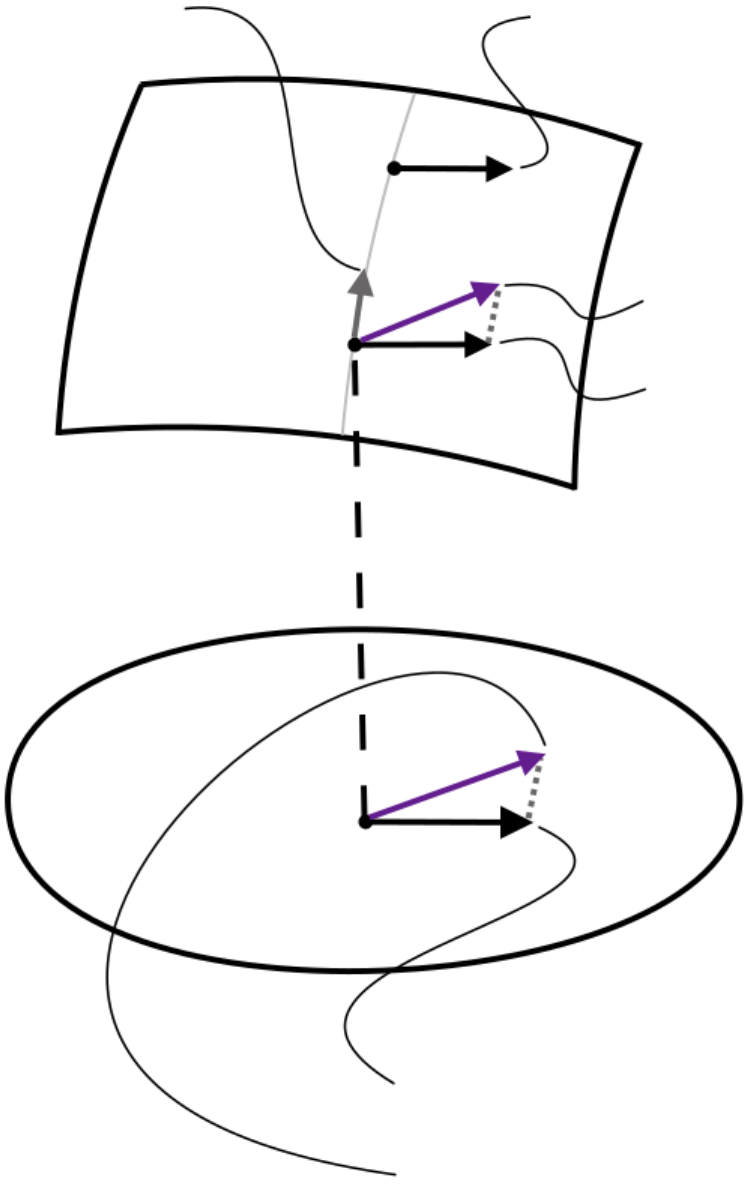}};
\node at (-2.8,7.0) {${\rm d}_A\phi$};
\node at (-0.2,5.0) {$^{\Phi}A$};
\node at (-0.6,3.4) {$A$};
\node at (3.2,6.9) {$\delta A = \delta y^a \mathfrak{D}_a A$};
\node at (5.5,3.5) {$^{\phi}\delta A = \delta y^a (\mathfrak{D}_a A + {\rm d}_A \phi_a)$};
\node at (4.7,2.5) {$^{\Phi}\delta A = \Phi \delta A \Phi^{-1}$};
\node at (-4.5,3.0) {${\mathcal A}$};
\node at (-4.5,-4.3) {$M$};
\node at (1.5,-6.0) {$\Lambda = {\rm d} y^a \Lambda_a$};
\node at (2.0,-7.0) {$^{\phi}\Lambda = {\rm d} y^a (\Lambda_a - \phi_a)$};
\end{tikzpicture}

\caption{The moduli space $M$ and the space of all HYM connections ${\mathcal A}$. The change of base point $A \to {}^\Phi A$ is a background gauge transformation; the change in deformation $\delta A \to {}^{\phi}\delta A$ is a small gauge transformation. The choice of connection on~$M$ is associated with a choice of deformation~$\delta A$. The small gauge transformation is realised by deforming $\Lambda \to {}^\phi \Lambda$. The vertical ${\mathcal G}$-orbits are generated by ${\rm d}_A \phi_a$.}\label{fig:gaugeorbits}
\end{figure}

The geometric picture of this has a standard presentation, in this context see \cite{Itoh:1988} for details. We summarise the salient points here. In the space ${\mathpzc A}$ of all HYM connections, background gauge transformations move a basepoint $A\to {}^\Phi A$. The points $^\Phi A$ form the gauge orbit. The tangent space at $A\in {\mathpzc A}$ can be split into a subspace tangential to gauge orbits (called the vertical direction) and a subspace spanned by $\{\mathfrak{D}_a A\}$ (called horizontal subspace), which are the non-trivial deformations of~$A$. The definition of the vertical subspace is determined by the action of the gauge group; the horizontal subspace is determined by the connection $\Lambda$. If a~gauge fixing analogous to harmonic gauge is imposed, that is in which ${\rm d}_A^\dag (\mathfrak{D}_a A) = 0$, then the horizontal subspace is orthogonal with respect to the natural Weyl--Peterson metric. This idea is illustrated in Fig.~\ref{fig:gaugeorbits}. It should be noted that in~\cite{Itoh:1988} the horizontal subspace is in fact defined by the kernel of ${\rm d}_A^\dag$. We defined horizontal through $\Lambda$ as it more easily allows gauge choices beyond harmonic gauge.
The Kodaira--Spencer map~\cite{Kodaira:1981cx} identifies gauge inequivalent deformations of connections~$\delta A$ with the tangent space~${\mathcal T}_M$. In fact, the covariant derivative~$\mathfrak{D}_a A$ is an explicit realisation of this map. This is the same in spirit as the Donaldson map used in Donaldson theory. We now see that the ambiguity in the choice of deformation~$\delta A$ inside~${\mathpzc A}$ corresponds to a choice of connection on~$M$. Viewed in this language, it is not surprising that small gauge transformations correspond to deformations of the connection~$\Lambda_a$.

One can combine the two connections into a connection ${\mathbb A} = A + \Lambda$ for a vector bundle ${\mathcal U} \to M \times X$ which we refer to as the universal bundle. This has field strength ${\mathbb F} = \Id {\mathbb A} + {\mathbb A}^2$. Under decomposition of legs, there is $F$ with legs purely along $X$; a mixed term ${\mathbb F}_{am} {\rm d} x^m = \mathfrak{D}_a A$, which is exactly the covariant derivative; finally there is a third term ${\mathbb F}_{ab}$ whose legs lie purely along $M$
\[
 {\mathbb F}_{ab} = \partial_a \Lambda_b - \partial_b \Lambda_a + [\Lambda_a, \Lambda_b] .
\]
Deforming the connection $\Lambda_a$ as in \eqref{eq:smallgaugeLambda}, in general, modifies this field strength
\begin{gather}\label{eq:FshTransf}
 {\mathbb F}_{ab} \to {\mathbb F}_{ab}-\mathfrak{D}_a\phi_b + \mathfrak{D}_b\phi_a, \qquad \mathfrak{D}_a\phi_b = \partial_a\phi_b + [\Lambda_a,\phi_b].
\end{gather}
As the Kodaira--Spencer map between parameters $y^a$ and deformations $\mathfrak{D}_a A$ depends on the choice of connection $\Lambda_a$, it clearly is related to the choice of curvature ${\mathbb F}_{ab}$. It is not clear to what extent this curvature is physical, in the sense of affecting the physical string theory, however we will make some progress in this direction in the following section. A complete treatment remains as future work.

We now introduce holomorphy. In \cite{Candelas:2016usb, Candelas:2018lib} we took this to mean that the bundle ${\mathcal U}$ is itself a~holomorphic bundle, and so ${\mathbb F}^{(0,2)} = 0$. The mixed component of this equation is $\mathfrak{D}_{\overline{\alpha}} {\mathcal A}= 0$. We now know this is actually part of holomorphic gauge \eqref{eq:finalGauge}. This is not surprising: asserting~${\mathcal U}$ is a holomorphic bundle involves a choice of connection $\Lambda_{\overline{\alpha}}$ and the space of such connections is related to gauge fixing. Indeed, if we deform $\Lambda_{\overline{\alpha}} \to \Lambda_{\overline{\alpha}} - \phi_{\overline{\alpha}}$, then we see that holomorphy is preserved if
\[
{\overline{\partial}}_{\mathcal A} \phi_{\overline{\alpha}} = 0, \qquad \mathfrak{D}_{\overline{\alpha}} \phi_{\overline\beta} = \mathfrak{D}_{\overline\beta} \phi_{\overline{\alpha}} .
\]
The first equation is discussed in Section~\ref{s:gaugeFixing} and the only possible solutions are parameter dependent constants. In particular, it means that $\phi_{\overline{\alpha}}$ must commute with the algebra ${\mathfrak g}$. The second equation we will interpret in the next section. It is a further constraint on the parameter dependence of gauge transformations $\phi_{\overline{\alpha}}$, which are associated with gauge fixing second order deformations. We note that $\phi_{\overline{\alpha}} \sim \phi_{\overline{\alpha}} + \mathfrak{D}_{\overline{\alpha}} \psi$ is a symmetry of this equation.

We make further comments on gauge transformations and structure group of ${\mathcal U}$. In full generality, the connection ${\mathbb A}$ transforms as
\begin{gather}
{\mathbb A}\to {}^{\Phi_{\mathcal U}}{\mathbb A} = {\Phi_{\mathcal U}}( {\mathbb A} - {\mathbb Y}) {\Phi_{\mathcal U}}^{-1} ,\qquad {\mathbb Y} = {\Phi_{\mathcal U}}^{-1} \Id {\Phi_{\mathcal U}},
\label{eq:IAtransf}\end{gather}
where ${\Phi_{\mathcal U}}$ is an element of the structure group of the universal bundle ${\mathcal U}$. If we fix $y\in M$ then the universal bundle ${\mathcal U}$ is such that it reduces to ${\mathcal E} \to X$ with structure algebra ${\mathfrak g}$. We take the structure algebra of ${\mathcal U}$ to be ${\mathfrak g}_{\mathcal U} = {\mathfrak g} \oplus {\mathfrak g}_b$, where we are leaving ${\mathfrak g}_b$ arbitrary apart from $[{\mathfrak g}_{\mathcal U},{\mathfrak g}] = {\mathfrak g}$ and that ${\mathfrak g}_b$ is real, semi-simple and compact Lie algebra. More general forms for ${\mathfrak g}_{\mathcal U}$ can be explored, such as the algebra not being a direct sum, or relaxing the semi-simple assumption, however we do not do so here. The connection $A$ is taken to be valued in ${\mathfrak g}$, while $\Lambda$ is valued in ${\mathfrak g}_{\mathcal U}$. We also demand that $\mathfrak{D}_a A \in \Omega^1(X, {\rm End}{\mathcal E})$ and so is in particular valued in ${\rm ad}_{\mathfrak g}$. The covariant derivative involves a term $[\Lambda_a, A]$ and so a necessary condition for this to be the case is $[{\mathfrak g}_{\mathcal U},{\mathfrak g}] = {\mathfrak g}$.

We see that \eqref{eq:IAtransf} decomposes into two separate transformation laws
\[
A\to {}^{\Phi} A = \Phi( A - Y) \Phi^{-1} ,\qquad \Lambda\to {} ^{\Phi_{\mathcal U}}\Lambda_a = {\Phi_{\mathcal U}}( \Lambda_a - Y_a) {\Phi_{\mathcal U}}^{-1} , \qquad Y_a = {\Phi_{\mathcal U}}^{-1} \partial_a {\Phi_{\mathcal U}},
\]
where $\Phi$ is exponentiation of the Lie algebras ${\mathfrak g}$. If we take the limit $\Phi = 1 - \phi$ and ${\Phi_{\mathcal U}} = 1- \phi_{\mathcal U}$, then we can gain some intuition for small gauge transformations. The transformation law reduces to
\[
A \to A + {\rm d}_A \phi, \qquad \Lambda_a \to \Lambda_a + \mathfrak{D}_a \phi_{\mathcal U}.
\]
We have already studied the deformations in the first equation. So let us for now restrict ourselves to those of the form ${\rm d}_A \phi = {\rm d}_A \phi_{\mathcal U} = 0$. We take this to mean that $\phi$, $\phi_{\mathcal U}$ depend only on parameters and commute with ${\mathfrak g}$. This means they are ${\mathfrak g}_b$-valued. If we denote $\phi_a = \mathfrak{D}_a \phi_{\mathcal U}$, then we see that ${\mathbb F}$ transforms as
\[
F \to F, \qquad \mathfrak{D}_a A \to \mathfrak{D}_a A, \qquad {\mathbb F}_{ab} \to {\mathbb F}_{ab} + [\mathfrak{D}_a, \mathfrak{D}_b] \phi_{\mathcal U} = {\mathbb F}_{ab} + [{\mathbb F}_{ab}, \phi_{\mathcal U}],
\]
where we use ${\rm d}_A( \mathfrak{D}_a \phi_{\mathcal U}) = \mathfrak{D}_a( {\rm d}_A\phi_{\mathcal U} )- [ \mathfrak{D}_a A, \phi_{\mathcal U} ] = 0$. Hence, there are deformations of $\Lambda_a$ that do not modify $A$ nor $\mathfrak{D}_a A$; these are like gauge-for-gauge transformations.

A curious aside: recall the space of deformations of holomorphic bundles ${\mathcal E}\to X$ is a moduli space~$M$. We see here that the space of deformations of ${\mathcal U}$ with $\delta A =0$ reduces to the space of deformations of holomorphic $\mathfrak{g}_b$-bundles on~$M$. We leave this for further work.

\subsection{Small gerbes}

Just as ${\mathbb F}_a = \mathfrak{D}_a A$ came from the mixed term of a field strength, the form ${\mathcal B}_a$ in \eqref{eq:ccBdef} is most elegantly defined as the mixed component of~${\mathbb H}$:
\begin{gather*}%\label{eq:ccBdef3}
{\mathcal B}_a = {\mathbb H}_a = \mathfrak{D}_a B + \tfrac{\alpha^{\backprime}}{4}\operatorname{Tr}{(A \mathfrak{D}_a A)}-{\rm d}{\mathbb B}_a , \qquad \mathfrak{D}_a B = \partial_a B -\tfrac{\alpha^{\backprime}}{4} \operatorname{Tr} (\Lambda_a\, {\rm d} A) .
\end{gather*}
This gives an interpretation to the one-form ${\mathbb B}_a$ as the mixed component ${\mathbb B}_{am}$ of the universal ${\mathbb B}$-field. Under background gauge transformations ${\mathbb B} \sim {\mathbb B} + \frac{\alpha^{\backprime}}{4} \operatorname{Tr} {\mathbb Y} {\mathbb A} + \frac{\alpha^{\backprime}}{4} {\mathbb U}$, and applying this to~${\mathbb B}_a$ and ${\mathcal B}_a$ we find exactly the transformation law \eqref{eq:ccBmixedBackground}.

We immediately see that small gerbe transformations ${\mathcal B}_a \to {\mathcal B}_a + {\rm d} {\mathfrak b}_a$ are realised by a~deformation
\begin{gather*}%\label{eq:smallGerbeIB}
{\mathbb B}_a \to {\mathbb B}_a - {\mathfrak b}_a,
\end{gather*}
and so we view ${\mathbb B}_a$ in an analogous fashion to $\Lambda_a$ for the gauge transformations: it is a `gerbe-connection' on the moduli space $M$. The tangibility $[2,1]$ of ${\mathbb H}$ is also deformed
\begin{gather*}%\label{eq:IHabTransformation}
 {\mathbb H}_{ab} \to {\mathbb H}_{ab} - \mathfrak{D}_a {\mathfrak b}_b + \mathfrak{D}_b {\mathfrak b}_a.
\end{gather*}
If we compare this with the gauge transformation law for the field strength~${\mathbb F}_{ab}$ in~\eqref{eq:FshTransf}, then we see that ${\mathbb B}_a$ is naturally viewed as a connection for gerbe transformations and~${\mathbb H}_{ab}$ is its related field strength.

In the previous section we deduced that small gauge transformations, parameterised by $\phi_a$, result in a transformation law for ${\mathcal B}_a$ in~\eqref{eq:ccBtransf} and that for $\mathfrak{D}_a B$ in~\eqref{eq:CovDerivB}. The realisation of this in universal geometry is more subtle. While a deformation $\Lambda_a \to \Lambda_a - \phi_a$ gives the correct law for~$\mathfrak{D}_a A$, we have
\[
\mathfrak{D}_a B \to \mathfrak{D}_a B + \tfrac{\alpha^{\backprime}}{4} \operatorname{Tr} (\phi_a {\rm d} A) ,
\]
which does not agree with \eqref{eq:ccBtransf}. We are missing the fact that ${\mathbb B}_a$ is to be simultaneously deformed. This is a manifestation of the fact the $B$-field is charged under the gauge symmetry. The correct deformation turns out to be{\samepage
\begin{gather}\label{eq:smallgaugedefIB}
\Lambda_a \to \Lambda_a - \phi_a, \qquad {\rm and} \qquad {\mathbb B}_a \to {\mathbb B}_a - \tfrac{\alpha^{\backprime}}{4} \operatorname{Tr} ( A \phi_a ).
\end{gather}
It is an exercise to see that ${\mathbb H}_a$ transforms in the same manner as \eqref{eq:ccBtransf}.}

To reconcile with \eqref{eq:CovDerivB} one could redefine $\mathfrak{D}_a B$ as follows
\begin{gather*}
 {\mathcal B}_a = \mathfrak{D}_a B +\tfrac{\alpha^{\backprime}}{4}\operatorname{Tr}{(A \mathfrak{D}_a A)}-{\rm d}\big( {\mathbb B}_a + \tfrac{\alpha^{\backprime}}{4}( \operatorname{Tr} \Lambda_a A) \big), \qquad \mathfrak{D}_a B = \partial_a B -\tfrac{\alpha^{\backprime}}{4} \operatorname{Tr} (A \, {\rm d}\Lambda_a ).
\end{gather*}
If we do so, we see that under \eqref{eq:smallgaugedefIB} the derivative $\mathfrak{D}_a B$ has the correct transformation law. We note that this definition of the covariant derivative of the B-field differs from~\cite{Candelas:2016usb} by a ${\rm d}$-exact term. The symmetry property~\eqref{eq:CovDerivB} gave us no choice in this matter. Physically, there is no difference as $\mathfrak{D}_a B$ is only ever defined up to ${\rm d}$-closed forms, a manifestation of its parent ${\rm d} B$.

Higher tangibilities of the three-form ${\mathbb H}$ also transform. For example,
 \begin{gather*}
 {\mathbb H}_{ab} \to {\mathbb H}_{ab} + \tfrac{\alpha^{\backprime}}{2}\operatorname{Tr}{(\phi_a \mathfrak{D}_b A-\phi_b \mathfrak{D}_a A)} ,
\end{gather*}
where in analogy with \eqref{eq:smallgaugedefIB} we defined the rule
\begin{gather*}
 {\mathbb B}_{ab} \to {\mathbb B}_{ab}-\tfrac{\alpha^{\backprime}}{4}\operatorname{Tr}{(\phi_a \Lambda_b-\phi_b \Lambda_a)}.
\end{gather*}

\subsection{Small diffeomorphisms}

We now suppose that $X$ and the bundle ${\mathcal E}$ can simultaneously deform. As in~\cite{Candelas:2018lib} we interpret diffeomorphisms as a gauge symmetry. This necessitates introducing an invariant basis of forms~$e^m$,~${\rm d} y^a$ and vectors $\partial_m$, $e_a$ where
\[
e^m = {\rm d} x^m + c_a{}^m {\rm d} y^a, \qquad e_a = \partial_a - c_a{}^m \partial_m,
\]
which factorise the Hermitian form on ${\mathbb X}$ into a block diagonal form
\begin{gather}\label{eq:hermitianFormonIX}
\Iomega = \tfrac{1}{2} \omega_{mn} e^m e^n + \tfrac{1}{2} \omega^{\sharp}_{ab} {\rm d} y^a {\rm d} y^b.
\end{gather}
The symbol $c_a{}^m$ transforms like a connection under diffeomorphisms and its role will be analogous to~$\Lambda_a$. As described in~\cite{Candelas:2018lib} we need to express quantities in the `e'-basis, and if there is any ambiguity will denote the result by a~superscript ${\sharp}$ symbol. A key example is the gauge field ${\mathbb A}$. When expressed in the `e'-basis it has a transformed component
\[
{\mathbb A} = A_m e^m + A^{\sharp}_a {\rm d} y^a, \qquad A^{\sharp}_a = \Lambda_a - A_m c_a{}^m.
\]
Consider a form $\eta$ without any additional gauge symmetries and whose legs lie purely along the manifold~$X$. Then a result derived in~\cite{Candelas:2018lib} is that its deformation $\delta \eta$ is a Lie derivative with respect to the vectors~$ e_a$:
\begin{gather}\label{eq:smallDef}
\delta \eta =\delta y^a {\mathcal L}_{e_a} \eta.
\end{gather}
Due to the fibration structure of ${\mathbb X}$, the Lie derivative acts as a directional derivative, $\mathfrak{D}_a \eta = {\mathcal L}_{e_a} \eta$ and it is helpful to write out the expression explicitly
\begin{gather}\label{eq:covderiveta}
\mathfrak{D}_a \eta = \tfrac{1}{q!}\big(e_a(\eta_{m_1\dots m_q}) -
c_a{}^n{}_{,m_1}\eta_{n m_2\dots m_q} - c_a{}^n{}_{,m_2} \eta_{m_1 n\dots m_q} - \cdots\big) e^{m_1\dots m_q}.
\end{gather}
If we have a tensor of the form
\begin{gather*}
 \xi = \xi_{n_1 \dots n_s}{}^{m_1 \dots m_r} e^{n_1}\otimes\cdots \otimes e^{n_s}\otimes\partial_{m_1} \otimes\cdots \otimes\partial_{m_r},
\end{gather*}
then its derivative is defined as
\begin{gather*}
\mathfrak{D}_{a} \xi_{n_1 \dots n_s}{}^{m_1 \dots m_r}
 = e_a(\xi_{n_1 \dots n_s}{}^{m_1 \dots m_r})+
c_a{}^{m_1}{}_{,k} \xi_{n_1 \dots n_s}{}^{k m_2\dots m_r} + \cdots +
c_a{}^{m_r}{}_{,k} \xi_{n_1 \dots n_s}{}^{m_1 \dots m_{r-1}k} \nonumber\\
\hphantom{\mathfrak{D}_{a} \xi_{n_1 \dots n_s}{}^{m_1 \dots m_r}=}{} - c_a{}^k{}_{,n_1} \xi_{k n_2 \dots n_s}{}^{m_1 \dots m_r} - \cdots -
c_a{}^k{}_{,n_s} \xi_{n_1 \dots n_{s-1} k}{}^{m_1 \dots m_r}.%\label{eq:TensorDeriv}
\end{gather*}
As we introduce additional gauge symmetries and associated connections we will generalise this covariant derivative. For ease of notation, we will use the same symbol except where a possible ambiguity may arise.

A small diffeomorphism, $\delta \eta \to \delta \eta + {\mathcal L}_\varepsilon \eta$ is realised by a deformation of the connection~$c_a$:
\begin{gather}\label{eq:shiftc}
 c_a \to c_a-\varepsilon_a, \qquad \varepsilon_a = \varepsilon_a{}^m\partial_m.
\end{gather}
It is elementary to see that
\begin{gather*}%\label{eq:shiftDethsh}
 \mathfrak{D}_a\eta \to \mathfrak{D}_a\eta + {\mathcal L}_{\varepsilon_a}\eta.
\end{gather*}
The connection $c_a$ has a curvature tensor defined as the Lie bracket of two vertical vectors $S_{ab}{}^m= [e_a, e_b]^m$. Under the deformation \eqref{eq:shiftc} we find
\[
 S_{ab}{}^m \to S_{ab}{}^m + \mathfrak{D}_a\varepsilon_b{}^m - \mathfrak{D}_b\varepsilon_a{}^m.
\]
As derived in~\cite{Candelas:2018lib}, this curvature tensor appears in forms on ${\mathbb X}$ with two or more legs along the moduli space. For example, in complex coordinates
\begin{gather*}
 (\Id^c\Iomega)_{\alpha\beta} = -{\rm i} S{}_{\alpha\beta}{}^\mu \omega_\mu , \qquad (\Id^c\Iomega)_{{\overline{\alpha}}{\overline{\beta}}}= {\rm i} S{}_{{\overline{\alpha}}{\overline{\beta}}}{}^{\overline{\mu}} \omega_{\overline{\mu}} ,\qquad (\Id^c\Iomega)_{\alpha\overline\beta} = -{\rm i} S{}_{\alpha{\overline{\beta}}}{}^{{\overline{\mu}}} \omega_{{\overline{\mu}}}+
{\rm i} S{}_{\alpha\overline\beta}{}^\mu \omega_\mu .
\end{gather*}
As discussed in \cite{Candelas:2018lib}, we make the assumption that the curvature vanishes so that $S = 0$. This leads to simplifications in intermediate calculations. The situation in which $S \ne 0$ is an interesting though far more complicated case and involves a thorough study of second order and higher order derivatives. This is currently work in progress which we hope to report on soon. That being said, if we start with $S = 0$ and demand it is preserved then
\[
 \mathfrak{D}_a\varepsilon_b{}^m - \mathfrak{D}_b\varepsilon_a{}^m = 0.
\]
Locally, this amounts to $\varepsilon_a{}^m=\mathfrak{D}_a v{}^m$ for some vertical vector field $v^m \partial_m$. See \cite{Candelas:2018lib} for the definition of $\mathfrak{D}_a$ acting on mixed component tensors.

The connection $c_a$ determines tangibility, and so is also modified by deformations. In the following discussion it is convenient to compare the deformed frame, denoted by a tilde with the undeformed frame. The basis of 1-forms deforms as
\begin{gather}\label{eq:framedeformation}
 e^m \to {\widetilde{e}}^m = e^m - {\rm d} y^a \varepsilon_a{}^m.
\end{gather}
This has interesting consequences. Consider the field strength ${\mathbb F}$. Firstly
\begin{gather*}
 {\mathbb F} = \tfrac{1}{2} F_{mn} e^{mn} + {\mathbb F}_{an} {\rm d} y^a e^n + \tfrac{1}{2} {\mathbb F}_{ab} {\rm d} y^{ab} \\
\hphantom{{\mathbb F}}{} = \tfrac{1}{2} F_{mn} {\widetilde{e}}^{mn} + ({\mathbb F}_{an} + \varepsilon_a{}^m F_{mn}) {\rm d} y^a {\widetilde{e}}^n + \tfrac{1}{2} ({\mathbb F}_{ab} + {\mathbb F}_{an} \varepsilon_b{}^n - {\mathbb F}_{bn} \varepsilon_a{}^n){\rm d} y^{ab}.
\end{gather*}
Comparing with ${\mathbb F}$ expressed in terms of its tilded components:
\[
{\mathbb F} = \tfrac{1}{2} F_{mn} {\widetilde{e}}^{mn} + \widetilde {\mathbb F}_{an} {\rm d} y^a {\widetilde{e}}^n + \tfrac{1}{2} \widetilde {\mathbb F}_{ab} {\rm d} y^{ab},
\]
and identifying ${\mathbb F}_{an} = \mathfrak{D}_a A_n$ and $\widetilde{ {\mathbb F}}_{an} = \widetilde {\mathfrak{D}}_a A_n$ we find a transformation law
\[
\widetilde {\mathfrak{D}}_a A = \mathfrak{D}_a A + \varepsilon_a{}^m F_{m}, \qquad \widetilde {{\mathbb F}}_{ab} = {\mathbb F}_{ab} - \varepsilon_a{}^n \mathfrak{D}_b A_n + \varepsilon_b{}^n \mathfrak{D}_a A_n.
\]
The second equation implies the curvature on $M$ is also deformed in a non-trivial way. The first equation is exactly the transformation law under a small diffeomorphisms as derived in supergravity. When applied to $\mathfrak{D}_\alpha {\mathcal A}$ and combined with $A^{\sharp}_\alpha \to A^{\sharp}_\alpha - \phi_\alpha$~-- note that we use $A^{\sharp}$ and not $\Lambda$ as we are in the `e'-basis~-- we recover exactly~\eqref{eq:fDAtransf}
\[
\widetilde {\mathfrak{D}}_\alpha {\mathcal A} = \mathfrak{D}_\alpha {\mathcal A} + \varepsilon_\alpha{}^\mu F_{\mu} + {\overline{\partial}}_{\mathcal A} \phi_\alpha.
\]

In the same way, we determine that field strength ${\mathbb H}$ transforms and from this we deduce the transformation law for ${\mathbb H}_a = {\mathcal B}_a$ under small diffeomorphisms
\[
\widetilde {\mathcal B}_a = {\mathcal B}_a + \varepsilon_a{}^m H_m.
\]
Again this is in exact agreement with supergravity.

There is a complex structure on ${\mathbb X}$
\[
 {\mathbb J} = J_m{}^n e^m\otimes \partial_n + J^{\sharp}{}_a{}^b\,{\rm d} y^a\otimes e_b.
\]
The manifold ${\mathbb X}$ being complex means the indices must be pure~\cite{Candelas:2018lib}
\begin{gather*}%\label{eq:cpure}
 c_\alpha{}^{{\overline{\mu}}} = 0.
\end{gather*}
It is interesting to study how ${\mathbb J}$ transforms under the change of `e'-basis in \eqref{eq:framedeformation}:
\[
 {\mathbb J} = J_m{}^n {\widetilde{e}}^m \otimes \partial_n + J^{\sharp}_a{}^b \,{\rm d} y^a\otimes \widetilde e_b + \big(\varepsilon_a{}^m J_{m}{}^n-J^{\sharp}{}_a{}^b \varepsilon_b{}^n\big) {\rm d} y^a \partial_n.
\]
At first we see the structure is no longer block diagonal, which is not surprising considering we have deformed~$c_a$. The condition that ${\mathbb J}$ remains block-diagonal, expressed in terms of the unperturbed complex structure, is $\varepsilon_\alpha{}^{\overline{\mu}} = 0$. These are exactly the deformations that preserve integrability of ${\mathbb J}$ so that the two are identified. This has an echo in supergravity. In the last section, we saw that part of holomorphic gauge, \eqref{eq:finalGauge}, is $\Delta_{{\overline{\alpha}}{\overline\nu}}{}^\mu = 0$. Small diffemorphisms that preserve holomorphic gauge are of the form $\varepsilon_{\overline{\alpha}}{}^\mu = 0$. So we learn that holomorphic gauge amounts to ${\mathbb J}$ being integrable and block diagonal.

Furthermore, in \cite{Candelas:2018lib} it was shown that $\Delta_{\alpha{\overline\nu}}{}^\mu = - \partial_{\overline\nu} c_\alpha{}^\mu$. Hence, under a deformation $c_\alpha{}^\mu \to c_\alpha{}^\mu - \varepsilon_\alpha{}^\mu$ we have exactly the correct transformation law
\[
 \Delta_{\alpha{\overline\nu}}{}^\mu \to \widetilde \Delta_{\alpha{\overline\nu}}{}^\mu = \Delta_{\alpha{\overline\nu}}{}^\mu + \partial_{\overline\nu}\varepsilon_\alpha{}^\mu.
\]

Before we proceed we pause to introduce some technical details, which while maybe mildy painful, but nonetheless are necessary to account for the diffeomorphism symmetries correctly. Recall a form with $p$ legs along $M$ and $q$ legs along $X$ is said to have tangibility $[p,q]$. We denote the set of such forms by $\Omega^{[p,q]}({\mathbb X})$. The diffeomorphic invariant way to distinguish legs in this way is through the `e'-basis. This means however, we need to introduce a generalisation of the de Rham operator that acts like the de Rham operator on $X$, but annihilates the bases~$e^m$. Neither of the operators $e^m \partial_m$ or ${\rm d} x^m \partial_m$ achieve this. So with a slight abuse of notation,\footnote{In \cite{Candelas:2018lib} we used the Iceland `deth' symbol ${\text{-}}\hspace{-3.5pt}{\rm D}$ for derivatives of corpus while here we denote this by the usual ${\rm d}$-operator; here we use $\mathfrak{D}$ which was denoted ${\text{-}}\hspace{-3.5pt}{\rm D}^{\sharp}$ in~\cite{Candelas:2018lib}. We hope by simplifying the number of derivative symbols used, the paper will be easier to digest.} we redefine~${\rm d}$ and $\mathfrak{D}$ to act as
 \begin{alignat*}{4}
& {\rm d} f = (\partial_m f) e^m,\qquad && {\rm d} e^m = 0,\qquad && {\rm d} ({\rm d} y^a) = 0,& \nonumber\\
& \mathfrak{D} f= e_a(f)\,{\rm d} y^a,\qquad && \mathfrak{D} e^m = - c_a{}^m{}_{,n}\,{\rm d} y^a e^n,
\qquad && \mathfrak{D}({\rm d} y^a)= 0 .&%\label{eq:dSharp}
\end{alignat*}
It is straightforward to check that
\begin{gather*}%\label{eq:ddandfD}
{\rm d}^2 = 0\qquad \text{and}\qquad \{{\rm d}, \mathfrak{D}\}= 0.
\end{gather*}
Note that ${\rm d}$ now maps a $[p,q]$-form to a $[p,q+1]$-form and $\mathfrak{D}$ maps a $[p,q]$-form to a $[p+1,q]$ form. There are strong similarities with complex manifolds. A form of definite holomorphic type, say $(r,s)$, is acted upon by the $\partial$- and ${\overline{\partial}}$-operators that map this form to $(r+1,s)$ and $(r,s+1)$. In the limit where $c_a \to 0$, which includes the case of a fixed manifold $X$, the operators above reduce to what has been used in previous sections.

The de Rham operator $\Id$ acting on a vertical form $\eta$ decomposes
\begin{gather*}
 \Id\eta= {\rm d}\eta + \mathfrak{D}\eta -
 \frac{1}{2 (q-1)!} \eta_{m_1\dots m_{q-1}\,n} S{}_{ab}{}^n \, {\rm d} y^{ab} e^{m_1\dots m_{q-1}}.%\label{eq:Idsplit}
\end{gather*}
The term involving $S_{ab} = [e_a, e_b]$ measures the lack of integrability of tangibility.
In general a~tangibility $[p,q]$ form with $q\ge 1$ is mapped to a form with three different tangibilities divided up into branches as shown
\[
\begin{tikzcd}[row sep=2em, column sep=3em]
&\Omega^{[p, q+1]}({\mathbb X}) \\
\Id\colon \ \Omega^{[p,q]}({\mathbb X}) \arrow[ru, "{\rm d}", xshift=2.5ex] \arrow[r, "\mathfrak{D}"] \arrow[rd, "S", xshift=2.5ex]
&\Omega^{[p+1, q]}({\mathbb X}) \\
&\Omega^{[p+2, q-1]}({\mathbb X}).
\end{tikzcd}
\]

The Hermitian form is given in \eqref{eq:hermitianFormonIX} and its deformation, derived from~\eqref{eq:smallDef}, is given by $\mathfrak{D}_a \omega = {\mathcal L}_{e_a} \omega$. Just as ${\mathbb F}$ includes~$\mathfrak{D}_a A$ as its mixed component, the deformation of $\omega$ is a mixed component of~$\Id \Iomega$:
\begin{gather*}
\Id \Iomega = \tfrac{1}{3!} ({\rm d} \omega)_{mnp} e^{mnp} + \tfrac{1}{2} \mathfrak{D}_a \omega_{mn}\, {\rm d} y^a e^{mn} - \tfrac{1}{2} S_{bc}{}^n \omega_{nm} {\rm d} y^{bc} e^m , %\label{eq:dIomega1}
\end{gather*}
where
\[
({\rm d} \omega)_{mnp} = \partial_m \omega_{np} + \partial_n \omega_{pm} + \partial_p \omega_{mn},
\]
and $\mathfrak{D}_a \omega$ is of the form~\eqref{eq:covderiveta} with $q=2$. The transformation law under $c_a \to \widetilde c_a = c_a - \varepsilon_a$ follows directly from~\eqref{eq:covderiveta}:
\begin{gather*}%\label{eq:shiftDethom}
 \mathfrak{D}_a\omega \to \widetilde{\mathfrak{D}}_a\omega = \mathfrak{D}_a\omega + \varepsilon_a{}^m\,({\rm d}\omega)_m + {\rm d}(\varepsilon_a{}^m \omega_m),
\end{gather*}
where the notation here means
\[
({\rm d} \omega)_m = \tfrac{1}{2} (\partial_m \omega_{np} + \partial_n \omega_{pm} + \partial_p \omega_{mn}) e^{np}, \qquad {\rm d} (\varepsilon_a{}^m \omega_m) = \partial_p (\varepsilon_a{}^m \omega_{mn}) e^{pn}.
\]
This is exactly the transformation law of a small diffeomorphism. It is also possible to deduce this transformation law by studying how $\Id \Iomega$ transforms under the change of frame.

\subsection{Holomorphy}

This section opened with a discussion of the role of holomorphy for the gauge field. In universal geometry this is the fact ${\mathbb F}^{(0,2)} =0$ and results in gauge fixing $A$ to holomorphic gauge,~\eqref{eq:finalGauge}. We now extend this to other universal fields.

We begin with complex structure $J$. Holomorphic gauge in supergravity includes the condition $\Delta_{{\overline{\alpha}}{\overline\nu}}{}^\mu = 0$. As described above this amounts to the connection~$c_a$ being pure in indices, or equivalently, the fibration ${\mathbb X}$ being complex.

As for the moduli space of Calabi--Yau manifolds the deformations of $B$ and $\omega$ are combined and holomorphy is gauge fixed to be
\[
 {\overline{\mathcal Z}}_{\overline{\alpha}}^{(2,0)} = 0 , \qquad {\mathcal Z}_{\overline{\alpha}}^{(1,1)} = 0 , \qquad {\mathcal Z}_{\overline{\alpha}}^{(0,2)} = {\overline{\mathcal Z}}_{\overline{\alpha}}^{(0,2)} = 0 .
\]
For the Hermitian form $\Iomega$ and ${\mathbb B}$-field, holomorphy is captured by
\[
{\mathbb H} - \Id^c \Iomega = 0.
\]
 To see this we need only consider the tangibility $[1,2]$ component, which has one leg along the moduli space $M$ and the remainder along $X$. Then, note that $\tfrac{1}{2} {\mathbb H}_{{\overline{\alpha}}mn} e^{mn}= {\mathcal B}_{\overline{\alpha}}$ and $\tfrac{1}{2}(\Id^c \Iomega)_{{\overline{\alpha}}mn} e^{mn} = - {\rm i} \mathfrak{D}_{\overline{\alpha}} \omega^{(1,1)} + {\rm i} \mathfrak{D}_{\overline{\alpha}} \omega^{(2,0)}$. Hence, we find
\[
\Delta_{\overline{\alpha}}{}^\mu = 0, \qquad \mathfrak{D}_{\overline{\alpha}} {\mathcal A} = 0, \qquad {\overline{\mathcal Z}}_{\overline{\alpha}}^{(2,0)} = 0, \qquad {\mathcal Z}_{\overline{\alpha}}^{(1,1)} = 0, \qquad {\mathcal Z}_{\overline{\alpha}}^{(0,2)} = {\overline{\mathcal Z}}_{\overline{\alpha}}^{(0,2)} = 0.
\]
As promised this is precisely the Hermitian moduli part of holomorphic gauge \eqref{eq:finalGauge}.

In Section~\ref{s:gaugeFixing} we observed there is a residual gauge freedom which can be used to set $\delta \Omega^{(3,0)}$ to be harmonic. Equivalently, $\omega^{\mu{\overline\nu}} \mathfrak{D}_a \omega_{\mu{\overline\nu}}$ is a constant on~$X$. There is a geometric echo of this gauge fixing in the universal geometry. We can consider the Lee form of~$\Iomega$ for the universal geometry
\begin{gather*}
 {\mathbb W}(\Iomega) = \tfrac{1}{2} \Iomega^{mn}(\Id \Iomega)_{mn} + \tfrac{1}{2} \Iomega^{ab}(\Id \Iomega)_{ab}\\
\hphantom{{\mathbb W}(\Iomega)}{} = W(\omega) + \tfrac{1}{2} {\rm d} y^a \omega^{mn} \mathfrak{D}_a \omega_{mn} - \tfrac{1}{2} \omega^{{\sharp} ab} \big(S_{ab}{}^m \omega_m\big) ,
\end{gather*}
where for us $W(\omega) = \tfrac{1}{2} \omega^{mn} ({\rm d} \omega)_{mn} = 0$ due to our assumption that the dilaton is constant Section~\ref{s:dilatongauge} and~$X$ is balanced. The gauge fixing implies the second term, which has a leg along the moduli space, is a constant over manifold~$X$ and depends only on parameters
\[
 {\rm d} {\mathbb W}_a^{\sharp} = \tfrac{1}{2} \, {\rm d} \big(\omega^{mn} \mathfrak{D}_a\omega_{mn}\big) = 0 .
\]
The third term has legs along the manifold and is an extrinsic contribution coming from the embedding of $X$ inside ${\mathbb X}$. If we do as in~\cite{Candelas:2018lib} and set $S_{ab} = 0$ then this term also vanishes. Noting in addition that ${\mathbb W}_a^{\sharp} = \mathfrak{D}_a \log{\sqrt{g}}$ we see that the residual gauge fixing of Section~\ref{s:gaugeFixing} amounts to the Lee form on the universal bundle being closed $\Id{\mathbb W} = 0$. It would be interesting to explore this geometric condition in the case where the dilaton is not constant.

\section{A taste of second order universal geometry}\label{s:Second}
The study of higher order deformations is necessary if one wants to fully understand the moduli space. They can capture quantities such as the metric $g^\#_{ab}$ or the Yukawa couplings~$\kappa_{abc}$, and will be crucial to any heterotic generalisation of special geometry. Furthermore, the work of \cite{Ashmore:2018ybe} suggests they govern the underlying algebra of heterotic moduli spaces.

Our aim here is to give a taste of what is to be learnt at second order with an example which is the study of holomorphic bundles on a fixed manifold, a situation which occurs at the standard embedding or at zeroth order in heterotic. We will find that studying this moduli space in the context of universal geometry will actually allow us to learn something useful about the role of curvature ${\mathbb F}_{ab}$ on the moduli space. We will not achieve the ultimate aim of describing the theory of higher order covariant derivatives for heterotic theories. This we hope to publish on soon. Indeed, even at first order when one fleshes out the details of gauge fixing, this is already extremely complicated as demonstrated in Section~\ref{s:HodgeHeterotic}.

We study the moduli of holomorphic stable vector bundles ${\mathcal E}\to X$ over a fixed complex manifold with a balanced metric ${\rm d} \big(\omega^2\big) = 0$. The structure group of~${\mathcal E}$ is a compact real group~$G$, with Lie algebra ${\mathfrak g}$. This example is discussed in Section~\ref{eq:smallgaugeonX}. It is relevant to the mathematics of the Kobayashi--Hitchin correspondence, see, e.g.,~\cite{Kobayashi:1987}. It is also relevant to heterotic theories to lowest order in $\alpha^{\backprime}$. To next order in $\alpha^{\backprime}$, the example becomes more complicated. As we have shown in Section~\ref{s:HodgeHeterotic} to first order in~$\alpha^{\backprime}$ the heterotic equations of motion demand the complex structure and Hermitian structure of $X$ also vary with a change in the bundle.

The manifold $X$ is fixed so we do not worry about $c_a$ or $S_{ab}$. Therefore, we consider a universal bundle ${\mathcal U} \to M \times X$. We are ignoring $\alpha^{\backprime}$-corrections, which as discussed in Section~\ref{s:HodgeHeterotic} will force~$X$ to vary, and so introduce a connection~$c_a$. In this limit, the bundle is equipped with a connection:
\[
 {\mathbb A} = \Lambda_a \, {\rm d} y^a + A ,\qquad {\mathbb F} = \tfrac{1}{2} {\mathbb F}_{ab} \, {\rm d} y^{ab} + {\rm d} y^a {\mathbb F}_{a} + F , \qquad {\mathbb F}_a = \mathfrak{D}_a A ,
\]
with Bianchi identity
\begin{gather}\label{eq:BianchiIF}
\mathfrak{D}_a F = {\rm d}_A\left( \mathfrak{D}_a A\right) , \qquad [\mathfrak{D}_a,\mathfrak{D}_b] A = - {\rm d}_A {\mathbb F}_{ab} .
\end{gather}
A deformation of the gauge field can be written in real coordinates as
\[
\delta {\mathcal A} = \delta y^a \mathfrak{D}_a {\mathcal A} + \big(\delta y^a \otimes \delta y^b \big) \mathfrak{D}_a \mathfrak{D}_b {\mathcal A} + \cdots.
\]
Here the second order derivative is defined as
\[
\mathfrak{D}_a \mathfrak{D}_b {\mathcal A} = \partial_a (\mathfrak{D}_b {\mathcal A}) + [\Lambda_a, \mathfrak{D}_b {\mathcal A}],
\]
and it transforms homogeneously under \eqref{eq:Atransf}: $\mathfrak{D}_a \mathfrak{D}_b {\mathcal A} \to \Phi( \mathfrak{D}_a \mathfrak{D}_b {\mathcal A} )\Phi^{-1}$. The $\delta y^a$ form a basis for ${\mathcal T}_M^*$ and so their tensor product a basis for ${\mathcal T}_M^*\otimes {\mathcal T}_M^*$ in the usual way. Just as $\delta y^a \mathfrak{D}_a A$ describes the first order deformation of the gauge field $A$, $ \big(\delta y^a \otimes \delta y^b\big)\mathfrak{D}_a\mathfrak{D}_b {\mathcal A}$ describes the second order deformation. The second order derivative $\mathfrak{D}_a \mathfrak{D}_b {\mathcal A}$ is constructed precisely so that it transforms in the same way as $\mathfrak{D}_a A$ under~\eqref{eq:Atransf} and~\eqref{eq:Ltransf}. However, the map is not yet well-defined as we have not fixed small gauge transformations. Before we address this issue, we must bring up holomorphy.

Recall the first order analysis in Sections~\ref{s:holomorphicgauge} and~\ref{s:HodgeHeterotic}. Holomorphy on a fixed manifold means that $\mathfrak{D}_{\overline{\alpha}} {\mathcal A} = {\overline{\partial}}_{\mathcal A}\Phi_{\overline{\alpha}}$ and fixing to holomorphic gauge means $\mathfrak{D}_{\overline{\alpha}} {\mathcal A} = 0$. This completely fixes the gauge at first order. The deformation above becomes
\[
\delta {\mathcal A} = \delta y^\alpha \mathfrak{D}_\alpha {\mathcal A} + \big(\delta y^\alpha \otimes \delta y^\beta\big) \mathfrak{D}_\alpha \mathfrak{D}_\beta {\mathcal A} + \big(\delta y^{\overline{\alpha}}\otimes \delta y^\beta\big) \mathfrak{D}_{\overline{\alpha}} \mathfrak{D}_\beta {\mathcal A} + \cdots.
\]
The commutators imply $[\mathfrak{D}_\beta, \mathfrak{D}_\alpha]{\mathcal A} = {\overline{\partial}}_{\mathcal A} {\mathbb F}_{\alpha\beta}$ and $[\mathfrak{D}_{\overline\beta}, \mathfrak{D}_\alpha]{\mathcal A} = {\overline{\partial}}_{\mathcal A} {\mathbb F}_{\alpha{\overline{\beta}}}$ and we immediately see that deformations do not necessarily commute. However, a key point is that the choice of connection $\Lambda_a$ is closely related to gauge fixing. What we show here that under the assumption $X$ is fixed we can find a gauge in which ${\mathbb F}_{\alpha\beta} = 0$. But the term ${\mathbb F}_{\alpha{\overline{\beta}}}$ is not necessarily zero. So deformations of the gauge field do not necessarily commute.

To see this we start by defining holomorphy at second order in same way as first order:
\begin{gather*}%\label{eq:holDDA}
 \mathfrak{D}_{\overline{\alpha}}\mathfrak{D}_{{\overline{\beta}}}{\mathcal A} = {\overline{\partial}}_{{\mathcal A}}\Phi_{{\overline{\alpha}}{\overline{\beta}}} , \qquad \mathfrak{D}_\alpha\mathfrak{D}_{{\overline{\beta}}}{\mathcal A} = {\overline{\partial}}_{{\mathcal A}}\Phi_{\alpha{\overline{\beta}}} .
\end{gather*}
As $\mathfrak{D}_{\overline\beta} \mathfrak{D}_\alpha {\mathcal A}$ is ${\overline{\partial}}_{\mathcal A}$-exact and we find two additional equations
\[
\mathfrak{D}_{\overline\beta} \mathfrak{D}_\alpha {\mathcal A} = {\overline{\partial}}_{\mathcal A} \Phi_{{\overline{\beta}}\alpha}, \qquad {\mathbb F}_{\alpha{\overline{\beta}}} = -\Phi_{\alpha{\overline{\beta}}} + \Phi_{{\overline{\beta}}\alpha}.
\]
That ${\mathbb F}$ is antihermitian implies $ \Phi^\dag_{\alpha{\overline{\beta}}} + \Phi_{\alpha{\overline{\beta}}} = \Phi^\dag_{{\overline{\beta}}\alpha} + \Phi_{{\overline{\beta}}\alpha}$.

Finally, there is a purely holomorphic derivative. We write down a Hodge decomposition
\begin{gather}\label{eq:HodgeDDAhol}
 \mathfrak{D}_\alpha\mathfrak{D}_\beta {\mathcal A} = \mathfrak{D}_\alpha\mathfrak{D}_\beta {\mathcal A}^{\rm harm} + {\overline{\partial}}_{{\mathcal A}} \Phi_{\alpha\beta} + {\overline{\partial}}_{{\mathcal A}}^\dag \Psi_{\alpha\beta} ,
\end{gather}
where the harmonic term will be proportional to the first order variations $\mathfrak{D}_\alpha{\mathcal A}$ through some parameter-dependent coefficients.

Now we return to the issue of gauge fixing. The action of small gauge transformations is described by $\Lambda_a\to\Lambda_a - \phi_a$. The field $\phi_a$ evaluated at $y_0\in M$ gauge fixes the first order derivative $\mathfrak{D}_a {\mathcal A}$, which as described previously, is holomorphic gauge $\mathfrak{D}_{\overline{\alpha}}{\mathcal A} = 0$. Given this gauge fixing, the second order derivatives transform under $\mathfrak{D}_a\phi_b|_{y_0} = \partial_a \phi_b + [\Lambda_a, \phi_b]|_{y_0}$ as follows
\[
 \mathfrak{D}_b\mathfrak{D}_a{\mathcal A} \sim \mathfrak{D}_b \mathfrak{D}_a{\mathcal A} + {\overline{\partial}}_{{\mathcal A}}\left(\mathfrak{D}_b \phi_a\right) .
\]
Holomorphic gauge at second order is given by setting $\mathfrak{D}_{\overline{\alpha}}\phi_{{\overline{\beta}}} = -\Phi_{{\overline{\alpha}}{\overline{\beta}}}$ and $\mathfrak{D}_\alpha\phi_{{\overline{\beta}}} = - \Phi_{\alpha{\overline{\beta}}}$ so that
\begin{gather}\label{eq:holgDDA}
 \mathfrak{D}_{{\overline{\beta}}}\mathfrak{D}_{\overline{\alpha}}{\mathcal A} = 0 , \qquad \mathfrak{D}_\beta\mathfrak{D}_{\overline{\alpha}}{\mathcal A} = 0 .
\end{gather}
Note that antihermiticity requires $\mathfrak{D}_{\overline\beta}\phi_\alpha = - (\mathfrak{D}_{\beta}\phi_{\overline{\alpha}})^\dag = \Phi_{{\overline{\beta}}\alpha}^\dag$. This, together with the fact that there are no ${\overline{\partial}}_{{\mathcal A}}$-closed scalars, completely fixes the gauge freedom at this order.

From the commutator of the first equation in \eqref{eq:holgDDA} we gather that ${\mathbb F}_{{\overline{\alpha}}{\overline{\beta}}}=0$ and the universal bundle is holomorphic. In passing we observe that this is only true after gauge-fixing second order deformations to holomorphic gauge. There is not enough freedom, however, to get rid of~${\mathbb F}_{\alpha{\overline{\beta}}}$. A perhaps useful intuition for this is a simple counting. Second derivatives are given by four independent quantities
\begin{gather}\label{eq:DDAarefour}
 \mathfrak{D}_{\alpha} \mathfrak{D}_{\beta} {\mathcal A}_{{\overline{\mu}}} , \qquad \mathfrak{D}_{\overline{\alpha}} \mathfrak{D}_{{\overline{\beta}}} {\mathcal A}_{{\overline{\mu}}} , \qquad \mathfrak{D}_{\alpha} \mathfrak{D}_{{\overline{\beta}}} {\mathcal A}_{{\overline{\mu}}} , \qquad \mathfrak{D}_{\overline{\alpha}} \mathfrak{D}_{\beta} {\mathcal A}_{{\overline{\mu}}} ,
\end{gather}
while small gauge transformations amount to two independent quantities
\[
 \mathfrak{D}_\alpha\phi_\beta = \partial_\alpha \phi_\beta + [\Lambda_\alpha, \phi_\beta], \qquad \mathfrak{D}_\alpha\phi_{{\overline{\beta}}} = \partial_\alpha \phi_\beta + [\Lambda_\alpha, \phi_\beta].
\]
Hence, we can only gauge-fix only two among \eqref{eq:DDAarefour}, which is what we did in \eqref{eq:holgDDA}.

In this gauge we have
\begin{gather}\label{eq:DbbDaAgf}
 \mathfrak{D}_{{\overline{\beta}}}\mathfrak{D}_\alpha {\mathcal A} = {\overline{\partial}}_{{\mathcal A}} {\mathbb F}_{\alpha{\overline{\beta}}} , \qquad {\mathbb F}_{\alpha{\overline{\beta}}} = \Phi_{{\overline{\beta}}\alpha} + \Phi^\dag_{{\overline{\beta}}\alpha} = \Phi_{\alpha{\overline{\beta}}} + \Phi^\dag_{\alpha{\overline{\beta}}} .
\end{gather}

We now turn to the equations of motion. A real variation of the Bianchi identity \eqref{eq:BianchiIF} is
\[
 \mathfrak{D}_a\mathfrak{D}_b F = {\rm d}_A \mathfrak{D}_a \mathfrak{D}_b A + \{ \mathfrak{D}_a A , \mathfrak{D}_b A \} .
\]
The manifold $X$ has fixed complex structure. This implies $\mathfrak{D}_a \mathfrak{D}_b F^{(0,2)}=0$ and projecting the previous equation onto type we find
\begin{gather*}%\label{eq:MCDDA}
 {\overline{\partial}}_{{\mathcal A}} ( \mathfrak{D}_a \mathfrak{D}_b {\mathcal A} )+ \{ \mathfrak{D}_a {\mathcal A}, \mathfrak{D}_b {\mathcal A} \} = 0 .
\end{gather*}
When indices of mixed holomorphic type are considered, we do not learn anything new from this. When $a$, $b$ are holomorphic, we recognise the second order Maurer--Cartan equation
\begin{gather*}
 {\overline{\partial}}_{{\mathcal A}} ( \mathfrak{D}_\alpha \mathfrak{D}_\beta {\mathcal A} )+ \{ \mathfrak{D}_\alpha {\mathcal A}, \mathfrak{D}_\beta {\mathcal A} \} = 0 .%\label{eq:MCDDA2}
\end{gather*}
From this we can read the coefficients $\Psi_{\alpha\beta}$ in the Hodge decomposition~\eqref{eq:HodgeDDAhol}
\[
 \Box_{{\mathcal A}} \Psi_{\alpha\beta} = - \{\mathfrak{D}_\alpha{\mathcal A},\mathfrak{D}_\beta{\mathcal A}\} .
\]
Unobstructedness of deformations means that the bracket does not contain zero-modes and this is the condition that allows to invert the Laplacian and solve for $\Psi_{\alpha\beta}$.

There is another equation to consider, the HYM equation $\omega^2 F = 0$. A second variation of this, while keeping $X$ fixed, gives
\[
 \omega^2 \big( {\rm d}_A \mathfrak{D}_a \mathfrak{D}_b A + \{\mathfrak{D}_a A,\mathfrak{D}_b A\} \big) = 0 .
\]
In holomorphic gauge \eqref{eq:holgDDA} this gives two equations
\begin{gather*}
{\overline{\partial}}_{{\mathcal A}}^\dag \mathfrak{D}_\alpha\mathfrak{D}_\beta {\mathcal A} = 0 , \qquad {\overline{\partial}}_{\mathcal A}^\dag \mathfrak{D}_{{\overline{\beta}}}\mathfrak{D}_\alpha {\mathcal A} = -{\rm i} \omega^{\mu{\overline\nu}} \big\{\mathfrak{D}_\alpha {\mathcal A},\mathfrak{D}_{{\overline{\beta}}}{\mathcal A}^\dag\big\}_{\mu{\overline\nu}}.
\end{gather*}
Substituting \eqref{eq:HodgeDDAhol} and \eqref{eq:DbbDaAgf} inside these and using stability to invert the Laplacian, we finally obtain
\begin{gather*}
 \Phi_{\alpha\beta} = 0 , \qquad {\mathbb F}_{\alpha{\overline{\beta}}} = \Box_{{\overline{\partial}}_{{\mathcal A}}}^{-1} \big( {-}{\rm i} \omega^{\mu{\overline\nu}} \{\mathfrak{D}_\alpha {\mathcal A},\mathfrak{D}_{{\overline{\beta}}}{\mathcal A}^\dag\}_{\mu{\overline\nu}} \big) .
\end{gather*}
The non-primitive part of the bracket of mixed variations acts as a source for ${\mathbb F}_{\alpha{\overline{\beta}}}$. Therefore, deformations of stable bundles over a fixed manifold do not necessarily commute.

It would be very interesting to apply this second order analysis with the beautiful work in~\cite{Ashmore:2018ybe} who described `holomorphic' data in terms of a functional of~${\rm SU}(3)$ structures. In that work the starting point at first order in deformations implicitly assumes holomorphic gauge (together with the residual gauge constraint $\delta \Omega^{(3,0)}$ is harmonic) and at second order assumes holomorphic derivatives commute. The latter condition may or may not be true. Determining this requires a complete analysis of gauge fixing second order derivatives and determining if the curvature tensor $S_{ab}$, ${\mathbb H}_{ab}$, ${\mathbb F}_{ab}$ are non-zero. Doing so is therefore both interesting and important.

\section{Conclusions}

We started with the humble ambition of writing down the action of small gauge transformations together with clarifying gauge fixing in heterotic theories. We achieved this goal.

Along the way we clarified certain issues. We related the choice of gauge fixing to a choice of connection on the moduli space. We found holomorphic gauge corresponds to the universal bundle being holomorphic. We reiterated that taking a holomorphic deformation of fields is itself a gauge fixing, holomorphic gauge. There is a residual gauge freedom which we showed can be used to gauge fix the deformation of the holomorphic top form to be harmonic $\delta \Omega^{(3,0)} = k \Omega$ with $k$ a constant. This gauge fixing implies $\partial \chi_\alpha = 0$.

 We checked that the equations of motion are properly invariant under gauge transformations and they do not fix the gauge. They can then be used to write explicit expressions for terms in the Hodge decomposition of all the fields, which as we have gauge fixed, are the physical degrees of freedom. The role of quantum corrections is to generate exact and co-exact terms. The coupled nature of the equations makes it very hard to disentangle the heterotic moduli as `bundle' or `complex structure' or `Hermitian moduli'. For example, in heterotic theories one cannot deform the bundle without also deforming the manifold. Nonetheless we give a~prescription for how to compute through the mess applying it to the moduli space metric.

We showed the choice of gauge is related to a choice of connection on the moduli space $M$ corresponds to choice of small gauge. For example, holomorphic gauge in spacetime corresponds to the universal bundle being holomorphic ${\mathbb F}^{(0,2)} = 0$. At second order deformations do not obviously commute and gauge fixing is related to the field strengths ${\mathbb F}_{ab}$, $S_{ab}$ and ${\mathbb H}_{ab}$ on the moduli space. In a toy example, we found gauge fixing allowed us to eliminate some, but not all of these field strengths. Consequently, bundle deformations do not commute.

There are many interesting questions. We hope to publish a full analysis of second order universal geometry in the near future. This has important implications for any `heterotic special geometry' such as relations between the Yukawa couplings and the moduli space metrics. Is it possible to formulate a description of the moduli space in a fully gauge invariant way? The discussion in this paper and the literature makes use of a nice choice of gauge. Is there a more gauge invariant formulation of these results in which we do not need to gauge fix? Or can we show our results do not depend on our choice of gauge?

\appendix

\section{Complexified tangent spaces and gauge groups}\label{app:Complexification}

In this short appendix we write down some well-known mathematical facts, largely drawn from~\cite{Nakahara:2003nw}, about complexification of vector spaces and groups, and relate these facts to gauge transformations.

Let $X$ be a differentiable manifold of dimension $m$. Let $g,h \in {\mathcal F}(X)$ be two real-valued smooth functions. Then, $f = g+{\rm i} h \colon X \to {\mathbb C}$ is a complex valued function and is an element of the complexified set of smooth functions of $X$ denoted as ${\mathcal F}(X)^{{\mathbb C}}$. Its complex conjugate is $\overline{f} = g - {\rm i} h$ and it is real if and only if $\overline{f} = f$.

Let ${\mathcal V}$ be a real vector space with real dimension $m$. Its complexification ${\mathcal V}^{{\mathbb C}}$ is a vector space of complex dimension $m$ whose elements consist of $V+{\rm i} W$ with $V,W\in {\mathcal V}$, and multiplication by a~scalar is enhanced from real scalars to complex scalar quantities. ${\mathcal V} \subset {\mathcal V}^{{\mathbb C}}$ with the identification of $V + {\rm i} 0\in {\mathcal V}^{{\mathbb C}}$ and $V\in {\mathcal V}$. Such vectors are said to be real. The complex conjugate of $Z = V + {\rm i} W$ is denoted $\overline{Z} = V - {\rm i} W$ and $Z$ is real if and only if $Z = \overline{Z}$.
A linear operator~$e$ acts on ${\mathcal V}^{{\mathbb C}}$ is the obvious way: $e(V+{\rm i} W) = e(V) + {\rm i} e(W)$. If $e \to {\mathbb R}$ is a linear function, so that $e\in {\mathcal V}^*$, then its complexification acts on ${\mathcal V}^{{\mathbb C}}$ as $e\colon {\mathcal V}^{{\mathbb C}} \to {\mathbb C}$. The quantity $e$ is said to be real if $\overline{e(V+{\rm i} W)} = e(V-{\rm i} W)$. Tensors are complexified in the obvious way: $T=T_1+ {\rm i} T_2$ where~$T_1$,~$T_2$ are tensors of the same type. A tensor is real if $\overline{T} = T$.

Let $e_m$ be a basis for ${\mathcal V}$. If we regard these as complex vectors {\it the same basis} $\{ e_m\}$ becomes a basis for ${\mathcal V}^{{\mathbb C}}$. That is, if $V = V^m e_m$ and $W= W^m e_m$ are both real vectors then $V+{\rm i} W = (V^m + {\rm i} W^m) e_m \in {\mathcal V}^{{\mathbb C}}$, where $\overline{e_m} = e_m$. Hence, $\dim_{\mathbb R} {\mathcal V} = \dim_{\mathbb C} {\mathcal V}^{\mathbb C}$.

The tangent space ${\mathcal T}_p X=\{ V^m \partial_m \,|\, V^m \in {\mathbb R} \}$ is an example of a real vector space. Its complexification is therefore ${\mathcal T}_p X^{{\mathbb C}} = \{ (V^m + {\rm i} W^m)\partial_m\,|\, V^m, W^m \in {\mathbb R}\}$. As above, the basis elements for ${\mathcal T}_p X$ are regarded as basis elements for the complexified tangent space. The co-tangent space is complexified in the obvious way ${\mathcal T}^*_p X = \{ \omega + {\rm i} \eta \,|\, \omega,\eta \in {\mathcal T}^*_p X$. Tensors are similarly complexified, and the extension to the tangent bundle ${\mathcal T}_X^{\mathbb C}$ and co-tangent bundle ${\mathcal T}_X^*{}^{\mathbb C}$ follows.

If $X$ has $\dim_{\mathbb R} X = 2k$ and admits an integrable complex structure then we can find a split of the coordinates so that $\frac{\partial}{\partial x^1} , \dots, \frac{\partial}{\partial x^k} , \frac{\partial}{\partial y^1}, \dots, \frac{\partial}{\partial y^k}$ are a basis for ${\mathcal T}_p X$ so that $J\big(\frac{\partial}{\partial x^m}\big) = \frac{\partial}{ \partial y^m}$ and $J\big(\frac{\partial}{\partial y^m}\big) = -\frac{\partial}{ \partial x^m}$. These vectors are also a basis for the complexified tangent space ${\mathcal T}_p X^{\mathbb C}$, so that $\dim_{\mathbb C} {\mathcal T}_p X^{\mathbb C} = 2k$. Furthermore, we can redefine the basis so that
\[
\frac{\partial}{\partial z^\mu} = \frac{1}{2} \left(\frac{\partial}{\partial x^\mu} - {\rm i} \frac{\partial}{\partial y^\mu} \right), \qquad \frac{\partial}{\partial z^{\overline{\mu}}} = \frac{1}{2} \left(\frac{\partial}{\partial x^{\overline{\mu}}} + {\rm i} \frac{\partial}{\partial y^{\overline{\mu}}} \right), \qquad \mu,{\overline{\mu}} = 1, \dots, k,
\]
is also basis for ${\mathcal T}_p X^{\mathbb C}$ but one in which $J$ is constant and diagonal. Note that $\overline{\partial/\partial z^\mu} = \partial / \partial z^{\overline{\mu}}$. With this choice of basis we have a canonical split of the vector fields
\[
{\mathcal T}_X^{\mathbb C} = {\mathcal T}_X^{(1,0)} + {\mathcal T}_X^{(0,1)}.
\]
and so a vector $V\in {\mathcal T}_X^{\mathbb C}$ is
\[
V^\mu \partial_\mu + W^{\overline{\mu}} \partial_{\overline{\mu}},\qquad V^\mu, W^{\overline{\mu}} \in {\mathbb C}, \qquad \mu,{\overline{\mu}} = 1,\dots,k.
\]
This is real if and only $\overline{V^\mu} = W^{\overline{\mu}}$. We often write this condition as $\overline{V^\mu} = V^{\overline{\mu}}$ or as $\overline{V^{(1,0)} }= V^{(0,1)}$.
Note that $\dim_{\mathbb C} {\mathcal T}_X^{(1,0)} = \dim_{\mathbb C} {\mathcal T}_X^{(0,1)} = k$.

\section{Gauge fixing small diffeomorphisms on Calabi--Yau manifolds}\label{app:CYGaugeFix}

Suppose the manifold $X$ is K\"ahler so that ${\rm d} \omega = 0$ and $h^{(0,2)} = h^{(0,1)} = 0$. On a~K\"ahler manifold, there is a relation between the Laplacians
\begin{gather}\label{eq:KahLaplac}
\Box_{\partial} = \Box_{{\overline{\partial}}} = \tfrac{1}{2} \Box_{{\rm d}} .
\end{gather}

\subsection{Real deformations}\label{eq:appCYRealDef}

Consider a real deformation. By this we mean that $\delta = \delta y^\alpha \partial_\alpha + \delta y^{\overline\beta} \partial_{\overline\beta}$ so that when applied to the K\"ahler form, the deformation is manifestly real $(\delta \omega)^* = \delta \omega$.

Using \eqref{eq:KahLaplac}, $h^{0,2} = h^{2,0} = 0$ and the Hodge decomposition being unique, we can Hodge decompose the equation of motion ${\rm d} \delta \omega = 0$ with respect to the ${\rm d}$-operator
\[
\delta \omega = \gamma + {\rm d} \beta,
\]
where $\gamma$ is ${\rm d}$-harmonic $(1,1)$-form.

Small diffeomorphisms act as
\[
\delta \omega \sim \delta \omega + {\rm d} (\varepsilon^m \omega_m).
\]
We can partially fix this freedom by setting $\varepsilon^m \omega_m = - \beta + {\rm d} \psi$, where $\psi$ is an arbitrary function on~$X$. In this gauge $\delta \omega = \gamma^{(1,1)}$ is harmonic. There is a residual gauge freedom parameterised by~$\psi$.

There is also the holomorphic $(3,0)$ form $\Omega$. Its deformation decomposed into type is
\begin{gather}\label{eq:dO1}
\delta \Omega = \delta \Omega^{(3,0)} + \delta \Omega^{(2,1)} .
\end{gather}
The Hodge decomposition of $\delta \Omega^{(3,0)}$ with respect to the $\partial$-operator is
\begin{gather}\label{eq:dO2}
\delta \Omega^{(3,0)} = k \Omega + \partial \zeta^{(2,0)} = (k + k_\zeta(x))\Omega,
\end{gather}
where in the first equality we use $h^{(3,0)} = 1$ and so $\Omega$ is the unique harmonic $(3,0)$-form and $k$ is a parameter dependent constant. In the second equality we have shown the $\partial$-exact term is proportional to $\Omega$, except where the coefficient $k_\zeta(x) = \frac{1}{3! \|\Omega\|^2} (\partial \xi)_{\mu\nu\rho} {\overline{\Omega}}^{\mu\nu\rho} \Omega$ is not a constant. The equation of motion ${\rm d} \delta \Omega = 0$ implies
\[
\partial \chi = {\overline{\partial}} \partial \zeta^{(2,0)} , \qquad {\overline{\partial}} \chi = 0,
\]
where $\chi = \delta \Omega^{(2,1)}$.

We can also use $\Omega = \frac{1}{3!} f(x,y) \epsilon_{\mu\nu\rho} {\rm d} x^{\mu\nu\rho}$ where $f = e^{{\rm i} \phi} |f|$ and $|f|^2 = \|\Omega\|^2 \sqrt{g}$. The pure part of the deformation is
\[
\delta \Omega^{(3,0)} = \big( {\rm i} \delta \phi + \tfrac{1}{2} \big( \delta \log \|\Omega\|^2 + \omega^{\mu{\overline\nu}} \delta \omega_{\mu{\overline\nu}} \big) \big)\Omega,
\]
where $\delta \phi$ is a globally defined function on $X$. Supersymmetry implies $\|\Omega\|$ is a constant. Using the fact the Levi-Civita connection on a K\"ahler manifold is metric and Hermitian so that \mbox{$\nabla_\rho \omega = 0$} and torsionless
\begin{gather*}
 \partial_\rho \big( \omega^{\mu{\overline\nu}} \delta \omega_{\mu{\overline\nu}}\big) = \big(\nabla_\rho \omega^{\mu{\overline\nu}}\big)\delta \omega_{\mu{\overline\nu}} + \omega^{\mu{\overline\nu}} \big(\nabla_\rho \delta \omega_{\mu{\overline\nu}}\big)
= -{\rm i} \nabla^{\overline\nu} \delta \omega_{\rho{\overline\nu}} - {\rm i} g^{\mu{\overline\nu}} \big(\partial \delta \omega^{(1,1)}\big)_{\rho\mu{\overline\nu}}.%\label{eq:harmonicconstants}
\end{gather*}
With $\delta \omega$ harmonic, it is both $\partial$-closed and ${\overline{\partial}}$-coclosed and so both terms vanish. Hence $\omega^{\mu{\overline\nu}} \delta \omega_{\mu{\overline\nu}}$ is a constant.

We still have the residual gauge freedom parameterised by $\varepsilon^m \omega_m = {\rm d} \psi$. While $\delta \omega$ is invariant, $\delta \Omega^{(3,0)}$ transforms
\begin{gather*}%\label{eq:OmegaHolHarm}
\delta \Omega^{(3,0)} \sim \delta \Omega^{(3,0)} -{\rm i} (\nabla_\mu \nabla^\mu \psi) \Omega,
\end{gather*}
where we use $\partial (\varepsilon^m \Omega_m) = (\nabla_\mu \varepsilon^\mu) \Omega$ and $\varepsilon^\mu = - {\rm i} \nabla^\mu \psi$.
This implies an action on $\delta \phi$ given by
\begin{gather}\label{eq:smallgaugeonphi}
\delta \phi \sim \delta \phi - \nabla^\mu \nabla_\mu \psi.
\end{gather}
At this point, we pause to recount some properties of the Laplacian acting on a scalar~$\Psi$ with a source~$\Phi$ on the compact K\"ahler manifold~$X$:
\begin{gather}\label{eq:sourcedLaplacian}
\Box_\partial \Psi = \nabla^\mu \nabla_\mu \Psi = \Phi.
\end{gather}
We divide $\Psi$ into a formal sum of zero modes $\Box_\partial \Psi_0 = 0$ and non-zero modes $\Box_\partial \Psi_i = \lambda_i \Psi_i$, where $\lambda_i \ne 0$:
\[
\Psi = \Psi_0 + \sum_i \Psi_i.
\]
The obstruction to solving the equation \eqref{eq:sourcedLaplacian} is that the source $\Phi$ be orthogonal to the zero modes $\int_X \Psi_0 \Phi = 0$. On $X$ the zero modes are constants. Hence, pick a $\Psi_0 \ne 0$ and evaluate the following integral in two different ways
\[
\int \Psi_0 \nabla^\mu \nabla_\mu \Psi_i = \lambda_i\Psi_0 \int \Psi_i = - \int \nabla^\mu \Psi_0 \nabla_\mu \Psi_i = 0,
\]
we see that the non-zero modes have vanishing integral $\int_X \Psi_i = 0$. If $\Phi = k$ for some constant~$k$ then does \eqref{eq:sourcedLaplacian} have a solution? The integral of a total divergence on a compact manifold vanishes
\[
\int_X \nabla^\mu \nabla_\mu \Psi = k V = 0,
\]
where $V$ is the volume of $X$ and so we see that $k=0$.

We now apply these lessons to the gauge fixing~\eqref{eq:smallgaugeonphi}. We divide up the source $\delta \phi$ into zero modes $\delta \phi_0$ and non-zero modes $ \delta \phi_{nz}$. The facts above indicate that $\int_X \delta \phi_{nz} = 0$ and so we can find a $\psi$ so that $\nabla^\mu \nabla_\mu \psi = \delta \phi_{nz} $. However, we cannot find a $\psi$ in which $\nabla^\mu \nabla_\mu \psi = \delta \phi_0 $ for $\delta \phi_0$ a non-zero constant. Hence, by a choice of gauge, we can kill the non-zero modes of $\delta \phi$ but not the zero modes, which are constants.\footnote{We have completely fixed the gauge freedom as $\nabla^\mu \nabla_\mu \psi = k$ has solution only for $k=0$, in which case $\psi$ are constants and acts on the fields only through its derivatives.}

With the choice of gauge, we have $\delta \omega$ being harmonic and $\delta \Omega^{(3,0)} = k \Omega$ for some parameter-dependent constant~$k$. The equation ${\rm d} \delta\Omega = 0$ gives
\[
\partial \chi = \partial{\overline{\partial}} \zeta^{(2,0)} = 0, \qquad \text{and} \qquad {\overline{\partial}} \chi = 0,
\]
and we have $\nabla_\mu \Delta_\alpha^\mu = 0$.

\subsection{Holomorphic deformations}

Let us now repeat this analysis in which $\delta_h = \delta y^\alpha \partial_\alpha$ is a holomorphic deformation. This means~$\delta_h \omega$ is no longer real. The equation $ {\rm d} \delta_h \omega = 0$ gives three parts
\[
{\overline{\partial}} \delta_h \omega^{(0,2)} = 0, \qquad \partial \delta_h \omega^{(0,2)} + {\overline{\partial}} \delta_h \omega^{(1,1)} = 0, \qquad \partial \delta_h \omega^{(1,1)} = 0.
\]
The first equation is solved by
\[
 \delta_h \omega^{(0,2)} = {\overline{\partial}} \xi^{(0,1)}.
\]
The second equation becomes ${\overline{\partial}} \big( \delta_h \omega^{(1,1)} -\partial \xi^{(0,1)}\big) = 0$ which has a solution
\[
 \delta_h \omega^{(1,1)} = \gamma^{(1,1)} + \partial \xi^{(0,1)} + {\overline{\partial}} \beta^{(1,0)} ,
\]
where $\gamma^{(1,1)}$ is ${\overline{\partial}}$-harmonic. The final equation gives ${\overline{\partial}}\partial \beta^{(1,0)} = 0$ and after using $h^{1,0}=h^{2,0}=0$ we have $\beta^{(1,0)} = \partial \Phi$. Hence,
\[
 \delta_h \omega^{(1,1)} = \gamma^{(1,1)} + \partial\big( \xi^{(0,1)} - {\overline{\partial}} \Phi\big) .
\]
The gauge symmetries are small holomorphic diffemorphisms, in which $\varepsilon^{\overline\nu} = 0$ and $\varepsilon^\mu$ is free. These preserve $\delta_h \omega^{(2,0)} = 0$. The action on the remaining parts is
\begin{gather}\label{eq:smallHolDiffOmega}
\delta_h \omega^{(0,2)} \sim \delta_h \omega^{(0,2)} +{\rm i} {\overline{\partial}} \varepsilon^{(0,1)} , \qquad \delta_h \omega^{(1,1)} \sim \delta_h \omega^{(1,1)} +{\rm i} \partial \varepsilon^{(0,1)}.
\end{gather}
We can choose a gauge in which $\delta_h \omega^{(0,2)} = 0$ by setting
\[
{\rm i} \varepsilon^{(0,1)} = - \xi^{(0,1)} + {\overline{\partial}} \psi.
\]
The remaining term $\psi$ can be fixed by demanding
 $\delta_h \omega^{(1,1)}$ be harmonic
\[
 \psi = \Phi.
\]
While $\delta_h \omega$ is harmonic it is not necessarily real.

The holomorphic top form $\delta_h \Omega$ has a decomposition as before in \eqref{eq:dO1}--\eqref{eq:dO2}.
The utility of~$\delta$ being a holomorphic deformation is that $\delta_h {\overline{\Omega}} =0$ and so
\[
\delta_h \Omega = (\delta_h \log f) \Omega = \big(\delta_h \log |f|^2 \big)\Omega = \big(\delta_h \log \|\Omega\|^2 + \omega^{\mu{\overline\nu}} \delta_h \omega_{\mu{\overline\nu}}\big)\Omega.
\]
As before, the Levi-Civita connection satisfies $\nabla_\rho \omega = 0$ and $\partial$-closure of $\delta_h \omega^{(1,1)}$ implies $\nabla_\rho \delta_h \omega_{\mu{\overline\nu}}\allowbreak = \nabla_\mu \delta_h \omega_{\rho{\overline\nu}}$. Hence,
\[
\partial_\rho \big(\omega^{\mu{\overline\nu}} \delta_h \omega_{\mu{\overline\nu}}\big) = \omega^{\mu{\overline\nu}} \nabla_\mu \delta_h \omega_{\rho{\overline\nu}} = -{\rm i} \nabla^{\overline\nu} \delta_h \omega_{\rho{\overline\nu}}.
\]
In the gauge in which $\delta_h \omega^{(1,1)}$ is harmonic it follows that $\omega^{\mu{\overline\nu}} \delta_h \omega_{\mu{\overline\nu}}$ is a constant. Hence, $\delta_h \Omega^{(3,0)} = k \Omega$ for some parameter-dependent constant $k$ and $\partial \zeta^{(2,0)} = 0$. This is the same gauge as the previous subsection.

\subsection{A third way}\label{eq:hologaugefix}
To satisfy ourselves the previous calculation is correct, we repeat the calculation this time starting from a holomorphic variation of~$\Omega$. The decomposition of $\delta \Omega^{(3,0)}$ with respect to the $\partial$-operator gives
\[
\delta \Omega^{(3,0)} = k \Omega + \partial \xi^{(2,0)},
\]
where $\partial \xi^{(2,0)} = k_\xi \Omega$. A gauge transformation acts as
\[
\delta \Omega \to \delta \Omega + {\rm d} (\varepsilon^\nu \Omega_\nu) = \delta \Omega + (\nabla_\nu \varepsilon^\nu) \Omega.
\]
We choose
\[
\varepsilon^\nu = - \frac{1}{2\|\Omega\|^2} {\overline{\Omega}}^{\nu\rho\sigma} \big(\xi_{\rho\sigma} + \big(\partial \zeta^{(1,0)}\big)_{\rho\sigma}\big),
\]
where $\zeta^{(1,0)}$ is an arbitrary one form. With this choice $\nabla_\nu \varepsilon^\nu = -\frac{1}{3! \|\Omega\|^2} {\overline{\Omega}}^{\nu\rho\sigma}\big(\partial\xi^{(2,0)}\big)_{\nu\rho\sigma}$ and $\delta \Omega^{(3,0)} = k \Omega$ for some constant~$k$. We see that $\zeta^{(1,0)}$ is a residual gauge freedom that does not affect $\delta \Omega$.

The variation of the K\"ahler form is holomorphic, and so the equation of motion ${\rm d} \delta \omega = 0$ implies
\[
{\overline{\partial}} \delta \omega^{(0,2)} = 0, \qquad \partial \delta \omega^{(0,2)} + {\overline{\partial}} \delta \omega^{(1,1)} = 0, \qquad \partial \delta \omega^{(1,1)} = 0.
\]
The solution to these equations is
\[
\delta \omega^{(0,2)} = {\overline{\partial}} \alpha^{(0,1)}, \qquad \delta \omega^{(1,1)} = \gamma^{(1,1)} + \partial \alpha^{(0,1)},\qquad \delta \omega^{(2,0)} = 0,
\]
where $\gamma^{(1,1)}$ is a harmonic $(1,1)$-form. There is a condition on $\alpha^{(0,1)}$ that derives from the gauge fixing of $\delta \Omega$. To see this, we first turn to a calculation used several times above
\[
\partial \big(\omega^{\mu{\overline\nu}} \delta \omega_{\mu{\overline\nu}} \big) = {\rm i} {\overline{\partial}}^\dag \delta \omega^{(1,1)} = 0,\qquad {\overline{\partial}} \big(\omega^{\mu{\overline\nu}} \delta \omega_{\mu{\overline\nu}} \big) = -{\rm i} g^{\mu{\overline\nu}} \big({\overline{\partial}} \delta \omega^{(1,1)}\big)_{\mu{\overline\nu}} + {\rm i} \partial^\dag \delta \omega^{(1,1)} = 0.
\]
The vanishing of these equation is the gauge fixing $\delta \Omega^{(3,0)} = k \Omega$. Consider the first equation:
\[
{\overline{\partial}}^\dag \delta \omega^{(1,1)} =\big( \nabla^{\overline\nu} \nabla_\rho \alpha_{\overline\nu} \big) {\rm d} x^\rho = \partial\big(\nabla^{\overline\nu} \alpha_{\overline\nu}\big) = 0.
\]
There are no holomorphic functions on $X$ and so this implies $\nabla^{\overline\nu} \alpha_{\overline\nu}$ is a constant. In fact, the constant must vanish, as can be seen by integrating it over the manifold and using that a total divergance on a compact manifold has vanishing integral. Hence,
\[
{\overline{\partial}}^\dag \alpha^{(0,1)} = 0.
\]
On the other hand, we write this in terms of the Hodge dual
\[
{\overline{\partial}}^\dag \alpha^{(0,1)} = - \star \partial \star \alpha^{(0,1)} = - \star \partial \big( \alpha_{\overline\nu} \Omega^{\overline\nu}\big) \frac{{\rm i}{\overline{\Omega}}}{\|\Omega\|^2}= 0.
\]
The fact this vanishes implies $ \alpha_{\overline\nu} \Omega^{\overline\nu} = \partial \beta^{(1,0)}$ for some $(1,0)$-form. This equation is invertible
\[
\alpha^{(0,1)} = \frac{1}{2\|\Omega\|^2} {\overline{\Omega}}^{\rho\sigma} \big(\partial \beta^{(1,0)}\big)_{\rho\sigma}.
\]
We finally, see that $\alpha^{(0,1)}$ takes the same form as the residual gauge freedom described above, and that by setting $\zeta^{(1,0)} = -\beta^{(1,0)}$ we fix $\delta \omega^{(1,1)} = \gamma^{(1,1)}$ to be harmonic.

We have now illustrated how to gauge fix to harmonic gauge in three seperate ways, and this serves as a useful guide for gauge fixing the heterotic system.

\subsection{Conformally balanced condition}

A Calabi--Yau manifold is in particular a balanced manifold and satisfies
\[
{\rm d}\omega^2 = 0.
\]
A holomorphic variation gives a pair of independent equations
\[
\partial \big(\delta_h \omega^{(0,2)} \omega\big) + {\overline{\partial}} \big(\delta_h \omega^{(1,1)} \omega\big) = 0, \qquad \partial \big(\delta_h \omega^{(1,1)} \omega\big) = 0.
\]
Under a small diffeomorphism \eqref{eq:smallHolDiffOmega} these equations are preserved provided we use ${\rm d} \omega = 0$. Hence, a holomorphic variation of the balanced condition on a K\"ahler manifold is not a gauge fixing.

Using the following Hodge duals
\[
\star \delta_h \omega^{(1,1)} = \big(\omega^{\mu{\overline\nu}} \delta_h \omega_{\mu{\overline\nu}}\big) \omega^2 - \delta_h \omega^{(1,1)} \omega , \qquad \star \delta_h \omega^{(0,2)} = \delta_h \omega^{(0,2)} \omega ,
\]
we get
\[
\partial^\dag \delta_h \omega^{(1,1)} = {\overline{\partial}}^\dag \delta_h \omega^{(0,2)} + {\rm i} {\overline{\partial}} \big(\omega^{\mu{\overline\nu}} \delta_h \omega_{\mu{\overline\nu}}\big), \qquad {\overline{\partial}}^\dag \delta_h \omega^{(1,1)} = {\rm i} \partial\big(\omega^{\mu{\overline\nu}} \delta_h \omega_{\mu{\overline\nu}}\big).
\]

\section{Constant dilaton is a choice of gauge}\label{s:dilatongauge}
We have focussed primarily on the role of small deformations of the background corresponding to motions along the moduli space. For a fixed point in moduli space, the background is fixed. However, the fields still have a perturbative expansion in $\alpha^{\backprime}$ and so we need to gauge fix the $\alpha^{\backprime}$-corrections. This calculation has already been treated in the literature, see for example~\cite{Anguelova:2010ed, Gillard:2003jh,Witten:1985bz, Witten:1986kg}, and so we provide a summary of those results in our notation following primarily~\cite{Anguelova:2010ed}.

For a fixed background, there is an $\alpha^{\backprime}$-expansion. At zeroth order the metric is K\"ahler and $H=0$~\cite{Witten:1986kg}. The corrections in $\alpha^{\backprime}$ are assumed small and so we may apply the background field method and as a part of this gauge fix diffeomorphisms. The metric and dilaton we write as
\begin{gather}\label{eq:metricDilatonap}
g_{mn} = g^{(0)}_{mn} + \alpha^{\backprime} g^{(1)}_{mn} + {\alpha^{\backprime}}^2 g^{(2)}_{mn} + \cdots, \qquad \phi = \phi^{(0)} + \alpha^{\backprime} \phi^{(1)} + {\alpha^{\backprime}}^2 \phi^{(2)}_{mn} + \cdots.
\end{gather}
 The inverse metric has an $\alpha^{\backprime}$-expansion given by (suppressing indices for simplicity)
\begin{gather}\label{eq:inverseMetric}
g^{-1} = g^{(0)\,-1} - \alpha^{\backprime} g^{(0)\,{-}1} g^{(1)} g^{(0)\,{-}1} - {\alpha^{\backprime}}^2 \big( g^{(0)\,{-}1} g^{(2)} g^{(0)\,{-}1} - g^{(0)\,{-}1} g^{(1)} g^{(0)\,{-}1} g^{(1)} \big).
\end{gather}
We assume the large radius expansion applies so that these are guaranteed solutions of an underlying sigma model and so CFT. This means $g^{(0)}$ is K\"ahler and $H= \mathcal{O}(\alpha^{\backprime})$. The calculation in this appendix shows that a necessary consequence is that the dilaton is constant.

Diffeomorphisms are treated according to the background gauge principle, see for example. This means $g^0_{mn}$ as a background field while perturbations $g^{(1)}_{mn}$ are subject to small diffeomorphisms
\[
g^{(i)}_{mn} \to \widetilde g^{(i)}_{mn} = g^{(i)}_{mn} + \nabla_m \epsilon^{(i)}_n + \nabla_n \epsilon^{(i)}_m,
\]
where $\nabla_n$ is the Levi-Civita connection with respect to~$g^{(0)}$ and $i=1,2,\dots$ denotes order in the $\alpha^{\backprime}$-expansion.

The issue of gauge fixing is discussed in~\cite{Anguelova:2010ed} which we now recount. The gauge fixing described in that paper is
\begin{gather}
g^{np} \nabla_n g_{mp} = \tfrac{1}{2} (1-2\xi) \nabla_m (\log \det g ) + \alpha^{\backprime} \zeta P_m,
\label{eq:gaugefix}
\end{gather}
where $\xi$, $\zeta$ are gauge parameters and
\[
P_{m} = 3 e^{2\phi} \nabla^{(-)\, q} \big(e^{-2\phi} {\rm d} H\big)_{mnpq} g^{np}.
\]
The Bianchi identity for $H$ implies that $P_m$ is at least order $\alpha^{\backprime}$.
Substituting \eqref{eq:metricDilatonap} into \eqref{eq:gaugefix} and using
\[
\log \det g = \log \det g^{(0)} + \alpha^{\backprime} g^{(1)}{}_n{}^n + {\alpha^{\backprime}}^2 \big(g^{(2)}{}_n{}^n -\tfrac{1}{2} g^{(1)}{}^{mn} g^{(1)}{}_{mn} \big) + \cdots ,
\]
the gauge fixing condition becomes
\begin{gather}
\nabla^n g^{(1)}_{mn} = \tfrac{1}{2} (1-2\xi) \nabla_m g^{(1)}{}_{n}{}^n,\label{eq:PerturbativegaugeFix}\\
\nabla^n g^{(2)}_{mn} - \tfrac{1}{2} (1-2\xi) \nabla_m g^{(2)}{}_{n}{}^n = - \tfrac{1}{4} (1-2\xi) \nabla_m \big(g^{(1)}{}_{np}g^{(1)}{}^{np} \big)+ g^{(1)\, np} \nabla_n g^{(1)}{}_{mp} + \alpha^{\backprime} \zeta P_m ,\nonumber
\end{gather}
where indices are raised by $g^{(0)}$. The first line with $\xi=0$ captures the gauge fixing described in~\cite{Witten:1986kg}.

Now consider the role of supersymmetry. The gravitino variation vanishing implies
\[
H_{mnp} = ({\rm d}^c \omega)_{mnp} + 2\alpha^{\backprime} P_{mnp} + \mathcal{O}\big({\alpha^{\backprime}}^3\big).
\]

The dilatino variation, when combined with the previous equation imposes
\[
\partial_m \phi =\tfrac{1}{2} ({\rm d}^c\omega - \alpha^{\backprime} P)_{mnp} g^{np} + \mathcal{O}\big({\alpha^{\backprime}}^3\big).
\]
At this point it is useful to introduce complex coordinates. The first term is
\[
({\rm d}^c \omega)_{\mu\nu{\overline{\rho}}} g^{\nu{\overline{\rho}}} = ( {\rm i} \partial \omega)_{\mu\nu{\overline{\rho}}} g^{\nu{\overline{\rho}}} = ( {\rm i} \nabla_\mu \omega_{\nu{\overline{\rho}}} - {\rm i} \nabla_\nu \omega_{\mu{\overline{\rho}}} ) g^{\nu{\overline{\rho}}},
\]
where at this point $\nabla$ is any symmetric connection. Note that this coincides with the $(1,0)$-component of the Lee form up to a sign:
\[
\theta_\mu = ({\rm d} \omega)_{\mu \nu{\overline{\rho}}} \omega^{\nu{\overline{\rho}}}= - ({\rm i} \nabla_\mu \omega_{\nu{\overline{\rho}}} -{\rm i} \nabla_\nu \omega_{\mu{\overline{\rho}}} ) g^{\nu{\overline{\rho}}} .
\]
Now, recall a connection is Hermitian if and only if $\nabla_\mu J = 0$ where $J$ is the complex structure and the complex structure does not receive $\alpha^{\backprime}$ corrections. It is metric compatible if $\nabla_\mu g = 0$. Finally, a metric is K\"ahler if and only if its Levi-Civita connection is Hermitian $\nabla_\mu J = 0$. Taken together, a convenient choice is if~$\nabla$ is the Levi-Civita connection with respect to~$g^0$, which is K\"ahler. $\nabla$ will now denote this connection unless we specify otherwise. Hence,
\[
({\rm d}^c \omega)_{\mu\nu{\overline{\rho}}} \, g^{\nu{\overline{\rho}}} = ( - \nabla_\mu g_{\nu{\overline{\rho}}} + \nabla_\nu g_{\mu{\overline{\rho}}} ) g^{\nu{\overline{\rho}}}.
\]
The three-form $P$ can be simplified using the fact ${\rm d} H = \mathcal{O}(\alpha^{\backprime})$ and we are working to order ${\alpha^{\backprime}}^2$:
\begin{gather*}
\alpha^{\backprime} P_\mu = \frac{3{\alpha^{\backprime}}^2}{2} \nabla^{ {\overline{\sigma}}} \big(\operatorname{Tr} F^2 - \operatorname{Tr} R^2\big)_{\mu {\overline\nu}\rho{\overline{\sigma}}} g^{\rho{\overline\nu}} = \frac{3{\alpha^{\backprime}}^2}{2} \partial_\mu \big( \operatorname{Tr} F_{\tau{\overline{\rho}}} F^{\tau{\overline{\rho}}} - \operatorname{Tr} R_{\tau{\overline{\rho}}} R^{\tau{\overline{\rho}}} \big)\\
\hphantom{\alpha^{\backprime} P_\mu}{} = \frac{3{\alpha^{\backprime}}^2}{2} \partial_\mu \big( \operatorname{Tr} |F|^2 - \operatorname{Tr} |R|^2 \big).
\end{gather*}
The indices in $|F|^2$, $|R|^2$ are raised by $g^{(0)}$. We have also used here the field equation ${\rm d}_A^\dag F = \mathcal{O}(\alpha^{\backprime})$ and Bianchi identity ${\rm d}_A F = 0$.

Putting these two results together
\begin{gather*}
 \partial_\mu \phi = \big( {-} \nabla_\mu g_{\nu{\overline{\rho}}} + \nabla_\nu g_{\mu{\overline{\rho}}}\big) g^{\nu{\overline{\rho}}} - \tfrac{3{\alpha^{\backprime}}^2}{2} \partial_\mu \big( \operatorname{Tr} |F|^2 - \operatorname{Tr} |R|^2 \big).
\end{gather*}
We now expand in $\alpha^{\backprime}$. Using \eqref{eq:metricDilatonap}--\eqref{eq:inverseMetric} we find
\begin{gather*}
 \partial_\mu \phi^{(0)} = 0, \qquad \partial_\mu \phi^{(1)} = \nabla^{\overline\nu} g^{(1)}_{\mu{\overline\nu}} - \nabla_\mu g^{(1)}{}_{\nu}{}^{\nu} ,\\
 \partial_\mu \phi^{(2)} =
 \nabla^{\overline\nu} g^{(2)}_{\mu{\overline\nu}} - \nabla_\mu g^{(2)}{}_{\nu}{}^\nu
+ \big( \nabla_\mu g^{(1)}_{\nu{\overline{\rho}}} -\nabla_\nu g^{(1)}_{\mu{\overline{\rho}}} \big) g^{(1)\,\nu{\overline{\rho}}} - \tfrac{3}{2} \partial_\mu \big( \operatorname{Tr} |F|^2 - \operatorname{Tr} |R|^2 \big) ,
\end{gather*}
where recall indices are raised with $g^{(0)}$ and so for example $g^{(1)\,\nu{\overline{\rho}}} = g^{(0)\,\nu{\overline{\lambda}}} g^{(1)}_{\tau{\overline{\lambda}}} g^{(0)\,\tau{\overline{\rho}}}$.

The gauge fixing condition \eqref{eq:PerturbativegaugeFix} in complex coordinates after a modicum of algebra is
\begin{gather*}
\nabla^{\overline\nu} g^{(1)}_{\mu{\overline\nu}} - \nabla_\mu g^{(1)}{}_{\nu}{}^\nu = -2\xi \nabla_\mu g^{(1)}{}_{\nu}{}^\nu,\\
\nabla^{\overline\nu} g^{(2)}_{\mu{\overline\nu}} - \nabla_\mu g^{(2)}{}_{\nu}{}^\nu + \big( \nabla_\mu g^{(1)}{}_{\nu{\overline{\rho}}} - \nabla_\nu g^{(1)}{}_{\mu{\overline{\rho}}} \big) g^{(1)}{}^{\nu{\overline{\rho}}} \\
\qquad{} = - 2\xi \nabla_\mu g^{(2)}{}_{\nu}{}^\nu + \xi \nabla_\mu \big(g^{(1)}{}_{\nu{\overline{\rho}}}g^{(1)}{}^{\nu{\overline{\rho}}} \big) + \tfrac{3}{2} \zeta \partial_\mu \left( \operatorname{Tr} |F|^2 - \operatorname{Tr} |R|^2 \right) .
\end{gather*}

The variation of the dilaton is then
\begin{gather*}
 \partial_\mu \phi^{(0)} = 0, \qquad \partial_\mu \phi^{(1)} = -2\xi \nabla_\mu g^{(1)}{}_\nu{}^\nu,\\
 \partial_\mu \phi^{(2)} = - 2\xi \nabla_\mu g^{(2)}{}_{\nu}{}^\nu + \xi \nabla_\mu \big(g^{(1)}{}_{\nu{\overline{\rho}}}g^{(1)}{}^{\nu{\overline{\rho}}} \big) + \tfrac{3}{2} (\zeta-1) \partial_\mu \big( \operatorname{Tr} |F|^2 - \operatorname{Tr} |R|^2 \big) .
\end{gather*}
If we choose $\xi = 0$ and $\zeta = 1$, the dilaton is constant to the order we work
\[
\phi = \phi_0 + \mathcal{O}\big({\alpha^{\backprime}}^3\big).
\]
As shown in \cite{Anguelova:2010ed, Witten:1986kg} these gauge condition exists on a compact~$X$. As the metric is K\"ahler at leading order, the Lee form $\theta$ vanishes to ${\alpha^{\backprime}}^3$. A consequence of the dilatino and gravitino variation is that $\nabla^B\big(e^{2\phi} \Omega\big)=0$ where ${\rm d} \Omega = 0$. Hence, $\big\|e^{2\phi} \Omega\big\|$ is a constant on $X$ and so
\begin{gather}\label{eq:OmDilaton}
{\rm d} \log \|\Omega\| = -2 {\rm d} \phi.
\end{gather}
From the discussion above, as $\phi$ is a constant so is $\|\Omega\|$. Finally, note that the dilatino variation implies $H_{\mu\nu}{}^\nu = 0$.

In sum, for backgrounds that arise as the low energy limits of non-linear sigma models, so that $H=\mathcal{O}(\alpha^{\backprime})$, this calculation shows the dilaton is constant to this order. Up to gauge transformations, if a supergravity background has a varying dilaton that cannot be gauge fixed to vanish then it must have $H=\mathcal{O}(1)$. In this case, however, some sickness in the limit $\alpha^{\backprime}\to 0$ occurs such as the background geometry degenerating to points. Hence, more work is needed to understand whether these are solutions of string theory or simply in the swampland.

\subsection{Checking deformations preserve the gauge fixing on the background}
In Section~\ref{s:Gauge} we gauge fix deformations and the final result is summarised in~\eqref{eq:finalGauge}. We check that this gauge fixing is compatible with the gauge fixing described in this section.

We check that $\|\Omega\|$ and the dilaton, see \eqref{eq:OmDilaton}, remain constant along the moduli space. The variation of the norm is
\[
\delta \|\Omega\|^2 = \tfrac{1}{3!} \delta \Omega_{mnp} {\overline{\Omega}}^{mnp} + \tfrac{1}{3!} \Omega_{mnp} \delta {\overline{\Omega}}^{mnp} + \tfrac{1}{2} \Omega_{\mu\nu\rho} {\overline{\Omega}}_{{\overline{\tau}}}{}^{\nu\rho} \delta g^{\mu{\overline{\tau}}}.
\]
The first two terms are constants by \eqref{eq:finalGauge}. The third term we can rewrite using the Hermitian form
\[
 \Omega_{\mu\nu\rho} {\overline{\Omega}}_{{\overline{\tau}}}{}^{\nu\rho} \delta g^{\mu{\overline{\tau}}} = \Omega_{\mu\nu\rho} {\overline{\Omega}}_{{\overline{\tau}}}{}^{\nu\rho} {\rm i} \delta \omega^{\mu{\overline{\tau}}}\propto \omega_{\mu{\overline\nu}} \delta \omega^{\mu{\overline\nu}},
\]
where in the last line we have suppressed an irrelevant constant and used the compatibility relation between $\omega$ and $\Omega$. The compatability relation also implies the trace of $\delta \omega^{(1,1)}$ is a~constant
\[
\delta \big({\rm i} \Omega {\overline{\Omega}}\big) = {\rm i} \delta k \Omega {\overline{\Omega}} = \|\Omega\|^2 \tfrac{1}{2} (\delta \omega) \omega^2 = \delta \omega^{\mu{\overline\nu}} \omega_{\mu{\overline\nu}} \|\Omega\|^2 \tfrac{1}{3!}\omega^3 = \delta \omega^{\mu{\overline\nu}} \omega_{\mu{\overline\nu}} \big({\rm i} \Omega {\overline{\Omega}}\big) .
\]
Hence $\delta \|\Omega\|$ is a constant on~$X$ and so the only deformations of the dilaton are also constants.

\section{Hodge duals and contractions}\label{s:appendixHodge}

We establish a set of notations and conventions, as well enumerate several useful relations.

\subsection*{Real Riemannian manifolds}

We start with a $D$-dimensional compact Riemannian manifold $X$ with metric $g$.

\subsection*{Inner products for forms}

The pointwise inner product on (exterior powers of) the cotangent space uses the inverse metric
$ g^{-1} = g^{mn} \partial_m \otimes \partial_n$.
Given $k$-forms $\eta$, $\xi$ it is defined as
\begin{gather}\label{g-1etaxi}
 g^{-1}(\eta,\xi) = \frac{1}{k!} \eta_{m_1\dots m_k} g^{m_1 n_1} \cdots g^{m_k n_k} \xi_{n_1\dots n_k} = \frac{1}{k!} \eta^{m_1 \dots m_k} \xi_{m_1\dots m_k} ,
\end{gather}
which is a real function over $X$, whose integral defines an inner product for forms
\[
 (\cdot,\cdot) \colon \ \Omega^k(X) \times \Omega^k(X) \to \mathbb{R} ,
\]
with
\[
 (\eta,\xi) = \frac{1}{V} \int_X \operatorname{vol} g^{-1}(\eta,\xi) = \frac{1}{V k!} \int_X \operatorname{vol} \eta^{m_1 \dots m_k} \xi_{m_1 \dots m_k} .
\]
This is positive definite and with the definition above the zero-form $1$ and the $D$-form $\operatorname{vol}$ have unit norm. Recall, the Riemannian volume form is
\[
 \operatorname{vol} = \frac{1}{D!} \sqrt{g} \epsilon_{m_1 \dots m_D} {\rm d} x^{m_1 \dots m_D} ,
\]
where $\epsilon_{m_1 \dots m_D}$ is the constant antisymmetric symbol.

\subsection*{Contraction of forms}

Given two forms $\eta_k$ and $\xi_l$, where the subscript denotes their degree and $k \leq l$, we can form their contraction, or interior product. The symbol that denotes this operation is
\[
 \lrcorner \colon \ \Omega^k(X) \times \Omega^l(X) \to \Omega^{l-k}(X) ,
\]
and acts on forms as follows
\[
 \eta_k \lrcorner \xi_l = \frac{1}{k!(l-k)!} \eta^{m_1 \dots m_k} \xi_{m_1 \dots m_k n_1 \dots n_{l-k}} {\rm d} x^{n_1 \dots n_{l-k}} = \frac{1}{k!} \eta^{m_1 \dots m_k} \xi_{m_1 \dots m_k} .
\]
The inner product \eqref{g-1etaxi} becomes then just a special case of contraction
\[
 g^{-1}(\eta,\xi) = \eta \lrcorner \xi .
\]
An interesting feature of this operator is that it is the adjoint of the wedge product
\begin{gather}\label{contradjwedg}
 (\sigma_{l-k} \lrcorner \xi_l , \eta_k) = (\xi_l , \sigma_{l-k} \wedge \eta_k) .
\end{gather}

\subsection*{Hodge star operator}

The Hodge dual operator
\[
 \star \colon \ \Omega^k(X) \to \Omega^{D-k}(X) ,
\]
defined as
\[
 \star \eta = \frac{\sqrt{g}}{k!(D-k)!} \eta^{m_1\dots m_k} \epsilon_{m_1\dots m_k n_1\dots n_{D-k}}\,{\rm d} x^{n_1\dots n_{D-k}} .
\]
Hence, $\eta \star \xi = (\eta \lrcorner \xi) \operatorname{vol}$.

The $\star$ operator satisfies the identities
\begin{gather}
 \star^2 \eta_k = (-)^{k(D-k)} \eta_k ,\nonumber\\
 g^{-1}(\eta,\xi) = g^{-1}(\star\eta , \star\xi) .\label{properstar}
\end{gather}
The first line means that $\star$ is invertible with possible eigenvalues~$\pm 1$ or~$\pm i$ according to the degree of the form and the number of dimensions. The second property means~$\star$ is an isometry.

\subsection*{Codifferential}

The codifferential is denoted
\[
 {\rm d}^\dag \colon \ \Omega^k(X) \to \Omega^{k-1}(X) ,
\]
and is defined as the adjoint of the de Rham operator. That is,
\[
 ({\rm d}\eta_{k-1},\xi_k) = \big(\eta_{k-1},{\rm d}^\dag\xi_k\big) .
\]
In order to find its explicit expression we need to perform an integration by parts and use the first line in \eqref{properstar}. Boundary terms will be neglected because we assume $X$ has no boundary. The calculation goes as follows
\begin{gather*}
 ({\rm d}\eta_{k-1} , \xi_k) = \frac{1}{V} \int_X {\rm d} \eta_{k-1} \star \xi_k = (-)^k \frac{1}{V} \int_X \eta_{k-1}\, {\rm d} \star \xi_k \\
\hphantom{({\rm d}\eta_{k-1} , \xi_k)}{} = (-)^k \frac{1}{V} \int_X \eta_{k-1} (-)^{(D-k+1)(k-1)} \star^2 {\rm d} \star \xi_k \\
\hphantom{({\rm d}\eta_{k-1} , \xi_k)}{} = \frac{1}{V} \int_X \eta_{k-1} \star \big( {-} (-)^{D(k+1)} \star {\rm d} \star \xi_k \big) ,
\end{gather*}
and we end up with
\[
 {\rm d}^\dag\xi_k = -(-)^{D(k-1)} \star {\rm d} \star \xi_k .
\]
Another expression, very useful in calculations, is written in terms of the LC connection. It can be obtained very quickly by reminding that ${\rm d}$ has a representation
\[
 {\rm d} = {\rm d} x^m \, \nabla^{\rm LC}_m ,
\]
using property \eqref{contradjwedg} and an integration by parts, as follows
\[
 ( {\rm d} \eta , \xi ) = \big( {\rm d} x^m \, \nabla^{\rm LC}_m \eta , \xi \big) = \big( \nabla^{\rm LC}_m \eta , \xi^m \big) = \big( \eta , - \nabla^{\rm LC}_m \xi^m \big) .
\]
We gather that
\[
 {\rm d}^\dag \xi_k = - \nabla^{\rm LC}_m \xi^m = - \frac{1}{(k-1)!} \nabla^{\rm LC}_n \xi^n{}_{m_1 \dots m_{k-1}} \, {\rm d} x^{m_1 \dots m_{k-1}} .
\]
Proving this identity without using \eqref{contradjwedg} involves rather cumbersome calculations.

\subsection*{Complex manifolds}\label{s:Hodgecomplex}

We now take $X$ to be complex with complex dimension~$N$. We denote holomorphic coordinates~$x^\mu$,~$x^{{\overline\nu}}$. The manifold is equipped with a Hermitian metric
\begin{gather*}%\label{hermg}
 g = g_{\mu{\overline\nu}} \big({\rm d} x^\mu \otimes {\rm d} x^{{\overline\nu}} +{\rm d} x^{{\overline\nu}} \otimes {\rm d} x^{\mu} \big) , \qquad {\rm d} s^2 = 2 g_{\mu{\overline\nu}} \, {\rm d} x^\mu {\rm d} x^{{\overline\nu}},
\end{gather*}
and compatible Hermitian form $\omega$
\[
 \omega = \omega_{\mu{\overline\nu}} \, {\rm d} x^{\mu{\overline\nu}} = J_m{}^p g_{pn} {\rm d} x^{mn} = {\rm i} g_{\mu{\overline\nu}} \, {\rm d} x^{\mu{\overline\nu}} ,
\]
where the last equality is evaluated in the appropriate complex coordinates.

The pointwise inner product for forms respects hermitianity: given forms~$\eta$,~$\xi$ of holomorphic type $(p,q)$ we define
\[
 g^{-1}\big(\eta,{\overline{\xi}}\big) = \eta \lrcorner {\overline{\xi}} = \frac{1}{p!q!} \eta^{{\overline{\mu}}_1 \dots {\overline{\mu}}_p \nu_1 \dots \nu_q} {\overline{\xi}}_{{\overline{\mu}}_1 \dots {\overline{\mu}}_p \nu_1 \dots \nu_q} .
\]
When integrated over the manifold this gives
\[
 \big(\eta, {\overline{\xi}}\big) = \frac{1}{V} \int_X \eta \star {\overline{\xi}} .
\]
Observe how the Hodge $\star$ operator acts on type
\[
 \star \colon \ \Omega^{(p,q)}(X) \to \Omega^{(N-q,N-p)}(X) .
\]
The de Rham differential splits into the sum of Dolbeault operators ${\rm d} = \partial + {\overline{\partial}}$. Analogously, the codifferential also splits ${\rm d}^\dag = \partial^\dag + {\overline{\partial}}^\dag$ where
\begin{gather*}
\partial^\dag \colon \ \Omega^{(p,q)}(X) \to \Omega^{(p-1,q)}(X) , \qquad \partial^\dag = - \star {\overline{\partial}} \star ,\nonumber\\
{\overline{\partial}}^\dag \colon \ \Omega^{(p,q)}(X) \to \Omega^{(p,q-1)}(X) , \qquad {\overline{\partial}}^\dag = - \star \partial \star .%\label{eq:deladjoint}
\end{gather*}

Let us discuss the volume form on~$X$. The holomorphic coordinates have a writing $x^\mu = u^\mu + {\rm i} v^\mu$ with~$u$,~$v$ real coordinates. In terms of the latter, the volume form is defined as
\[
 \operatorname{vol} = \sqrt{g} {\rm d} u^1 \,{\rm d} v^1 \cdots {\rm d} u^N {\rm d} v^N .
\]
Expressed in terms of the complex coordinates:
\begin{gather*}%\label{volholom}
 \operatorname{vol} = \frac{{\rm i}^{N^2}}{N!^2} \sqrt{|g|} \epsilon_{\mu_1\dots \mu_N} \epsilon_{{\overline\nu}_1\dots {\overline\nu}_N} \, {\rm d} x^{\mu_1\dots \mu_N} {\rm d} x^{{\overline\nu}_1\dots {\overline\nu}_N} .
\end{gather*}

We can also write this in terms of the holomorphic volume form
\[
 \operatorname{vol} = \frac{{\rm i} \Omega {\overline{\Omega}}}{\|\Omega\|^2}.
\]
It coincides with the $N$-th power of the Hermitian form
\begin{gather*}
 \frac{1}{N!}\omega^N = \frac{(-)^{\frac{N(N-1)}{2}} {\rm i}^{N}}{N!} g_{\mu_1{\overline\nu}_1} \dots g_{\mu_N{\overline\nu}_N} \epsilon^{\mu_1\dots \mu_N} \epsilon^{{\overline\nu}_1\dots {\overline\nu}_N} \, {\rm d} x^{12\dots N} {\rm d} x^{\bar{1}\bar{2}\dots \bar{N}} \\
\hphantom{\frac{1}{N!}\omega^N }{}
 = {\rm i}^{N^2}\det{g_{\mu{\overline\nu}}} \, {\rm d} x^{12\dots N} {\rm d} x^{\bar{1}\bar{2}\dots \bar{N}} = \frac{{\rm i}^{N^2}}{N!^2} \sqrt{|g|} \epsilon_{\mu_1\dots \mu_N} \epsilon_{{\overline\nu}_1\dots {\overline\nu}_N} \, {\rm d} x^{\mu_1\dots\mu_N} {\rm d} x^{{\overline\nu}_1\dots {\overline\nu}_N} .
\end{gather*}

\subsection*{Codifferential for $\boldsymbol{\Delta_\alpha}$}

Consider the space $\Omega^{(0,q)}\big(X,{\mathcal T}_X^{(1,0)}\big)$. We are mostly interested in $q=1$ but it is not much harder to work in generality. Elements of this space are
\[
 \eta^\mu = \frac{1}{q!} \eta_{{\overline\nu}_1 \dots {\overline\nu}_q}{}^\mu \, {\rm d} x^{{\overline\nu}_1 \dots {\overline\nu}_q}.
\]
There is a Hermitian metric
\begin{gather}\label{eq:0qmetric}
 \big(\eta^\mu , \xi^{{\overline\nu}} \big) = \frac{1}{V} \int_{X} \eta^\mu \star \xi^{{\overline\nu}} g_{\mu{\overline\nu}} , \qquad \xi^{{\overline\nu}} = \frac{1}{q!} \xi_{\rho_1 \dots \rho_q}{}^{{\overline\nu}} \, {\rm d} x^{\rho_1 \dots \rho_q} ,
\end{gather}
where we understand that $\star$ treats $\eta^\mu$, $\xi^{{\overline\nu}}$ as forms. For example, when $q=0,1$
\[
 \star v^\mu = v^\mu \operatorname{vol} , \qquad \star \Delta^\mu = {\rm i} \Delta^\mu \frac{\omega^2}{2} .
\]
There is also a differential operator{\samepage
\[
 {\overline{\partial}} \colon \ \Omega^{(0,q)}\big(X, {\mathcal T}_X^{(1,0)}\big) \to \Omega^{(0,q+1)}\big(X, {\mathcal T}_X^{(1,0)}\big) ,
\]
that raises the degree of one. We use holomorphic coordinates, and this acts covariantly.}

With the help of the metric \eqref{eq:0qmetric} we can define its adjoint
\[
 {\overline{\partial}}^\dag \colon \ \Omega^{(0,q)}\big(X, {\mathcal T}_X^{(1,0)}\big) \to \Omega^{(0,q-1)}\big(X, {\mathcal T}_X^{(1,0)}\big) ,
\]
which satisfies the property
\[
 \big(\eta^\mu , \partial \xi^{{\overline\nu}}\big) = \big({\overline{\partial}}^\dag \eta^\mu , \xi^{{\overline\nu}}\big) .
\]
The calculation uses an integration by parts -- in which we neglect boundary terms -- and standard properties of $\star$. We end up with
\[
 \big(\eta^\mu , \partial \xi^{{\overline\nu}}\big) = \frac{1}{V} \int_X \big( {-} \star \partial \big( {\star} \eta^\rho g_{\rho{\overline{\sigma}}}\big) g^{{\overline{\sigma}}\mu} \big) \star \xi^{{\overline\nu}} g_{\mu{\overline\nu}} .
\]
From this we read the expression for the codifferential
\[
{\overline{\partial}}^\dag \eta^\mu = - \star \partial\big( {\star} \eta^\rho g_{\rho{\overline{\sigma}}}\big) g^{{\overline{\sigma}}\mu} .
\]
For $q=0$ this vanishes trivially. When $q=1$, the case we are most interested in, we find using also the balanced condition
\begin{gather*}%\label{eq:codiffDelta}
 {\overline{\partial}}^\dag \Delta^\mu = - \star \frac{{\rm i} \omega^2}{2} \big( \partial \Delta^\mu + g^{\mu{\overline{\sigma}}}\partial g_{{\overline{\sigma}}\rho} \Delta^\rho \big) = - \nabla_{\rm Ch}^{{\overline\nu}} \Delta_{{\overline\nu}}{}^\mu .
\end{gather*}
Observe how this is different from the gauge-fixing condition \eqref{eq:gfDelta}, due to a different ordering of the indices. The two expressions coincide when $\Delta_{[{\overline{\mu}}{\overline\nu}]}=0$.

\subsection*{Hodge dual relations on a three-fold}

We can finally enumerate some Hodge dual relations for various types of forms on~$X$.

One-forms, type $(1,0)$:
\begin{gather*}%\label{eq:Hodge1form}
 \star \eta^{(1,0)} = -{\rm i} \eta^{(1,0)} \frac{\omega^2}{2} .
\end{gather*}
 Two-forms, types $(2,0)$ and $(1,1)$:
\begin{gather}
\star \eta^{(2,0)} = \eta^{(2,0)} \omega ,\qquad
\star \eta^{(1,1)} = -{\rm i} \eta_{\mu}{}^\mu \frac{\omega^2}{2} - \eta^{(1,1)} \omega = \big(\omega \lrcorner \eta^{(1,1)}\big) \frac{\omega^2}{2} - \eta^{(1,1)} \omega .\label{eq:Hodge2form}
\end{gather}
 Three-forms, types $(3,0)$ and $(2,1)$:
 \begin{gather*}
 \star \eta^{(3,0)} = - {\rm i} \eta^{(3,0)} ,\qquad
\star \eta^{(2,1)} = {\rm i} \eta^{(2,1)} - \eta_{\mu}{}^\mu{}^{(1,0)} \omega = {\rm i} \eta^{(2,1)} - {\rm i} \big(\omega \lrcorner \eta^{(2,1)}\big) \omega .%\label{eq:Hodge3form}
\end{gather*}
 Type $(2,3)$:
\begin{gather}\label{eq:Hodge23}
 \star \eta^{(2,3)} = \tfrac{{\rm i}}{2} \eta_{\mu\nu}{}^{\mu\nu}{}^{(0,1)} = \tfrac{{\rm i}}{2} \omega \lrcorner \big(\omega \lrcorner \eta^{(2,3)}\big) .
\end{gather}

A useful special case of \eqref{eq:Hodge2form} is $\tfrac{1}{2} \omega^2 = \star \omega$. Further relations such as that for $\eta^{(0,1)}$ can be easily determined using complex conjugation.

\section[Connection symbols on $X$]{Connection symbols on $\boldsymbol{X}$}

We enumerate some commonly used connection symbols in heterotic theories. We list the components in complex coordinates. We also give expressions for various divergences which are useful for calculations in the paper.

$X$ is a complex manifold with complex structure $J$ and Hermitian metric~$g$. A vector bundle ${\mathcal E}\to X$ is Hermitian if there is a Hermitian inner product on sections of the bundle. For example ${\mathcal T}_X$ has a Hermitian structure facilitated by the Hermitian metric ${\rm d} s^2 = 2 g_{\mu{\overline\nu}} {\rm d} x^\mu \otimes {\rm d} x^{\overline\nu}$. The bundle is holomorphic (or has a holomorphic structure) if the total space ${\mathcal E}$ is a complex manifold with complex structure ${\mathbb J}$ and the projection map $\pi\colon {\mathcal E} \to X$ is holomorphic. That is, fibres consist only of holomorphic sections according to ${\mathbb J}$. This is equivalent to the transition functions being purely holomorphic. For example, the bundle ${\mathcal T}_X$ is not a holomorphic bundle while ${\mathcal T}_X^{(1,0)}$ is a holomorphic bundle.

A connection $\nabla$ on ${\mathcal T}_X$ is metric compatible if $\nabla g = 0$. A connection $\nabla$ is Hermitian if it preserves the Hermitian structure. That is, it is metric compatible and $\nabla J = 0$. In terms of components, a Hermitian connection has $\Gamma_\mu{}^\nu{}_{\overline{\rho}} = \Gamma_\mu{}^{\overline\nu}{}_\rho = 0$. There may be more than one Hermitian connection.

If ${\mathcal E}$ is a holomorphic bundle then a connection $\nabla$ is compatible with its holomorphic structure if $\nabla^{(0,1)} = {\overline{\partial}}$. In terms of components $\Gamma_{\overline{\mu}}{}^\nu{}_\rho = 0$. For example, if $V$ is a section of ${\mathcal T}_X^{(1,0)}$ then $\nabla$ is compatible with the holomorphic structure if $\nabla^{(0,1)} V = {\rm d} x^{\overline{\mu}} \big(\partial_{\overline{\mu}} V^\nu + \Gamma_{\overline{\mu}}{}^\nu{}_\rho V^\rho \big) \partial_\nu = {\overline{\partial}} V$.

\subsubsection*{Levi-Civita}

Levi-Civita is the unique metric compatible connection with no torsion (symmetric in lower indices). It is Hermitian if the manifold is K\"ahler but not in general:
 \begin{gather*}
 \Gamma^{\rm LC}{}_{\mu}{}^{\nu}{}_\rho = \tfrac{1}{2} g^{\nu{\overline{\sigma}}}(\partial_\mu g_{\rho{\overline{\sigma}}} + \partial_\rho g_{\mu{\overline{\sigma}}}) = g^{\nu{\overline{\sigma}}} \partial_\mu g_{\rho{\overline{\sigma}}} - \tfrac{1}{2} H_{\mu}{}^{\nu}{}_{\rho} = g^{\nu{\overline{\sigma}}} \partial_\rho g_{\mu{\overline{\sigma}}} + \tfrac{1}{2} H_{\mu}{}^{\nu}{}_{\rho} ,\\
 \Gamma^{\rm LC}{}_{\mu}{}^{{\overline\nu}}{}_{\rho} = 0 ,\\
 \Gamma^{\rm LC}{}_{\mu}{}^{\nu}{}_{{\overline{\rho}}} = \tfrac{1}{2} g^{\nu{\overline{\sigma}}}(\partial_{{\overline{\rho}}} g_{\mu{\overline{\sigma}}} - \partial_{{\overline{\sigma}}} g_{\mu{\overline{\rho}}}) = \tfrac{1}{2} H_{\mu}{}^\nu{}_{{\overline{\rho}}} ,\\
 \Gamma^{\rm LC}{}_{\mu}{}^{{\overline\nu}}{}_{{\overline{\rho}}} = \tfrac{1}{2} g^{{\overline\nu}\sigma}(\partial_\mu g_{{\overline{\rho}}\sigma} - \partial_{\sigma} g_{\mu{\overline{\rho}}}) = -\tfrac{1}{2} H_{\mu}{}^{{\overline\nu}}{}_{{\overline{\rho}}} .
\end{gather*}

\subsubsection*{Bismut}

The supersymmetry Killing spinor of heterotic supergravity (to first order in $\alpha^{\backprime}$) is covariantly constant with respect to the connection $\Gamma_m^{\rm B} = \Gamma_m^{\rm LC} - \tfrac{1}{2} H_m$. Writing~$J$ as a spinor bilinear it follows that $\omega$ and $J$ are covariantly constant with respect to this connection $\nabla^{\rm B} J = \nabla^{\rm B} \omega = 0$ and so it follows $\Gamma^{\rm B}$ is metric compatible and Hermitian. The torsion of $\Gamma^{\rm B}$ is completely antisymetric and equal to~$H={\rm d}^c \omega$, i.e., $T^m{}_{np} = H^m{}_{np}$. A geometric statement is that there is a unique connection on ${\mathcal T}_X$ that is Hermitian with completely antisymmetric torsion. This is the Bismut connection:
 \begin{gather*}
 \Gamma^{\rm B}{}_{\mu}{}^{\nu}{}_\rho = g^{\nu{\overline{\sigma}}} \partial_\rho g_{\mu{\overline{\sigma}}} = g^{\nu{\overline{\sigma}}} \partial_\mu g_{\rho{\overline{\sigma}}} - H_{\mu}{}^{\nu}{}_{\rho} ,\qquad
 \Gamma^{\rm B}{}_{\mu}{}^{{\overline\nu}}{}_{\rho} = 0 ,\qquad
 \Gamma^{\rm B}{}_{\mu}{}^{\nu}{}_{{\overline{\rho}}} = 0 ,\\
 \Gamma^{\rm B}{}_{\mu}{}^{{\overline\nu}}{}_{{\overline{\rho}}} = g^{{\overline\nu}\sigma} \big(\partial_\mu g_{\sigma{\overline{\rho}}} - \partial_\sigma g_{\mu{\overline{\rho}}}\big) = - H_{\mu}{}^{{\overline\nu}}{}_{{\overline{\rho}}} .
\end{gather*}

\subsubsection*{Hull}

While the spinor in heterotic is covariantly constant with respect to~$\Gamma^{\rm B}$, a different connection~$\Gamma^{\rm H}$ appears in the heterotic action. It is not Hermitian but has completely antisymmetric torsion with opposite sign~$-H$. Hence, $\Gamma_m^{\rm H} = \Gamma_m^{\rm LC} + \tfrac{1}{2} H_m$. This we call the Hull connection:
\begin{gather*}
\Gamma^{\rm H}{}_{\mu}{}^{\nu}{}_\rho = g^{\nu{\overline{\sigma}}} \partial_\mu g_{\rho{\overline{\sigma}}} = g^{\nu{\overline{\sigma}}} \partial_\rho g_{\mu{\overline{\sigma}}} + H_{\mu}{}^{\nu}{}_{\rho} ,\qquad
 \Gamma^{\rm H}{}_{\mu}{}^{{\overline\nu}}{}_{\rho} = 0 ,\\
 \Gamma^{\rm H}{}_{\mu}{}^{\nu}{}_{{\overline{\rho}}} = g^{\nu{\overline{\sigma}}}(\partial_{{\overline{\rho}}} g_{\mu{\overline{\sigma}}} - \partial_{{\overline{\sigma}}} g_{\mu{\overline{\rho}}}) = H_\mu{}^\nu{}_{{\overline{\rho}}} ,\qquad
 \Gamma^{\rm H}{}_{\mu}{}^{{\overline\nu}}{}_{{\overline{\rho}}} = 0 .
\end{gather*}

\subsubsection*{Chern}

The Chern connection is the unique connection which is Hermitian ($\nabla g = \nabla J = 0$) and compatible with the holomorphic structure of ${\mathcal T}_X^{(1,0)}$. The connection has no mixed indices. If the manifold is non-K\"ahler then it has torsion:
 \begin{gather*}
 \Gamma^{\rm Ch}{}_{\mu}{}^{\nu}{}_\rho = g^{\nu{\overline{\sigma}}} \partial_\mu g_{\rho{\overline{\sigma}}} = g^{\nu{\overline{\sigma}}} \partial_\rho g_{\mu{\overline{\sigma}}} + H_{\mu}{}^{\nu}{}_{\rho} ,\\
 \Gamma^{\rm Ch}{}_{\mu}{}^{{\overline\nu}}{}_{\rho} = 0 ,\qquad
 \Gamma^{\rm Ch}{}_{\mu}{}^{\nu}{}_{{\overline{\rho}}} = 0 ,\qquad
 \Gamma^{\rm Ch}{}_{\mu}{}^{{\overline\nu}}{}_{{\overline{\rho}}} = 0 .
\end{gather*}

\subsection*{Divergences}

The divergence of a vector $\varepsilon^\mu$ taken with respect to a generic connection
\[
 \nabla_\mu \varepsilon^\mu = \partial_\mu\varepsilon^\mu + \varepsilon^\mu \Gamma_\nu{}^{\nu}{}_{\mu} .
\]
To compute this we need the following contraction
\begin{gather*}
 \Gamma^{\rm LC}{}_\nu{}^\nu{}_\mu = \partial_\mu\log{\sqrt g} + \tfrac{1}{2} H_{\mu\nu}{}^{\nu} ,\\
 \Gamma^{\rm B}{}_\nu{}^\nu{}_\mu = \partial_\mu\log{\sqrt g} ,\\
 \Gamma^{\rm H}{}_\nu{}^\nu{}_\mu = \Gamma^{\rm Ch}{}_\nu{}^\nu{}_\mu = \partial_\mu\log{\sqrt g} + H_{\mu\nu}{}^\nu .
\end{gather*}
The four choices above give
 \begin{gather*}
 \nabla_\mu^{\rm LC}\varepsilon^\mu = \partial_\mu\varepsilon^\mu + \varepsilon^\mu \partial_\mu\log{\sqrt g} + \tfrac{1}{2} \varepsilon^\mu H_{\mu\nu}{}^\nu ,\nonumber\\
 \nabla_\mu^{\rm B}\varepsilon^\mu = \partial_\mu\varepsilon^\mu + \varepsilon^\mu \partial_\mu\log{\sqrt g} ,\nonumber\\
 \nabla_\mu^{\rm H}\varepsilon^\mu = \nabla_\mu^{\rm Ch}\varepsilon^\mu = \partial_\mu\varepsilon^\mu + \varepsilon^\mu \partial_\mu\log{\sqrt g} + \varepsilon^\mu H_{\mu\nu}{}^\nu.%\label{divve}
\end{gather*}
As for the vector-valued form $\Delta_{{\overline\nu}}{}^\mu$ we have, for a generic connection
\[
 \nabla_\mu\Delta_{{\overline\nu}}{}^\mu = \partial_\mu\Delta_{{\overline\nu}}{}^\mu + \Delta_{{\overline\nu}}{}^\mu \Gamma_\rho{}^\rho{}_\mu - \Gamma_\mu{}^{{\overline{\rho}}}{}_{{\overline\nu}} \Delta_{{\overline{\rho}}}{}^\mu ,
\]
and the four choices above give
\begin{gather*}
 \nabla_\mu^{\rm LC}\Delta_{{\overline\nu}}{}^\mu = \partial_\mu\Delta_{{\overline\nu}}{}^\mu + \Delta_{{\overline\nu}}{}^\mu \partial_\mu \log{\sqrt g} + \tfrac{1}{2} \Delta_{{\overline\nu}}{}^\mu H_{\mu\rho}{}^\rho - \tfrac{1}{2} \Delta^{\mu\rho} H_{\mu\rho{\overline\nu}} ,\nonumber\\
 \nabla_\mu^{\rm B}\Delta_{{\overline\nu}}{}^\mu = \partial_\mu\Delta_{{\overline\nu}}{}^\mu + \Delta_{{\overline\nu}}{}^\mu \partial_\mu \log{\sqrt g} - \Delta^{\mu\rho} H_{\mu\rho{\overline\nu}} ,\nonumber\\
 \nabla_\mu^{\rm H}\Delta_{{\overline\nu}}{}^\mu = \nabla_\mu^{\rm Ch}\Delta_{{\overline\nu}}{}^\mu = \partial_\mu\Delta_{{\overline\nu}}{}^\mu + \Delta_{{\overline\nu}}{}^\mu \partial_\mu \log{\sqrt g} + \Delta_{{\overline\nu}}{}^\mu H_{\mu\rho}{}^\rho .%\label{divD}
\end{gather*}

\subsection*{Acknowledgements}
JM would like to thank the University of Sydney, Australia and ICMAT, Madrid, Spain for very kind hospitality while this work was completed. We would like to acknowledge conversations with A.~Ashmore, M.~Garcia-Fernandez, M.~Graffeo, C. Strickland-Constable, E. Svanes, G.~Williamson, and M.~Wolf.

\pdfbookmark[1]{References}{ref}
\LastPageEnding


\begin{thebibliography}{99}
\footnotesize\itemsep=0pt

\bibitem{Anderson:2014xha}
Anderson L.B., Gray J., Sharpe E., Algebroids, heterotic moduli spaces and the
 Strominger system, \href{https://doi.org/10.1007/JHEP07(2014)037}{\textit{J.~High Energy Phys.}} \textbf{2014} (20104),
 no.~7, 037, 40~pages, \href{https://arxiv.org/abs/1402.1532}{arXiv:1402.1532}.

\bibitem{Anguelova:2010ed}
Anguelova L., Quigley C., Sethi S., The leading quantum corrections to stringy
 {K}\"ahler potentials, \href{https://doi.org/10.1007/JHEP10(2010)065}{\textit{J.~High Energy Phys.}} \textbf{2010} (2010),
 no.~10, 065, 27~pages, \href{https://arxiv.org/abs/1007.4793}{arXiv:1007.4793}.

\bibitem{Ashmore:2018ybe}
Ashmore A., de~la Ossa X., Minasian R., Strickland-Constable C., Svanes E.E.,
 Finite deformations from a heterotic superpotential: holomorphic
 {C}hern--{S}imons and an {$L_\infty$} algebra, \href{https://doi.org/10.1007/jhep10(2018)179}{\textit{J.~High Energy Phys.}}
 \textbf{2018} (2018), no.~10, 179, 59~pages, \href{https://arxiv.org/abs/1806.08367}{arXiv:1806.08367}.

\bibitem{Candelas:2016usb}
Candelas P., de~la Ossa X., McOrist J., A metric for heterotic moduli,
 \href{https://doi.org/10.1007/s00220-017-2978-7}{\textit{Comm. Math. Phys.}} \textbf{356} (2017), 567--612,
 \href{https://arxiv.org/abs/1605.05256}{arXiv:1605.05256}.

\bibitem{Candelas:2018lib}
Candelas P., de~la Ossa X., McOrist J., Sisca R., The universal geometry of
 heterotic vacua, \href{https://doi.org/10.1007/jhep02(2019)038}{\textit{J.~High Energy Phys.}} \textbf{2019} (2019), no.~2,
 038, 46~pages, \href{https://arxiv.org/abs/1810.00879}{arXiv:1810.00879}.

\bibitem{Candelas:1990pi}
Candelas P., de~la Ossa X.C., Moduli space of {C}alabi--{Y}au manifolds,
 \href{https://doi.org/10.1016/0550-3213(91)90122-E}{\textit{Nuclear Phys.~B}} \textbf{355} (1991), 455--481.

\bibitem{delaOssa:2014msa}
de~la Ossa X., Svanes E.E., Connections, field redefinitions and heterotic
 supergravity, \href{https://doi.org/10.1007/jhep12(2014)008}{\textit{J.~High Energy Phys.}} \textbf{2014} (2014), no.~12,
 008, 27~pages, \href{https://arxiv.org/abs/1409.3347}{arXiv:1409.3347}.

\bibitem{delaOssa:2014cia}
de~la Ossa X., Svanes E.E., Holomorphic bundles and the moduli space of {$N=1$}
 supersymmetric heterotic compactifications, \href{https://doi.org/10.1007/JHEP10(2014)123}{\textit{J.~High Energy Phys.}}
 \textbf{2014} (2014), no.~10, 123, 55~pages, \href{https://arxiv.org/abs/1402.1725}{arXiv:1402.1725}.

\bibitem{Donagi:2011va}
Donagi R., Guffin J., Katz S., Sharpe E., Physical aspects of quantum sheaf
 cohomology for deformations of tangent bundles of toric varieties,
 \href{https://doi.org/10.4310/ATMP.2013.v17.n6.a2}{\textit{Adv. Theor. Math. Phys.}} \textbf{17} (2013), 1255--1301,
 \href{https://arxiv.org/abs/1110.3752}{arXiv:1110.3752}.

\bibitem{Donagi:2011uz}
Donagi R., Guffin J., Katz S., Sharpe E., A mathematical theory of quantum
 sheaf cohomology, \href{https://doi.org/10.4310/AJM.2014.v18.n3.a1}{\textit{Asian~J. Math.}} \textbf{18} (2014), 387--417,
 \href{https://arxiv.org/abs/1110.3751}{arXiv:1110.3751}.

\bibitem{Garcia-Fernandez:2015hja}
Garcia-Fernandez M., Rubio R., Tipler C., Infinitesimal moduli for the
 {S}trominger system and {K}illing spinors in generalized geometry,
 \href{https://doi.org/10.1007/s00208-016-1463-5}{\textit{Math. Ann.}} \textbf{369} (2017), 539--595, \href{https://arxiv.org/abs/1503.07562}{arXiv:1503.07562}.

\bibitem{Garcia-Fernandez:2018ypt}
Garcia-Fernandez M., Rubio R., Tipler C., Holomorphic string algebroids,
 \href{https://doi.org/10.1090/tran/8149}{\textit{Trans. Amer. Math. Soc.}} \textbf{373} (2020), 7347--7382,
 \href{https://arxiv.org/abs/1807.10329}{arXiv:1807.10329}.

\bibitem{Gillard:2003jh}
Gillard J., Papadopoulos G., Tsimpis D., Anomaly, fluxes and {$(2,0)$}
 heterotic-string compactifications, \href{https://doi.org/10.1088/1126-6708/2003/06/035}{\textit{J.~High Energy Phys.}}
 \textbf{2003} (2003), no.~6, 035, 25~pages, \href{https://arxiv.org/abs/hep-th/0304126}{arXiv:hep-th/0304126}.

\bibitem{Itoh:1988}
Itoh M., Geometry of anti-self-dual connections and {K}uranishi map,
 \href{https://doi.org/10.2969/jmsj/04010009}{\textit{J.~Math. Soc. Japan}} \textbf{40} (1988), 9--33.

\bibitem{Ivanov:2009rh}
Ivanov S., Heterotic supersymmetry, anomaly cancellation and equations of
 motion, \href{https://doi.org/10.1016/j.physletb.2010.01.050}{\textit{Phys. Lett.~B}} \textbf{685} (2010), 190--196,
 \href{https://arxiv.org/abs/0908.292}{arXiv:0908.292}.

\bibitem{Kobayashi:1987}
Kobayashi S., Differential geometry of complex vector bundles,
 \textit{Publications of the Mathematical Society of Japan}, Vol.~15,
 \href{https://doi.org/10.1515/9781400858682}{Princeton University Press}, Princeton, NJ, 1987.

\bibitem{Kodaira:1981cx}
Kodaira K., Complex manifolds and deformation of complex structures, \textit{Classics
 in Mathematics}, \href{https://doi.org/10.1007/b138372}{Springer-Verlag}, Berlin, 2005.

\bibitem{McOrist:2010ae}
McOrist J., The revival of {$(0,2)$} sigma models, \href{https://doi.org/10.1142/S0217751X11051366}{\textit{Internat.~J. Modern
 Phys.~A}} \textbf{26} (2011), 1--41, \href{https://arxiv.org/abs/1010.4667}{arXiv:1010.4667}.

\bibitem{McOrist:2016cfl}
McOrist J., On the effective field theory of heterotic vacua, \href{https://doi.org/10.1007/s11005-017-1025-0}{\textit{Lett.
 Math. Phys.}} \textbf{108} (2018), 1031--1081, \href{https://arxiv.org/abs/1606.05221}{arXiv:1606.05221}.

\bibitem{McOrist:2007kp}
McOrist J., Melnikov I.V., Half-twisted correlators from the {C}oulomb branch,
 \href{https://doi.org/10.1088/1126-6708/2008/04/071}{\textit{J.~High Energy Phys.}} \textbf{2008} (2008), no.~4, 071, 19~pages,
 \href{https://arxiv.org/abs/0712.3272}{arXiv:0712.3272}.

\bibitem{McOrist:2008ji}
McOrist J., Melnikov I.V., Summing the instantons in half-twisted linear sigma
 models, \href{https://doi.org/10.1088/1126-6708/2009/02/026}{\textit{J.~High Energy Phys.}} \textbf{2009} (2009), no.~2, 026,
 61~pages, \href{https://arxiv.org/abs/0810.0012}{arXiv:0810.0012}.

\bibitem{McOrist:2011bn}
McOrist J., Melnikov I.V., Old issues and linear sigma models, \href{https://doi.org/10.4310/ATMP.2012.v16.n1.a6}{\textit{Adv.
 Theor. Math. Phys.}} \textbf{16} (2012), 251--288, \href{https://arxiv.org/abs/1103.1322}{arXiv:1103.1322}.

\bibitem{Melnikov:2019tpl}
Melnikov I.V., An introduction to two-dimensional quantum field theory with
 {$(0,2)$} supersymmetry, \textit{Lecture Notes in Physics}, Vol.~951,
 \href{https://doi.org/10.1007/978-3-030-05085-6}{Springer}, Cham, 2019.

\bibitem{Melnikov:2012hk}
Melnikov I.V., Sethi S., Sharpe E., Recent developments in {$(0,2)$} mirror
 symmetry, \href{https://doi.org/10.3842/SIGMA.2012.068}{\textit{SIGMA}} \textbf{8} (2012), 068, 28~pages,
 \href{https://arxiv.org/abs/1209.1134}{arXiv:1209.1134}.

\bibitem{Melnikov:2011ez}
Melnikov I.V., Sharpe E., On marginal deformations of {$(0,2)$} non-linear
 sigma models, \href{https://doi.org/10.1016/j.physletb.2011.10.055}{\textit{Phys. Lett.~B}} \textbf{705} (2011), 529--534,
 \href{https://arxiv.org/abs/1110.1886}{arXiv:1110.1886}.

\bibitem{Nakahara:2003nw}
Nakahara M., Geometry, topology and physics, 2nd~ed., \textit{Graduate Student Series
 in Physics}, \href{https://doi.org/10.1201/9781420056945}{Institute of Physics}, Bristol, 2003.

\bibitem{Strominger:1990pd}
Strominger A., Special geometry, \href{https://doi.org/10.1007/BF02096559}{\textit{Comm. Math. Phys.}} \textbf{133}
 (1990), 163--180.

\bibitem{Witten:1985bz}
Witten E., New issues in manifolds of {${\rm SU}(3)$} holonomy, \href{https://doi.org/10.1016/0550-3213(86)90202-6}{\textit{Nuclear
 Phys.~B}} \textbf{268} (1986), 79--112.

\bibitem{Witten:1986kg}
Witten L., Witten E., Large radius expansion of superstring compactifications,
 \href{https://doi.org/10.1016/0550-3213(87)90249-5}{\textit{Nuclear Phys.~B}} \textbf{281} (1987), 109--126.

\end{thebibliography}
\end{document}